\documentclass[aps,prl,superscriptaddress,twocolumn]{revtex4-2}

\usepackage{amssymb}
\usepackage{graphicx}
\usepackage{dcolumn}
\usepackage{bm}
\usepackage{amsmath}
\usepackage[normalem]{ulem}
\usepackage{textcomp}
\usepackage{float}
\usepackage[usenames]{color}

\usepackage[unicode=true,bookmarks=true,bookmarksnumbered=false,bookmarksopen=false,breaklinks=false,pdfborder={0 0 1},backref=false,colorlinks=true]{hyperref}

\hypersetup{linkcolor=magenta,urlcolor=blue,citecolor=blue,pdfstartview={FitH},hyperfootnotes=false,unicode=true}

\begin{document}
\title{Statistics-governed dynamical scaling in interacting anyonic chains}

\author{Xu-Chen Yang}
\affiliation{College of Science, National University of Defense Technology, Changsha 410073, P. R. China}
\affiliation{Hunan Key Laboratory of Extreme Matter and Applications, National University of Defense Technology, Changsha 410073, P. R.  China}
\affiliation{Hunan Research Center of the Basic Discipline for Physical States, National University of Defense Technology, Changsha 410073,  P. R. China}

\author{Botao Wang}
\affiliation{Center for Nonlinear Phenomena and Complex Systems, Universit\'e Libre de Bruxelles, Brussels, Belgium}
\affiliation{International Solvay Institutes, Brussels, Belgium}

\author{Jianpeng Liu}
\affiliation{College of Science, National University of Defense Technology, Changsha 410073, P. R. China}
\affiliation{Hunan Key Laboratory of Extreme Matter and Applications, National University of Defense Technology, Changsha 410073, P. R.  China}
\affiliation{Hunan Research Center of the Basic Discipline for Physical States, National University of Defense Technology, Changsha 410073,  P. R. China}

\author{Bing Yang}
\affiliation{State Key Laboratory of Quantum Functional Materials, Department of Physics and Guangdong Basic Research Center of Excellence for Quantum Science, Southern University of Science and Technology, Shenzhen 518055, P. R. China}

\author{Jianmin Yuan}
\affiliation{Institute of Atomic and Molecular Physics, Jilin University, Changchun 130012, P. R. China}
\affiliation{College of Science, National University of Defense Technology, Changsha 410073, P. R. China}

\author{Yongqiang Li}
\email{li\_yq@nudt.edu.cn}
\affiliation{College of Science, National University of Defense Technology, Changsha 410073, P. R. China}
\affiliation{Hunan Key Laboratory of Extreme Matter and Applications, National University of Defense Technology, Changsha 410073, P. R.  China}
\affiliation{Hunan Research Center of the Basic Discipline for Physical States, National University of Defense Technology, Changsha 410073,  P. R. China}

\begin{abstract}
Particle statistics impose fundamental constraints on nonequilibrium quantum dynamics, yet it remains an open question whether anyonic statistics can lead to emergent dynamical scaling beyond the conventional Bose-Fermi paradigm.
Here we investigate the far-from-equilibrium many-body relaxation of anyons in a one-dimensional lattice, uncovering a statistics-governed, robust scaling behavior that deviates from standard Bose–Fermi limits.
Based on large-scale numerical simulations and scaling analysis, we find that in the weakly interacting regime, anyonic statistics leads to emergent superdiffusive scaling in particle transport, while the entanglement entropy remains ballistic and is essentially insensitive to exchange statistics.
The anomalous dynamics can be interpreted intuitively from the statistical-phase-induced quantum interference that suppresses coherent holon-doublon propagation; in contrast, the entanglement growth is dominated by its configurational component, which maintains ballistic spreading regardless of the statistical phase.
Our results establish anyonic statistics as a distinct source of universal nonequilibrium dynamics beyond bosons and fermions, with direct relevance to current quantum simulation experiments.

\end{abstract}

\date{\today}

\maketitle

{\it Introduction---} Anyonic statistics provides a fundamental extension of quantum many-body physics beyond the conventional Bose-Fermi framework~\cite{leinaas77,goldin81,wilczek82,khare98,greiter24}. Its physical relevance has been established through the observation of exotic quantum phenomena, including the fractional quantum Hall effect~\cite{halperin84,arovas84,bartolomei20,nakamura20} and topological spin liquids~\cite{coldea01,kitaev06,semeghini21}, and has stimulated extensive interest in topological quantum computation~\cite{kitaev03,bravyi06,nayak08,google23}. 
While anyonic statistics are intrinsically associated with two-dimensional systems, anyonic behavior can also emerge in one dimension (1D). For instance, fractional exclusion statistics were first revealed through the generalized Pauli principle in the Haldane-Shastry model~\cite{haldane88,shastry88,haldane91}. This has triggered extensive studies of 1D anyonic models both in the continuum~\cite{kundu99,harshman20,bonkhoff21} and on discrete lattices~\cite{keilmann11,greschner15,nagies24,wang25}, leading to hallmark ground-state phenomena such as asymmetric momentum distributions~\cite{hao08,hao09,keilmann11,tang15}, Friedel oscillations~\cite{strater16}, and statistics-induced phase transitions~\cite{greschner15,zhang17,bonkhoff25}. Experimentally, recent advances in ultracold atoms have brought the engineering of anyonic properties within reach~\cite{frolian22,chisholm22,kwan24,dhar25,bakkali26}, making it both timely and essential to investigate the role of anyonic statistics in the nonequilibrium dynamics~\cite{delcampo08,wang14,wright14,piroli17,liu18,chen25} of quantum many-body systems.

Far-from-equilibrium dynamics in bosonic and fermionic settings has attracted considerable attention in recent years~\cite{kinoshita06,trotzky12,gring12,choi16,prufer18,erne18}.
These quantum many-body systems exhibit rich emergent dynamical scaling behavior, providing valuable insights into collective quantum phenomena~\cite{polkovnikov11,vasseur16,alessio16,essler16,ueda20,bertini21}.
Ballistic and diffusive transport have been observed experimentally in a broad range of quantum simulators~\cite{nichols19,jepsen20,joshi22,wienand24}. 
Beyond these conventional regimes, qualitatively new universality classes have also been identified, such as Kardar-Parisi-Zhang scaling in 1D quantum magnets~\cite{ljubotina19,das19}.
This behavior arises from the interplay of integrability and internal non-Abelian symmetries~\cite{gopalakrishnan19,nardis19,ye22} and has been observed in quantum gas microscopy~\cite{wei22} and programmable quantum circuits~\cite{rosenberg24}.
These developments raise a fundamental question: can particle statistics itself give rise to emergent scaling in far-from-equilibrium quantum dynamics beyond the Bose-Fermi paradigm?

To address this question, we investigate the many-body relaxation dynamics of a 1D anyon-Hubbard model (AHM) [Fig.~\ref{figure-1}(a)], starting from far-from-equilibrium initial states. 
We demonstrate that anyonic statistics can induce emergent dynamical scaling that goes beyond conventional bosonic and fermionic limits.
This effect is particularly prominent in the weak-interaction regime: while the density correlations in these conventional limits exhibit ballistic transport, those in anyonic systems undergo a distinct crossover to superdiffusive behavior as the statistical angle $\theta$ is varied [Fig.~\ref{figure-1}(b)].
The anomalous scaling can be understood in terms of quantum interference between exchange paths of the holon-doublon pair, which suppresses its coherent propagation.
While anyonic interference suppresses particle transport, the entanglement entropy remains robustly ballistic.
This contrast arises because the configuration entropy dominates the growth and propagates ballistically, whereas the number entropy is primarily constrained by the statistical phase.
Finally, we show that this anomalous dynamical scaling is robust against onsite disorder and initially density configurations, offering a promising route for experimental verification in state-of-the-art quantum simulators.

\begin{figure}[t!]
\includegraphics[trim = 0mm 0mm 0mm 0mm, clip=true, width=0.475\textwidth]{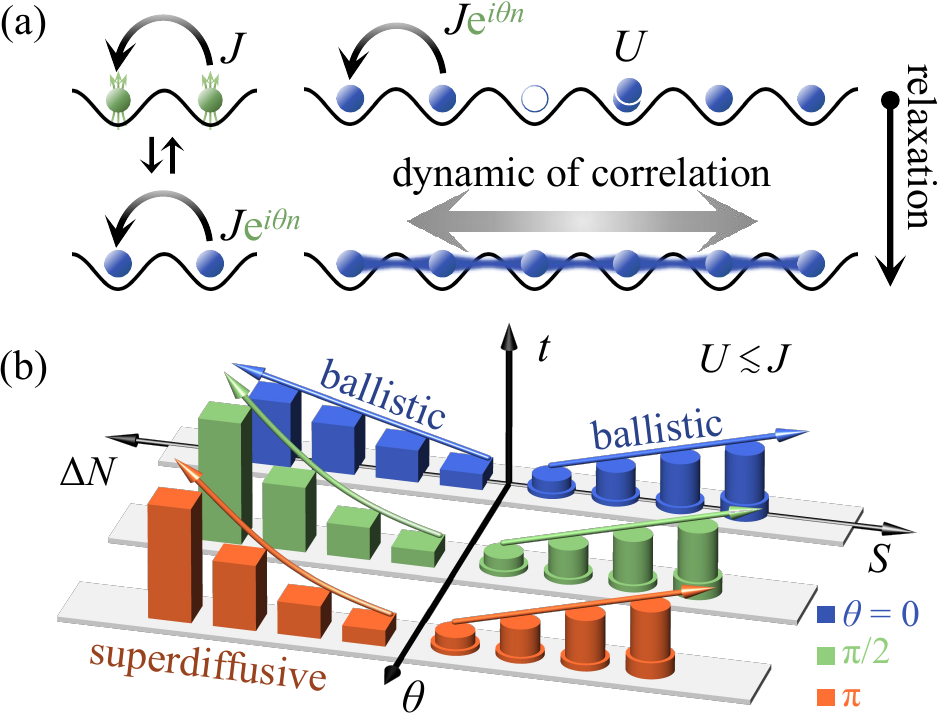}
\caption{Emergent far-from-equilibrium dynamical scaling in a 1D interacting anyonic system.
(a) Schematic of a 1D anyonic lattice with hopping $J$, mapped onto a BHM with density-dependent tunneling $Je^{i\theta n}$ and interactions $U$, where $n$ and $\theta$ denote the local density and statistical angle, respectively. A quantum quench from an out-of-equilibrium state triggers the spreading of spatial correlations and entanglement throughout the system.
(b) Distinct scaling of the half-chain particle-number fluctuations $\Delta N$ and the von Neumann entanglement entropy $S$. Both quantities are shown as functions of statistical angle $\theta$ and time $t$: for $U\leq J$, $\Delta N$ crosses over from ballistic to superdiffusive as $\theta$ increases from $0$ to $\pi$, while $S$ remains ballistic.} 
\label{figure-1}
\end{figure}

{\it Model and method---}We consider interacting anyons confined in a 1D optical lattice of $L$ sites, described by the AHM~\cite{keilmann11}
\begin{align}
\label{eq:Hama}
\hat{H}_\mathrm{AHM} =& -J\sum_{j=1}^{L-1}\left(\hat{a}_j^\dagger\hat{a}_{j+1}+\text{h.c.}\right)+\frac{U}{2}\sum_{j=1}^L\hat{n}_j(\hat{n}_j-1),
\end{align}
where $J$ denotes the nearest-neighbor hopping amplitude, $U$ onsite interaction strength, and $\hat{n}_j=\hat{a}_j^\dagger\hat{a}_j$ the anyon number operator on site $j$. The anyonic creation and annihilation operators $\hat{a}_j^\dagger$, $\hat{a}_j$ obey generalized commutation relations ~\cite{kundu99,batchelor06},
\begin{subequations}\label{eq:comm}
\begin{align}
\hat{a}_j\hat{a}_k^\dagger-e^{-i\theta\text{sgn}(j-k)}\hat{a}_k^\dagger \hat{a}_j=& \delta_{jk}, \label{eq:comm_a} \\
\hat{a}_j\hat{a}_k-e^{i\theta\text{sgn}(j-k)}\hat{a}_k\hat{a}_j=&0.
\label{eq:comm_b}
\end{align}
\end{subequations}
An experimentally relevant route to simulate interacting anyons is to map the AHM onto a Bose-Hubbard model (BHM) with a density-dependent Peierls phase~\cite{wilczek90,clark18,gorg19,schweizer19}, as illustrated in the top panel of Fig.~\ref{figure-1}(a). The mapped Hamiltonian reads~\cite{kwan24,bakkali26}
\begin{align}
\label{eq:Ham}
\hat{H} =& -J\sum_{j=1}^{L-1}\left(\hat{b}_j^\dagger e^{i\theta\hat{n}_j}\hat{b}_{j+1}+\text{h.c.}\right)+\frac{U}{2}\sum_{j=1}^L\hat{n}_j(\hat{n}_j-1),
\end{align}
where $\hat{b}_j^\dagger$, $\hat{b}_j$ are bosonic operators related to the anyons via a generalized Jordan-Wigner transformation, $\hat{a}_k=\hat{b}_ke^{i\theta\sum_{j<k}\hat{n}_j}$.
The statistical angle $\theta$ continuously interpolates between bosonic and pseudofermionic limits, reshaping the ground-state momentum distribution from a sharp condensate peak to a nearly flat fermionic profile~\cite{hao08,hao09,keilmann11,tang15} and giving rise to rich anyonic dynamics, including dynamical fermionization~\cite{delcampo08,piroli17}, asymmetric transport~\cite{liu18,chen25}, and two-particle correlations~\cite{wang14}. Here we focus on the emergent scaling governing the far-from-equilibrium dynamics of interacting many-body anyonic systems.

The system is initialized in a far-from-equilibrium product state $\lvert \psi_0 \rangle$, and its time evolution
$\left|\psi(t)\right\rangle=e^{-i\hat{H}t}\left|\psi_0\right\rangle$ 
is computed using the time-dependent variational principle~\cite{haegeman11} as implemented in TeNPy~\cite{tenpy24}.
To characterize density correlation spreading, we consider the two-particle correlation transport distance (CTD)~\cite{rispoli19,duenas25},
\begin{equation}
\label{eq:l}
l(t)=\sum_{d=1}^{L-1} d \, \left\langle\big| C_{k,k+d}(t) \big|\right\rangle_k,
\end{equation}
where $C_{j,k}=\langle \hat{n}_j \hat{n}_k \rangle - \langle \hat{n}_j \rangle \langle \hat{n}_k \rangle$ 
is the density-density correlation function, and $\langle\cdots\rangle_k$ denotes averaging over $k$. The CTD quantifies the mean correlation distance during nonequilibrium relaxation. It has been used to probe many-body localization in disordered BHM~\cite{rispoli19} and to characterize ballistic spreading of density correlations in standard BHM~\cite{duenas25}. We further examine the half-chain density fluctuations,
\begin{align}
\label{eq:dN}
\Delta N(t)=\sqrt{\langle\hat{N}^2\rangle(t)-\langle\hat{N}\rangle^2(t)},
\end{align}
where $\hat{N}=\sum_{j=1}^{L/2}\hat{n}_j$ denotes the total particle number operator of the left half-chain. This quantity provides a probe of nonequilibrium particle transport and relaxation in lattice Hubbard models~\cite{wienand24,bhakuni24,kwon26}. 
To quantify the spreading of quantum information, we compute the von Neumann entanglement entropy,
\begin{align}
S(t) = -\mathrm{Tr}\big[\rho_A(t)\log\rho_A(t)\big],
\end{align}
where $\rho_A(t)=\mathrm{Tr}_B[\lvert\psi(t)\rangle\langle\psi(t)\rvert]$ is the reduced density matrix of the left half-chain $A$, obtained by tracing out its complement $B$.
To extract the dynamical scaling of the observable dynamics, we rescale time according to $t \to t / L^z$ and define 
\begin{subequations}
\begin{align}
\Delta\widetilde{N}(t)&=\Delta N(t/L^z)/\Delta N_\text{sat},\label{eq:scaling_a}\\
\widetilde{S}(t)&=S(t/L^z)/S_\text{sat}.\label{eq:scaling_b}
\end{align}
\end{subequations}
where $\Delta N_{\mathrm{sat}}$ and $S_{\mathrm{sat}}$ denote the long-time saturation values. Here, $z$ denotes the dynamical exponent characterizing different scaling regimes of the dynamics. The dynamics is ballistic for $z=1$, diffusive for $z=2$, and superdiffusive for $1<z<2$~\cite{edwards82,vicsek84,family85,kardar86,fujimoto20}.

\begin{figure}[t!]
\includegraphics[trim = 0mm 0mm 0mm 0mm, clip=true, width=0.475\textwidth]{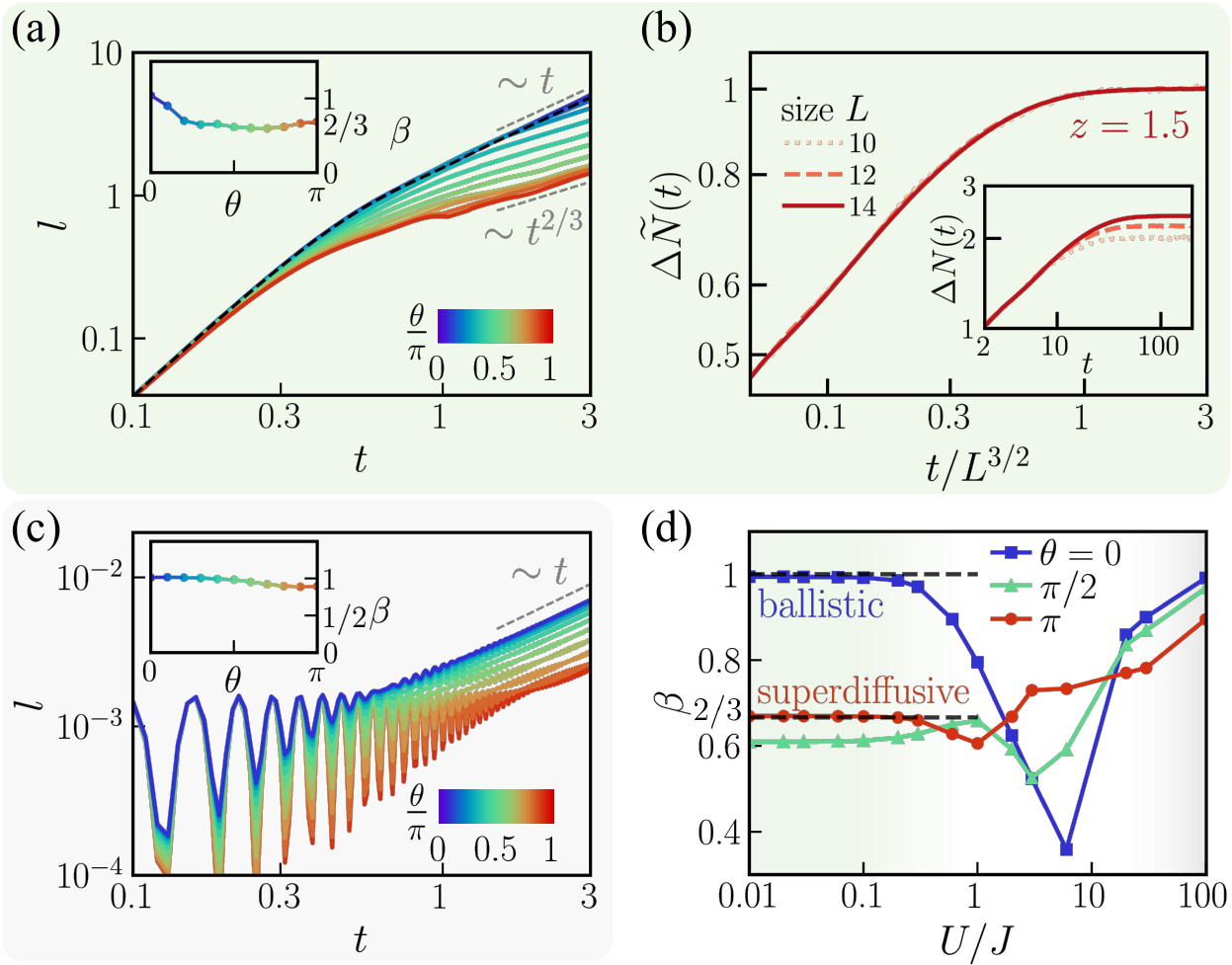}
\caption{Dynamical scaling of the correlation distance and particle-number fluctuations. (a)(b) In the noninteracting regime ($U=0$), (a) time evolution of the correlation distance $l(t)$ for statistical angles $\theta\in[0,\pi]$, revealing a crossover from ballistic growth ($\beta=1$) to superdiffusive scaling ($\beta=2/3$) (see inset); (b) half-chain particle-number fluctuations $\Delta \widetilde{N}(t)$, exhibiting a scaling collapse with dynamical exponent $z=3/2$ (inset: unscaled data for $\theta=\pi$).
(c) Time evolution of $l(t)$ in the strongly interacting regime ($U/J=100$) for $\theta\in[0,\pi]$, exhibiting an approximately ballistic scaling with $z\approx 1$. (d) Scaling exponent $\beta$ extracted from $l(t)$ over a broad range of interaction strengths $U/J=0.01$--$100$ at statistical angles $\theta=0$, $\pi/2$, and $\pi$, demonstrating a universal scaling regime for $U\leq J$. Numerical results of $l(t)$ are obtained for system size $L=100$ with bond dimensions up to $1500$.}
\label{figure-2}
\end{figure}

{\it Anomalous scaling of density correlation---}
We first investigate the far-from-equilibrium dynamics of density correlations in the anyonic chain in the noninteracting limit.
We find that tuning the statistical angle $\theta$ drives a crossover from ballistic to superdiffusive spreading, in contrast to the ballistic dynamics of integrable bosonic and fermionic chains.  
This anomalous behavior is revealed by the CTD following a quench from the unit-filled product state $|\psi_0\rangle = |\cdots111\cdots\rangle$.
As shown in Fig.~\ref{figure-2}(a), following a short $\theta$-independent period, the characteristic length $l(t)$ exhibits robust power-law scaling, $l(t)\sim t^\beta$.
The growth exponent $\beta$ decreases monotonically with increasing $\theta$, from $\beta=1$ in the bosonic limit ($\theta=0$) to $\beta=2/3$ at $\theta=\pi$ [inset of Fig.~\ref{figure-2}(a)]. 
In a 1D system, the corresponding dynamical exponent satisfies $z=1/\beta$~\cite{SM}, establishing superdiffusive dynamics ($1<z<2$) induced by anyonic statistics, distinct from the bosonic and fermionic limits ($z=1$)~\cite{fujimoto22,duenas25,SM}. 
In finite systems, boundary effects can affect the scaling behavior. 
Here we find that a lattice size of $L=100$ already suffices to capture the anomalous scaling~\cite{SM}. 
This scaling is independently confirmed by half-chain density fluctuations $\Delta N(t)$, which collapse onto a universal curve when time is rescaled as $t/L^z$, yielding $z=3/2$ at $\theta=\pi$ [Fig.~\ref{figure-2}(b)].
Although larger systems cannot be evolved long enough for $\Delta N(t)$ to reach the $L$-dependent saturation plateau, a small system of $L=14$ readily saturates, allowing accurate extraction of $z$ (inset of Fig.~\ref{figure-2}(b),~\cite{kwon26}).

We next examine the interplay between anyonic statistics and interactions by analyzing the CTD dynamics across a broad range of interaction strengths $U$. 
In the weakly interacting regime, the scaling exponent $\beta$ remains robust. 
As summarized in Fig.~\ref{figure-2}(d), the dynamics at $\theta=\pi$ is superdiffusive with $\beta \approx 2/3$ ($U\leq J$), while the bosonic limit ($\theta=0$) remains ballistic. 
By contrast, in the strong-interaction regime, the growth of $l(t)$ is nearly ballistic for all $\theta$, with $\beta\approx 1$, as shown in Fig.~\ref{figure-2}(c). This reflects the suppression of particle-exchange processes by strong interactions, rendering the effect of anyonic statistics negligible and recovering conventional ballistic dynamics.

The anomalous scaling is found to be robust against disorder and initial-state variations. We study $l(t)$ in the presence of finite onsite disorder and for different initial states, such as $|\cdots110211\cdots\rangle$ and the density wave $|\cdots0202\cdots\rangle$~\cite{SM}.
In all cases, we find that the growth exponent $\beta$ remains largely unchanged, demonstrating that the anomalous scaling is intrinsically determined by anyonic statistics.

\begin{figure}[t!]
\includegraphics[trim = 0mm 0mm 0mm 0mm, clip=true, width=0.475\textwidth]{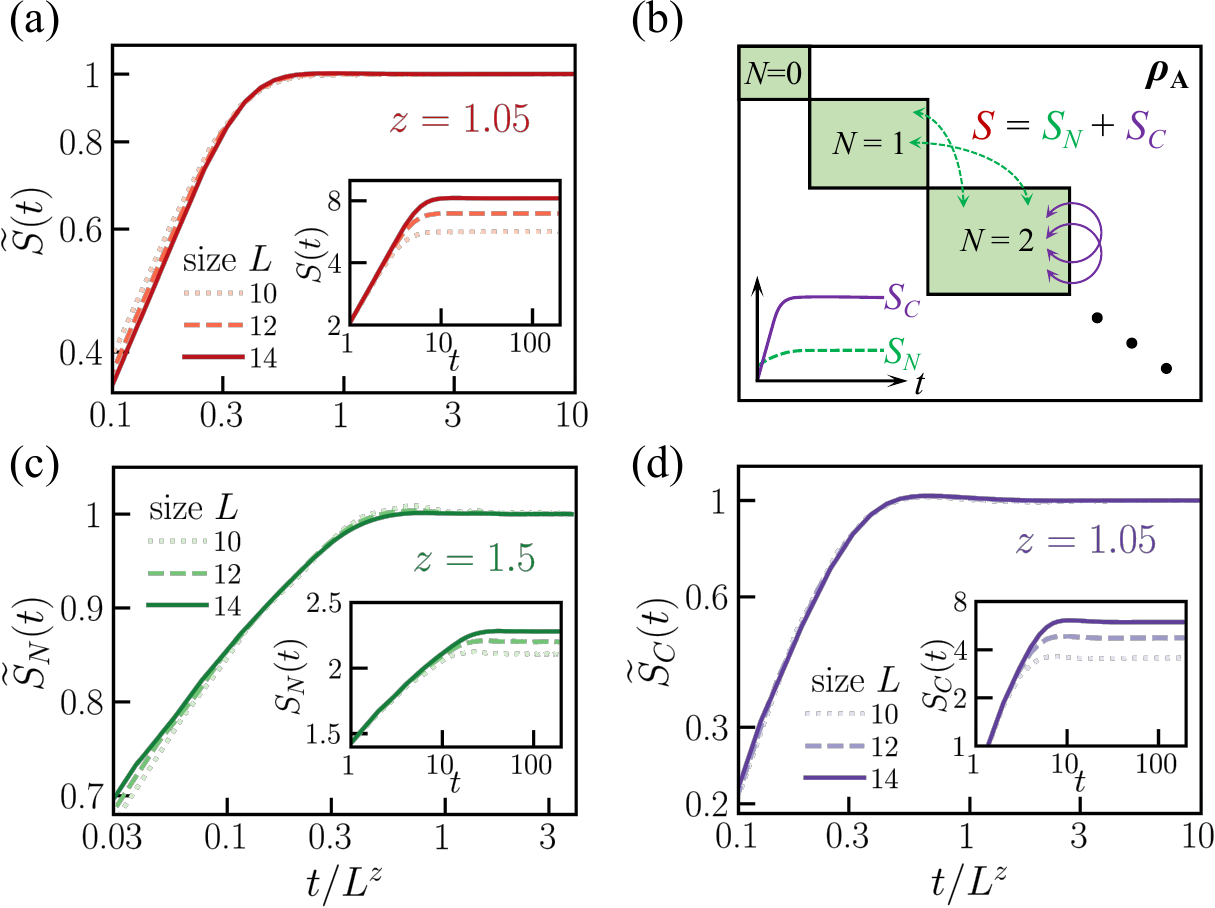}
\caption{Dynamical scaling of entanglement entropy in the relaxation dynamics of the anyonic chain.
(a) Rescaled von Neumann entropy $\widetilde{S}(t)$ as a function of time $t$ for $\theta=\pi/2$, with the inset showing the unrescaled data. The nearly ballistic scaling with $z\simeq1$ demonstrates a clear separation between entanglement propagation and particle transport [Fig.~\ref{figure-2}(d)].
(b) Schematic illustration of the reduced density matrix $\rho_A$, where the total entanglement entropy $S$ is decomposed into the number entropy $S_N$, associated with particle-number fluctuations between sectors, and the configuration entropy $S_C$, arising from superpositions within each particle-number sector. (c) Number entropy $\widetilde{S}_N(t)$ and (d) configuration entropy $\widetilde{S}_C(t)$ for $\theta=\pi/2$. Throughout the evolution, the entanglement growth is dominated by $S_C$ (insets). Results are obtained for system sizes $L=10$--$14$ at interaction strength $U/J=0.2$, using bond dimensions up to 4000.
}
\label{figure-3}
\end{figure}

{\it Ballistic dynamical scaling in entropy---}
We now turn to the dynamics of the entanglement entropy ${S}(t)$. 
While information propagation typically follows the same ballistic scaling as particle transport in the noninteracting bosonic~\cite{bhakuni24} and fermionic limits~\cite{PhysRevLett.127.090601}, we find that in anyonic chains, the entropy remains ballistic and essentially independent of the statistical angle $\theta$, despite the significant modification of the particle dynamics.
Focusing on an anyonic case with $\theta=\pi/2$ in the weakly interacting regime ($U/J=0.2$) and starting from the product state $|\psi_0\rangle$, $S(t)$ as expected grows from zero to an $L$-dependent saturation value, as shown in the inset of Fig.~\ref{figure-3}(a). Remarkably, data for different system sizes collapse when time is rescaled as $t\rightarrow t/L^{z}$, yielding $z\approx1.05$ (see Fig.~\ref{figure-3}(a) and Ref.~\cite{SM}). This ballistic growth contrasts with the anomalous scaling of density fluctuations $\Delta N(t)$ [Fig.~\ref{figure-2}(d), $z\approx1.62$], highlighting the different dynamical scales associated with information and particle spreading. Unlike localized systems where both particle and information spreading are simultaneously suppressed by disorder~\cite{PhysRevB.77.064426,bardarson12,huse14}, the phenomenon predicted here is driven by anyonic statistics.

To understand this behavior, we decompose $S(t)$ into number and configurational contributions. This decomposition was introduced as a symmetry-resolved entanglement diagnostic to distinguish many-body localization from thermalization in disordered systems~\cite{lukin19,emmanouilidis20,luitz20,chenj25}.
Particle-number conservation enforces the half-chain reduced density matrix to be block diagonal, leading to 
\begin{align}
S(t) = S_N(t) + S_C(t).
\end{align}
Here, the number entropy $S_N(t)$ quantifies inter-sector entanglement, while the configuration entropy $S_C(t)$ captures entanglement within each fixed-$N$ sector~\cite{SM}, as shown in Fig.~\ref{figure-3}(b). 
We find that while $\widetilde{S}_N(t)$ exhibits an anomalous dynamical exponent $z = 1.5$ [Fig.~\ref{figure-3}(c)], $\widetilde{S}_C(t)$ grows ballistically with $z = 1.05$ [Fig.~\ref{figure-3}(d)]. Throughout the evolution, $S_C(t) \gg S_N(t)$, and thus the total entropy inherits ballistic scaling from the configurational degrees of freedom. This accounts for the faster spread of entanglement compared to density fluctuations [Fig.~\ref{figure-2}(b)(d)].

\begin{figure*}[t!]
\includegraphics[trim = 0mm 0mm 0mm 0mm, clip=true, width=0.75\textwidth]{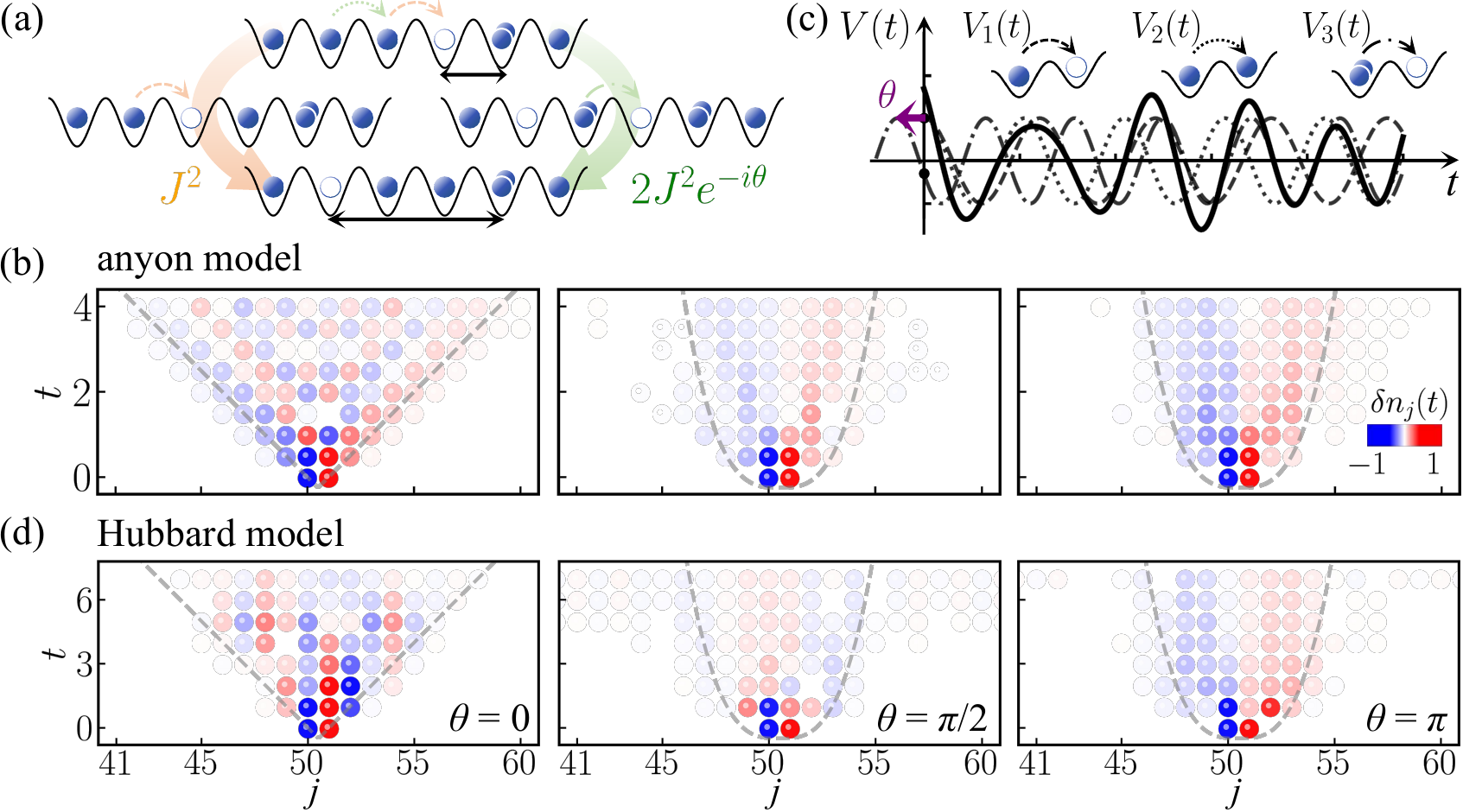}
\caption{Diagram of interference picture of holon-doublon pairs under anyonic statistics and experimental realization. (a) 
Schematic illustration of holon-doublon propagation starting from an initially adjacent holon-doublon pair ($d=1$) to separation $d=3$ via two distinct hopping paths. Paths along these trajectories acquire $\theta$-dependent phases, leading to destructive interference that suppresses the propagation of the pairs.
(b) Density deviation dynamics $\delta n_j(t)$ governed by Eq.~(\ref{eq:Ham}), after a central-site excitation for the bosonic (left, $\theta=0$), intermediate (middle, $\theta=\pi/2$), and pseudofermionic (right, $\theta=\pi$) cases. Dashed lines indicate the propagation front of the pair. (c) Floquet protocol realized in Ref.~\cite{kwan24}: three-tone modulation $V = \sum V_i$ (solid) resonantly restores tunneling in a tilted lattice; $V_{1,2,3}$ (dashed, dotted, dot-dashed) selectively drive singly-to-empty, singly-to-singly, and doubly-to-empty transitions. The phase offset $\theta$ of $V_3$ implements the density-dependent Peierls phase. (d) Density deviation dynamics $\delta n_j(t)$ for the setup in (c). The agreement demonstrates the robust realization of anyonic dynamics in the many-body regime. Time $t$ is in units of $1/J$ in (b) and $1/J'$ in (d), with $J'$ the effective hopping strength~\cite{SM}.
Results are obtained for a system of size $L=100$ in the weakly interacting regime ($U/J=0.2$).}
\label{figure-4}
\end{figure*}

{\it Quantum interference in anyonic lattice---}
We can understand the anomalous dynamical scaling by appealing to the propagation of a holon-doublon pair on a unit-filled background, which reflects the spreading of density correlations~\cite{cheneau12}.
We find that propagation of the pair is suppressed due to quantum interference between hopping paths, revealing that the anomalous correlation scaling arises from anyonic statistics.
Fig.~\ref{figure-4}(a) schematically illustrates holon-doublon propagation, where different hopping paths acquire $\theta$-dependent phases. Starting from a holon-doublon pair contributing density correlations at distance $d=1$, we consider its propagation to distance $d=3$, where $d$ denotes the holon-doublon separation. A representative contribution involves two interfering paths: (i) two adjacent particles tunnel sequentially to the right (orange arrows) without exchanging, yielding an amplitude $J^2$ in the noninteracting limit; (ii) the left particle tunnels twice (green arrows), exchanging with its neighbor, acquiring an anyonic phase factor $2J^2 e^{-i\theta}$. The resulting amplitude arises from the coherent sum of such processes. In the bosonic limit ($\theta=0$), the two paths interfere constructively, enhancing the propagation of the holon-doublon pair. In contrast, at $\theta=\pi$, the interference becomes destructive, strongly suppressing the propagation. 
Additional exchange processes give rise to more complex interference patterns, which can suppress correlation spreading over a broad range of $\theta>0$~\cite{SM}.

To quantify the effect of anyonic statistics on holon-doublon propagation, we compute the time evolution of the density deviation $\delta n_j(t) = n_j(t) - \bar{n}$ in the weakly interacting regime, starting from an initial state $|\cdots110211\cdots\rangle$ containing a central holon-doublon pair. Here, $\bar{n}$ denotes the average density. As shown in Fig.~\ref{figure-4}(b), the pair spreads ballistically in the bosonic limit ($\theta=0$). By contrast, the propagation is strongly suppressed for intermediate ($\theta=\pi/2$) and pseudofermionic ($\theta=\pi$) statistics. 

{\it Experimental accessibility---}
The identified many-body features are within reach of current ultracold atom experiments, leveraging the Floquet protocol recently demonstrated in Ref.~\cite{kwan24,bakkali26}.
While this protocol was originally demonstrated in a two-particle setting, the underlying interference mechanism remains robust in the many-body regime.
As detailed in Fig.~\ref{figure-4}(c), a three-tone modulation of the lattice depth resonantly restores tunneling otherwise suppressed by the static tilt. Crucially, the phase offset $\theta$ of the $V_3$ component (dot-dashed line) implements the density-dependent Peierls phase. The numerical results in Fig.~\ref{figure-4}(d) exhibit agreement between the full Floquet-driven dynamics and the direct evolution of Eq.~(\ref{eq:Ham}). 
Note that the difference at $\theta=\pi/2$ originates from the opposite statistical angle acquired during the Floquet realization in a many-body setting~\cite{SM}. 
Despite such a discrepancy, the overall dynamics exhibits qualitative agreement. This indicates that the additional complexities inherent in the Floquet scheme, such as higher-order tunneling, would not alter the essential physics, thereby establishing the predicted phenomena as an experimentally accessible feature.
Furthermore, single-site-resolved quantum-gas microscopy provides direct access to both correlation functions~\cite{endres11,cheneau12,schweigler17,kwon26} and entanglement entropy~\cite{islam15,kaufman16,lukin19}, offering a powerful experimental probe into nonequilibrium dynamics governed by anyonic statistics.

{\it Conclusion---}
In summary, we have demonstrated that anyonic statistics fundamentally modifies far-from-equilibrium quantum dynamics, driving a robust dynamical scaling that extends beyond the conventional bosonic and fermionic limits.
In contrast to previous studies focusing on asymmetric transport~\cite{liu18,chen25,kwan24,bakkali26} or dynamical fermionization~\cite{delcampo08,piroli17}, our work identifies a distinct regime of scaling dynamics: while particle transport enters a superdiffusive regime, the entanglement entropy growth remains ballistic. 
This anomalous scaling is connected to quantum interference effects induced by the anyonic statistical phase, which suppress the propagation of coherent holon-doublon pairs.
The scaling exponents exhibit a weak dependence on initial states within the explored regime, indicating that this statistics-induced dynamical scaling is a characteristic feature of anyonic systems.
Our results provide new insights into nonequilibrium physics governed by anyonic statistics and advance the control of transport and entanglement in synthetic quantum matter, realizable with current experimental techniques~\cite{kwan24,bakkali26}.
Future work may clarify the universality classes underlying the predicted scaling using high-order cumulants in the long-time dynamics of larger quantum systems~\cite{wei22,rosenberg24}. It is an interesting question to explore richer nonequilibrium dynamics and entanglement growth beyond one dimension~\cite{nayak08,mazza18,rossini19,greiter24,helluin25}.

We acknowledge useful discussions with Yinghai Wu, and Felix Helluin. This work is supported by the National Natural Science Foundation of China (Grants No. 12374252, No. 12074431, No. 12174130, No. 12304076, and No. 12404332), and the Excellent Youth Foundation of Hunan Scientific Committee under Grant No. 2021JJ10044. Numerical calculations were performed on the TianHe-1A cluster of the National Supercomputer Center at Tianjin, and HPC resources on the ChinaHPC.

\onecolumngrid
\clearpage

\setcounter{figure}{0}
\setcounter{table}{0}
\setcounter{equation}{0}
\renewcommand{\thefigure}{S\arabic{figure}}
\renewcommand{\thetable}{S\arabic{table}}
\renewcommand{\theequation}{S\arabic{equation}}

\begin{center}
\Large\bf
Supplemental Material for\\
``Statistics-governed dynamical scaling in interacting anyonic chains''
\end{center}

\maketitle

\renewcommand{\theenumi}{\Roman{enumi}}  %
\renewcommand{\theenumii}{\Alph{enumii}} %
\renewcommand{\theenumiii}{\arabic{enumiii}} %

\section*{Contents}
\addcontentsline{toc}{chapter}{Table of Contents}
\begin{enumerate}
    \item \hyperref[chapter:CTD]{Correlation transport distance in a free bosonic chain \hfill 1}
    \item \hyperref[chapter:numerical]{Numerical convergence \hfill 2}
    \item \hyperref[chapter:beta]{Extraction of growth exponent $\beta$ \hfill 3}
    \item \hyperref[chapter:initial]{Robustness of anomalous dynamical scaling \hfill 4}
    \item \hyperref[chapter:entropy]{Entanglement entropy in anyonic chain \hfill 5}
    \begin{enumerate}
        \item \hyperref[section:NCentropy]{Number and Configuration Entropy \hfill 5}
        \item \hyperref[section:boseentropy]{Entropy in bosonic chain \hfill 5}
        \item \hyperref[section:z]{Extraction of dynamical exponent $z$ \hfill 5}
    \end{enumerate}
    \item \hyperref[chapter:pairs]{Additional propagation processes of holon-doublon pairs \hfill 6}
    \item \hyperref[chapter:exp]{Experimental realization of the anyon hubbard model \hfill 7}
\end{enumerate}

\vspace{1em}


\section{I. Correlation transport distance in a free bosonic chain}
\label{chapter:CTD}

In this section, we derive the correlation transport distance (CTD) for the standard Bose-Hubbard model (BHM) in the noninteracting regime. The system is described by the tight-binding Hamiltonian
\begin{align}
\hat{H}_0 = -J \sum_j \left(\hat{b}_j^\dagger \hat{b}_{j+1} + \mathrm{h.c.} \right),
\end{align}
where $J$ denotes the nearest-neighbor hopping amplitude. The time evolution of the bosonic annihilation operator is governed by
\begin{align}
i \frac{d}{dt} \hat{b}_j(t) = [ \hat{b}_j(t), \hat{H}_0 ]
= -J \left[ \hat{b}_{j-1}(t) + \hat{b}_{j+1}(t) \right].
\end{align}
Introducing the Fourier transform
\begin{align}
\hat{b}_k = \frac{1}{\sqrt{L}} \sum_{j=1}^L e^{-i k j} \hat{b}_j ,
\end{align}
the equation of motion in momentum space becomes
\begin{align}
i \frac{d}{dt} \hat{b}_k(t)
= -J \left( e^{-i k} + e^{i k} \right) \hat{b}_k(t)
= -2J \cos(k)\, \hat{b}_k(t),
\end{align}
which yields the solution
\begin{align}
\hat{b}_k(t) = e^{2 i J t \cos(k)} \hat{b}_k(0).
\end{align}
Transforming back to real space, the annihilation and creation operators evolve as
\begin{align}
\hat{b}_j(t) &= \sum_{m=1}^L V_{j-m}(t)\, \hat{b}_m(0), \nonumber\\
\hat{b}_j^\dagger(t) &= \sum_{n=1}^L V_{j-n}^*(t)\, \hat{b}_n^\dagger(0),
\end{align}
where the single-particle propagator is
\begin{align}
V_l(t) = \frac{1}{L} \sum_k e^{i k l} e^{2 i J t \cos(k)} .
\end{align}

We consider an initial state with integer filling $\bar{n}$ per site. In this case, the four-point correlator satisfies~\cite{barmettler12}
\begin{align}
\langle \bar{n} | \hat{b}_p^\dagger \hat{b}_q \hat{b}_r^\dagger \hat{b}_s | \bar{n} \rangle
= \bar{n}^2 \delta_{p,q} \delta_{r,s}
+ \bar{n} (\bar{n}+1) (1-\delta_{p,q}) \delta_{p,s} \delta_{q,r}.
\end{align}
Using this identity, the density-density correlation function can be written as
\begin{align}
\langle \hat{n}_j \hat{n}_{j+d} \rangle
&= \sum_{m,n,p,q}
V_{j-m}^* V_{j-n} V_{j+d-p}^* V_{j+d-q}
\langle \hat{b}_m^\dagger \hat{b}_n \hat{b}_p^\dagger \hat{b}_q \rangle \nonumber\\
&= \bar{n}^2 - \bar{n} (\bar{n}+1)
\sum_m |V_{j-m}|^2 |V_{j+d-m}|^2 .
\end{align}
Subtracting the disconnected contribution, we obtain the connected density-density correlation
\begin{align}
C_d(t)
&= \langle \hat{n}_j \hat{n}_{j+d} \rangle
- \langle \hat{n}_j \rangle \langle \hat{n}_{j+d} \rangle \nonumber\\
&= -\bar{n} (\bar{n}+1)
\sum_m |V_{j-m}|^2 |V_{j+d-m}|^2 .
\end{align}
In the thermodynamic limit, the propagator reduces to Bessel functions,
$V_l(t) = i^l J_l(2Jt)$, yielding
\begin{align}
C_d(t)
= -\bar{n} (\bar{n}+1)
\sum_j
J_j^2\!\left( 2Jt \right)
J_{j+d}^2\!\left( 2Jt \right).
\end{align}

The two-particle CTD is defined as
\begin{align}
l(t)
= \sum_{d=1}^{L-1}
d \times \langle | C_{k,k+d}(t) | \rangle_k ,
\end{align}
where $\langle \cdots \rangle_k$ denotes averaging over all reference sites $k$.
This analytical result explains the ballistic scaling behavior shown in Fig.~2(a) of the main text.

\section{II. Numerical convergence}
\label{chapter:numerical}

\begin{figure}[t!]
    \centering
    \includegraphics[width=0.9\textwidth]{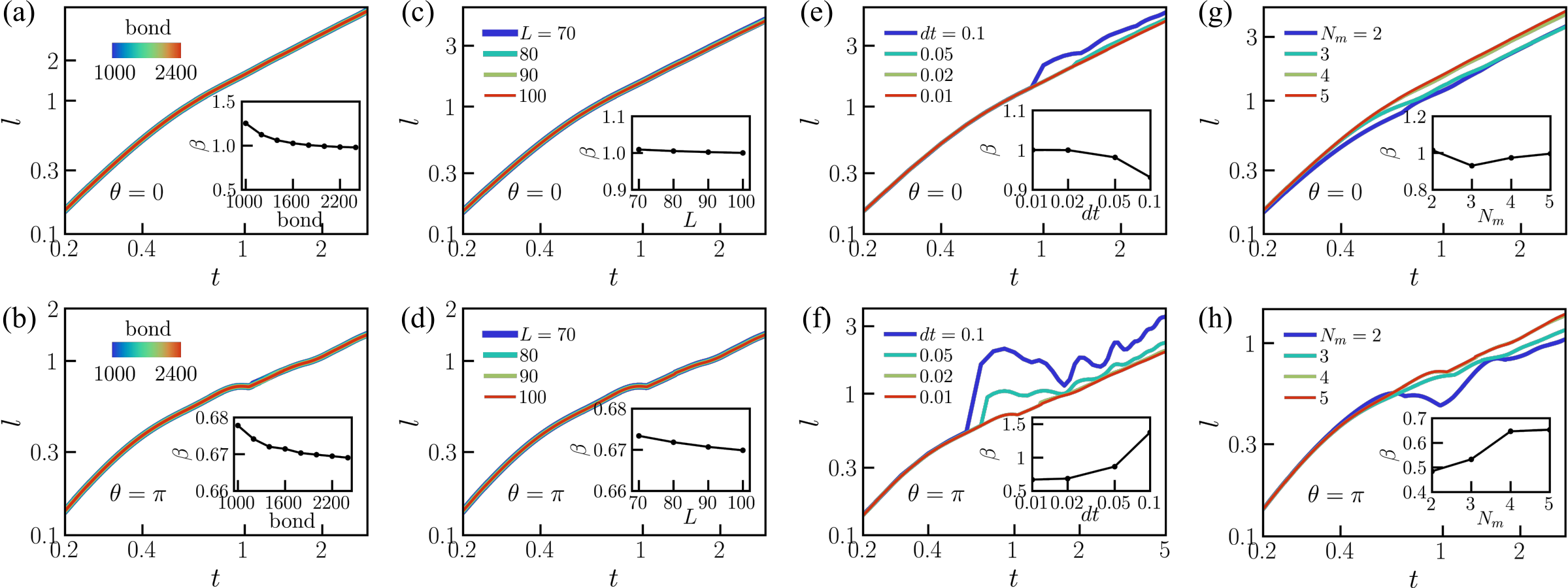}
    \caption{Dependence of the correlation transport distance $l(t)$ on numerical control parameters: (a,b) bond dimension, (c,d) system size $L$, (e,f) time step $dt$, and (g,h) local occupation cutoff $N_m$. The parameters used in the main text are a bond dimension of 1500 for $\theta=0$ and 2000 for $\theta=\pi$, with $L=100$, $dt=0.01$, and $N_m=5$.}
    \label{fig:convergence}
\end{figure}

In this section, we present convergence tests for the numerical results on the dynamics of the anyonic Hubbard chain, obtained using the time-dependent variational principle algorithm. 
Since matrix-product-state simulations are inherently variational, it is essential to verify convergence with respect to the numerical control parameters. We perform systematic convergence checks by varying the bond dimension, system size $L$, time step $dt$, and local occupation cutoff $N_m$, as shown in Fig.~\ref{fig:convergence}. The results demonstrate that, for the model parameters considered in this work, the dependence on these parameters is weak and does not significantly affect the system dynamics.

\section{III. Extraction of growth exponent $\beta$}
\label{chapter:beta}

In this section, we describe the numerical procedure used to extract the dynamical exponent $\beta$ from the time evolution of the CTD $l(t)$. As discussed in the main text, $l(t)$ exhibits a steady growth regime at intermediate times, characterized by a power-law scaling $l(t) \sim t^{\beta}$. To determine $\beta$ reliably, we perform power-law fits to $l(t)$ within an intermediate-time window $[t_l,t_r]$, where neither short-time transients nor finite-size saturation effects dominate the dynamics. Specifically, we vary the left boundary $t_l$ of the fitting window while keeping the right boundary $t_r$ fixed, and monitor the convergence of the extracted exponent. The early-time regime is excluded from the fitting window because the dynamics in this regime is dominated by free processes, which do not reflect the emergent many-body scaling behavior. At late times, $l(t)$ gradually approaches a saturation value due to the finite system size, and data points in this regime are also excluded from the fit. Therefore, the right boundary of the fitting window is chosen sufficiently before the onset of saturation.

\begin{figure}[t!]
    \centering
    \includegraphics[width=0.9\textwidth]{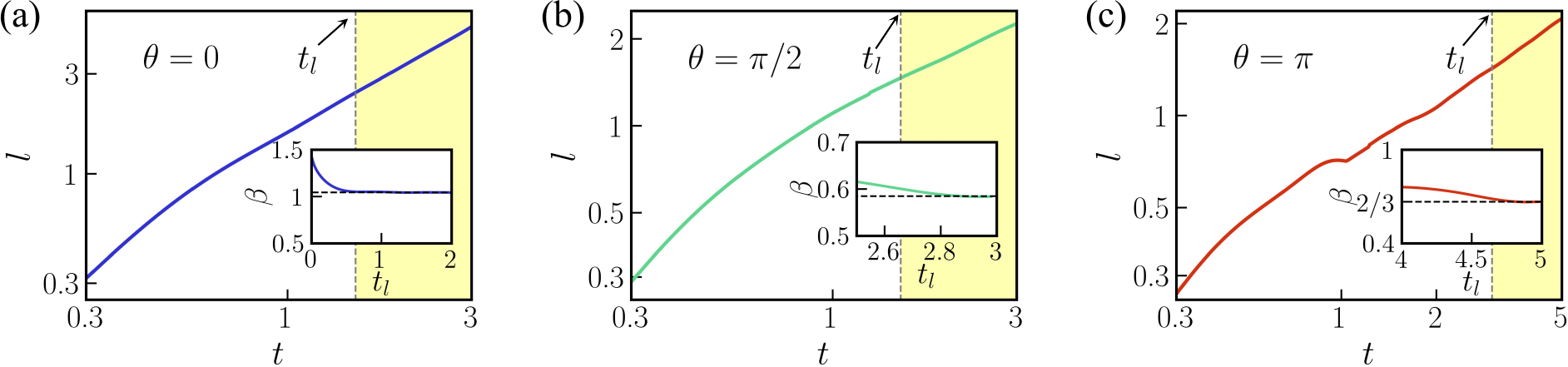}
    \caption{Convergence of the growth exponent $\beta$ extracted from the CTD $l(t)$ for system size $L=100$. The inset illustrates the growth exponent $\beta$ extracted from a power-law fit $l(t)\sim t^{\beta}$ within a fitting time window $[t_l,t_r]$ (highlighted in yellow). (a) Bosonic limit $\theta=0$ with $t_r=3$. (b) Anyonic regime $\theta=\pi/2$ with $t_r=3$. (c) Pseudofermionic limit $\theta=\pi$ with $t_r=5$.}
    \label{fig:beta}
\end{figure}

In practice, we take the right boundary of the fitting window to be $t_r = 3$ for smaller values of the statistical angle $\theta$, and $t_r = 5$ for larger $\theta$. Within these windows, the extracted exponent $\beta$ shows clear convergence as the left boundary $t_l$ approaches the deep intermediate-time regime. This convergence behavior demonstrates the robustness of the extracted $\beta$, as illustrated in Fig.~\ref{fig:beta}.

Here we clarify the relationship between the dynamical exponent $z$ and the growth exponent $\beta$ extracted from the CTD. In general, for a single-particle operator $\hat{O}$, the associated dynamical exponent satisfies the scaling relation $z=\alpha/\beta_O$, where $\alpha=1/2$ for one-dimensional (1D) transport. However, the CTD $l(t)$ considered here is constructed from a two-point density-density correlation function and therefore probes two-particle correlations. As a result, the growth exponent extracted from $l(t)$ satisfies $\beta = 2 \beta_O$, which leads to the relation $z = 1/\beta$ used in the main text.

\section{IV. Robustness of anomalous dynamical scaling}
\label{chapter:initial}

\begin{figure}[t!]
    \centering
    \includegraphics[width=0.7\textwidth]{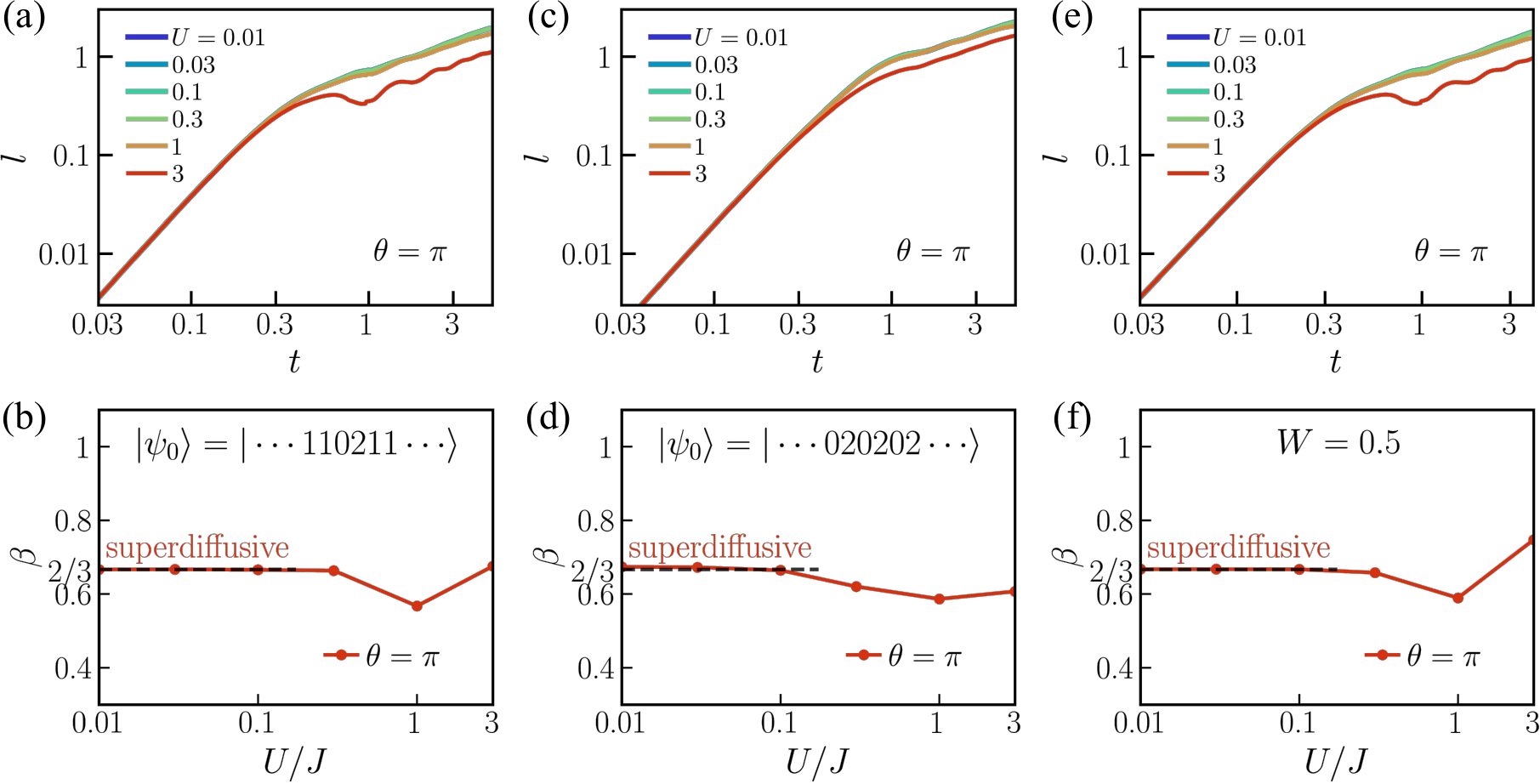}
    \caption{Robustness of the anomalous dynamical scaling. (a--d) Anomalous dynamical scaling starting from different initial states. (e,f) Robustness to the onsite disorder term $H_\mathrm{dis}$. For finite disorder strength $W=0.5$, $l(t)$ remains superdiffusive.}
    \label{fig:initial}
\end{figure}

To demonstrate that the anomalous dynamical scaling observed in the anyonic chain is robust and universal, we examine the dynamics starting from different initial states, specifically $|\cdots 110211 \cdots\rangle$ and $|\cdots 0202 \cdots\rangle$, as shown in Fig.~\ref{fig:initial}(a) and (c), respectively. Panels (a) and (c) show the time evolution of the CTD $l(t)$ under different interaction strengths, while panels (b) and (d) display the corresponding growth exponents $\beta$. In the weakly interacting regime, the anyonic chain exhibits superdiffusive scaling, indicating that the anomalous scaling does not require a homogeneous initial state.

To further test the robustness against disorder, we introduce an onsite disorder term
\begin{align}
\hat{H}_\mathrm{dis} = W\sum_j w_j \hat{n}_j,
\end{align}
where $w_j$ is drawn from a uniform distribution in the interval $[-1,1]$, and consider the total Hamiltonian $\hat{H}_\mathrm{tot} = \hat{H} + \hat{H}_\mathrm{dis}$. The resulting $l(t)$ averaged over 10 disorder samples with $W=0.5$ is shown in Fig.~\ref{fig:initial}(e), and the extracted growth exponent $\beta$ in Fig.~\ref{fig:initial}(f) indicates that the anomalous dynamical scaling persists in the weakly interacting regime even in the presence of finite disorder.


\section{V. Entanglement entropy in the anyonic chain}
\label{chapter:entropy}
\subsection{A. Number and Configuration Entropy}
\label{section:NCentropy}
For a subsystem $A$, the reduced density matrix $\rho_A$ can be block-diagonalized according to the particle number $N$ in $A$, $\rho_A = \bigoplus_N \rho_A^N$, where $\rho_A^N$ acts on the $N$-particle subspace $\mathcal{H}_A^N$. The probability to find $N$ particles in $A$ is
\begin{align}
p_N = \mathrm{Tr}\rho_A^N, \qquad \sum_N p_N = 1.
\end{align}
Normalized density matrices within each subspace are defined as
\begin{align}
\tilde{\rho}_A^N = \frac{\rho_A^N}{p_N}, \qquad \mathrm{Tr}\,\tilde{\rho}_A^N = 1.
\end{align}
The total von Neumann entropy of $A$ is
\begin{align}
S = -\mathrm{Tr}(\rho_A \log \rho_A) 
= -\sum_N \mathrm{Tr}(\rho_A^N \log \rho_A^N).
\end{align}
Substituting $\rho_A^N = p_N \tilde{\rho}_A^N$ and using $\log(p_N \tilde{\rho}_A^N) = (\log p_N)I + \log \tilde{\rho}_A^N$, we obtain
\begin{align}
-\mathrm{Tr}(\rho_A^N \log \rho_A^N) = -p_N \log p_N - p_N \mathrm{Tr}(\tilde{\rho}_A^N \log \tilde{\rho}_A^N).
\end{align}
Defining the conditional von Neumann entropy in the subspace $\mathcal{H}_A^N$ as
\begin{align}
S^N = -\mathrm{Tr}(\tilde{\rho}_A^N \log \tilde{\rho}_A^N),
\end{align}
we arrive at the standard decomposition
\begin{align}
S = S_N + S_C, \quad
S_N = -\sum_N p_N \log p_N, \quad
S_C = \sum_N p_N S^N.
\end{align}
Here, $S_N$ is the number entropy, reflecting particle-number fluctuations in $A$, and $S_C$ is the configuration entropy, quantifying quantum correlations within each fixed-$N$ subspace. This decomposition separates the total entanglement into contributions from particle-number fluctuations and configurational quantum correlations. The rescaled number entropy and rescaled configuration entropy are
\begin{subequations}
\begin{align}
\widetilde{S}_N(t)&=S_N(t/L^z)/S_{N,\text{sat}},\\
\widetilde{S}_C(t)&=S_C(t/L^z)/S_{C,\text{sat}},
\end{align}
\end{subequations}
where $S_{N,\text{sat}}$ and $S_{C,\text{sat}}$ are the long-time saturation values.

\subsection{B. Entropy in bosonic chain}
\label{section:boseentropy}
To further support our numerical results, we examine the entanglement entropy dynamics in the bosonic limit ($\theta=0$) to demonstrate its ballistic scaling behavior. We numerically simulate the evolution of the total entanglement entropy $S(t)$, number entropy $S_N(t)$, and configuration entropy $S_C(t)$ for different system sizes. We find that all contributions exhibit ballistic scaling, with the configuration entropy $S_C(t)$ remaining larger than the number entropy $S_N(t)$ throughout the evolution. These features are consistent with the behavior observed in the anyonic case ($\theta=\pi/2$) discussed in the main text.

\begin{figure}[t!]
    \centering
    \includegraphics[width=0.9\textwidth]{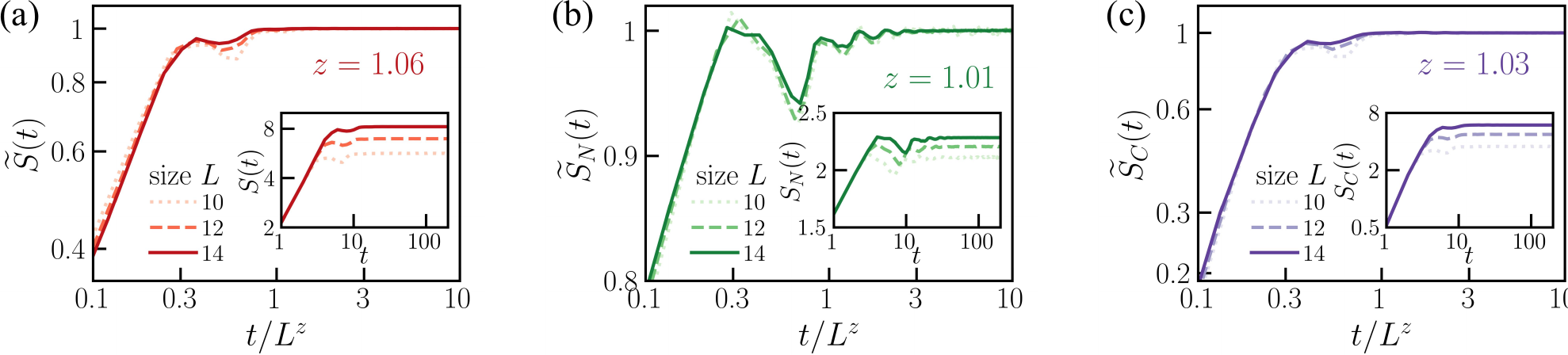}
    \caption{Dynamical scaling of entanglement entropy in the bosonic limit. (a) Scaled von Neumann entropy $\widetilde{S}(t)$, (b) number entropy $\widetilde{S}_N(t)$, (c) configuration entropy $\widetilde{S}_C(t)$ as a function of time $t$ for bosonic limit $\theta=0$, with the corresponding unscaled data shown in the insets. Numerical results are calculated for system sizes $L = 10$--$14$ with weak interaction strength $U/J = 0.2$.}
    \label{fig:entropy}
\end{figure}

\subsection{C. Extraction of dynamical exponent $z$}
\label{section:z}
\begin{figure}[t!]
    \centering
    \includegraphics[width=0.9\textwidth]{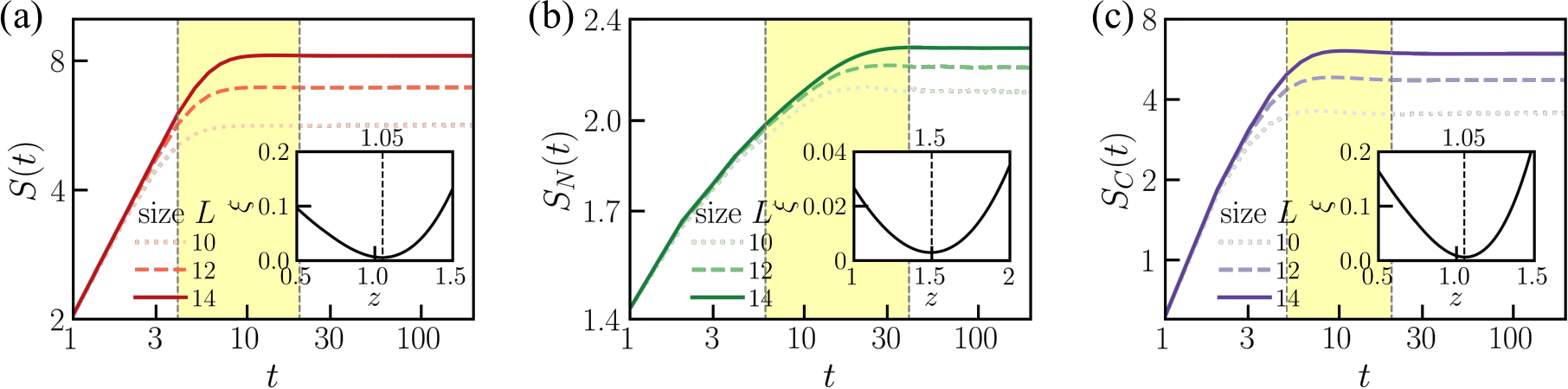}
    \caption{Extraction of the dynamical exponent $z$ from entanglement entropy. (a) Total entanglement entropy $S(t)$ for different system sizes $L$. (b) Number entropy $S_N(t)$. (c) Configuration entropy $S_C(t)$. The insets show the error function $\xi(z)$, with its minimum indicating the optimal dynamical exponent.}
    \label{fig:z}
\end{figure}

To illustrate the numerical extraction of the dynamical exponent $z$ from entanglement entropy, we show the evolution of $S(t)$ with $\theta=\pi/2$ and $U/J=0.2$ in Fig.~\ref{fig:z}. Entanglement growth for different system sizes $L$ is expected to follow the same scaling function. By rescaling time as $t \to t/L^z$ and normalizing the entropies as $\widetilde{S}(L,t)$, the saturation curves for different system sizes should collapse onto each other. We define an error function quantifying the deviation of each rescaled entropy from a reference size $L_\mathrm{ref}$ within a chosen time window $[t_1,t_2]$:
\begin{align}
\xi(z)=\sum_{L\neq L_\mathrm{ref}}\int_{t_1}^{t_2}\left|\widetilde{S}(L,t)-\widetilde{S}(L_\mathrm{ref},t)\right|dt.
\end{align}
The value of $z$ minimizing $\xi$ provides the optimal estimate of the dynamical exponent. Here we choose $L_\mathrm{ref}=14$.

Since early-time dynamics is dominated by free processes and does not reflect universal scaling, we focus on the late-time evolution approaching saturation. For the total entropy $S(t)$, we consider the time window $t \in [5,20]$, as shown in Fig.~\ref{fig:z}(a). The inset shows the error $\xi$ as a function of $z$, with a minimum at $z = 1.05$, where the rescaled entropies collapse onto a single curve, confirming ballistic scaling.

Applying the same procedure to the number entropy $S_N(t)$, using a later window $t \in [6,40]$ due to its slower saturation, yields a minimum at $z = 1.5$, as shown in Fig.~\ref{fig:z}(b), indicating superdiffusive behavior. For the configuration entropy $S_C(t)$, using a time window similar to that of $S(t)$, $t \in [5,20]$, we find $z=1.05$ in Fig.~\ref{fig:z}(c), consistent with ballistic scaling. These results show that the total entanglement growth is dominated by configuration entropy, while the number entropy reflects the anomalous particle transport.

\section{VI. Additional propagation processes of holon-doublon pairs}
\label{chapter:pairs}

\begin{figure}[t!]
    \centering
    \includegraphics[width=0.8\textwidth]{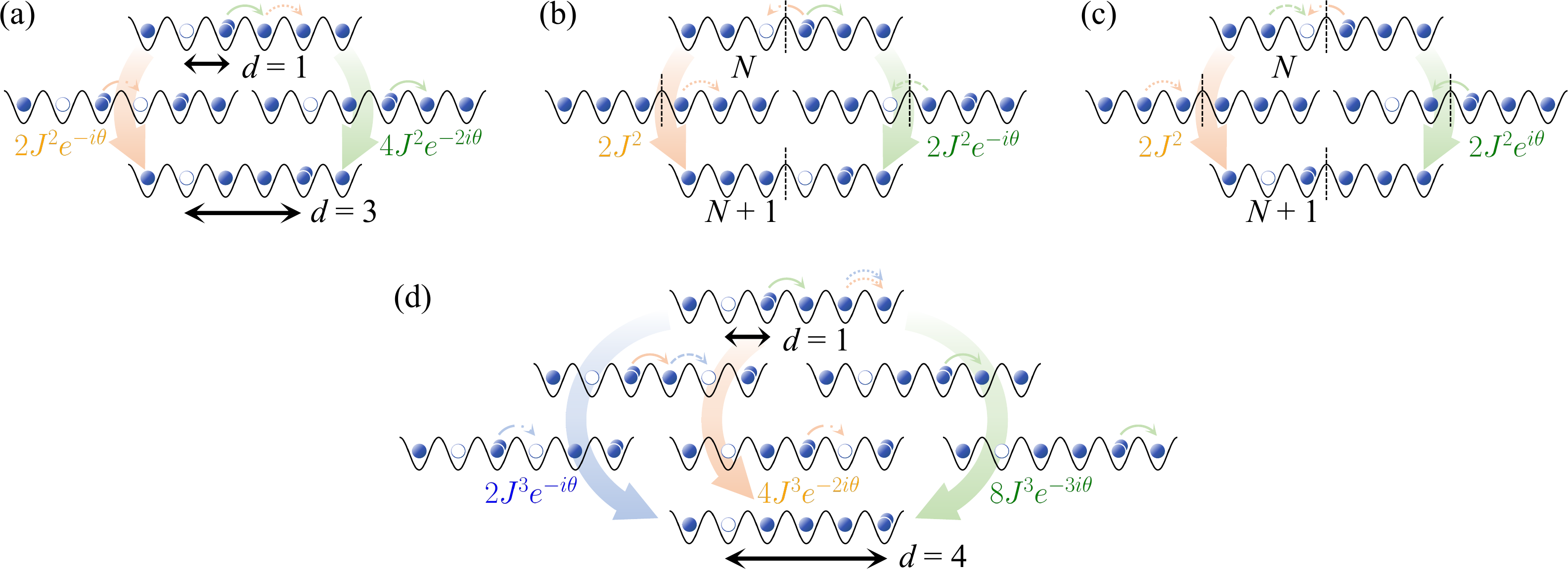}
    \caption{Interferometric processes for holon–doublon propagation. (a) Additional contribution to the increase of the separation from $d=1$ to $d=3$ via rightward motion of the doublon. Two tunneling paths interfere: sequential hopping of adjacent particles (orange) with amplitude $2J^2 e^{-i\theta}$ and hopping twice of a single particle (green) with amplitude $4J^2 e^{-i2\theta}$. (b,c) Coherent propagation of the holon–doublon pair to the right or left. Different hopping orders accumulate distinct anyonic exchange phases, leading to interference that suppresses coherent pair propagation  for $\theta>0$. (d) Representative processes contributing to $d=4$. Three typical paths (blue, orange, green) acquire amplitudes $2J^3 e^{-i\theta}$, $4J^3 e^{-i2\theta}$, and $8J^3 e^{-i3\theta}$.}
    \label{fig:pair}
\end{figure}

In the main text, we presented a representative example of the propagation of correlations from $d=1$ to $d=3$, corresponding to the leftward propagation of the holon. Here we provide the complementary process associated with the rightward propagation of the doublon in Fig.~\ref{fig:pair}(a). This contribution also involves two interfering paths: (i) two adjacent particles tunnel sequentially to the right (orange arrows), exchanging once and yielding an amplitude $2J^2 e^{-i\theta}$; (ii) the left particle tunnels twice (green arrows), exchanging twice with its neighbor and acquiring an anyonic phase $4J^2 e^{-i2\theta}$. The total amplitude arises from the coherent sum of these processes. In the bosonic limit ($\theta=0$), the two paths interfere constructively, enhancing the propagation of the holon-doublon pair, whereas in the fermionic limit ($\theta=\pi$), the interference becomes destructive, strongly suppressing the propagation.

The dynamics of the density deviation $\delta n_j(t)$ shown in the main text indicate that the site with $\delta n_j < 0$ remains on the left while the site with $\delta n_j > 0$ remains on the right. This implies that for $\theta>0$, the holon and doublon effectively propagate toward opposite sides of the chain. To understand this behavior, we analyze the motion of the holon-doublon pair. We consider cases where the pair propagates coherently to the right [Fig.~\ref{fig:pair}(b)] or to the left [Fig.~\ref{fig:pair}(c)]. For the rightward propagation [case (b)], the contribution again involves two interfering paths: (i) one particle of the doublon hops to the left, filling the holon site, followed by the other hopping to the right, yielding an amplitude $2J^2$; (ii) one particle of the doublon hops to the right first, then the other hops to the left, acquiring an anyonic phase $2J^2 e^{-i\theta}$. The hopping order determines the accumulated phase, and the total amplitude is given by the coherent superposition of these processes. Such coherent pair propagation is suppressed for $\theta>0$. Consequently, during the time evolution the holon-doublon pair tends either to remain localized or to separate toward opposite sides of the chain. As a result, the growth of particle-number fluctuations in the half-chain is suppressed for $\theta>0$, and the corresponding number entropy exhibits a strongly reduced growth rate, consistent with the main text.

To elucidate the role of the anyonic statistical angle in the coherent propagation of holon-doublon pairs at larger distances, we illustrate the representative processes contributing to a separation $d=4$, as shown in Fig.~\ref{fig:pair}(d). Three typical paths contribute to this process, each acquiring a distinct phase determined by the anyonic angle $\theta$, which controls the resulting interference pattern. (i) Three adjacent particles hop sequentially to the right (blue arrows), involving one exchange and yielding an amplitude $2J^3 e^{-i\theta}$; (ii) two particles separated by one site each hop once to the right, followed by a hop of one particle from the left doublon (orange arrows), resulting in an amplitude $4J^3 e^{-i2\theta}$; (iii) the leftmost particle hops three times sequentially from left to right (green arrows), acquiring an amplitude $8J^3 e^{-i3\theta}$. Additional hopping sequences are omitted for clarity. The total amplitude for propagation to $d=4$ arises from the coherent superposition of all contributing paths, each carrying a distinct exchange phase. 
As the propagation distance increases, the superposition of these multiple contributions generates increasingly intricate interference structures in $\theta$ space. 
Consequently, destructive interference can arise over a broad range of nonzero $\theta$, naturally accounting for the suppression of correlation spreading for general $\theta>0$.

\section{VII. Experimental realization of the anyon Hubbard model}
\label{chapter:exp}
\begin{figure}[t!]
    \centering
    \includegraphics[width=0.8\textwidth]{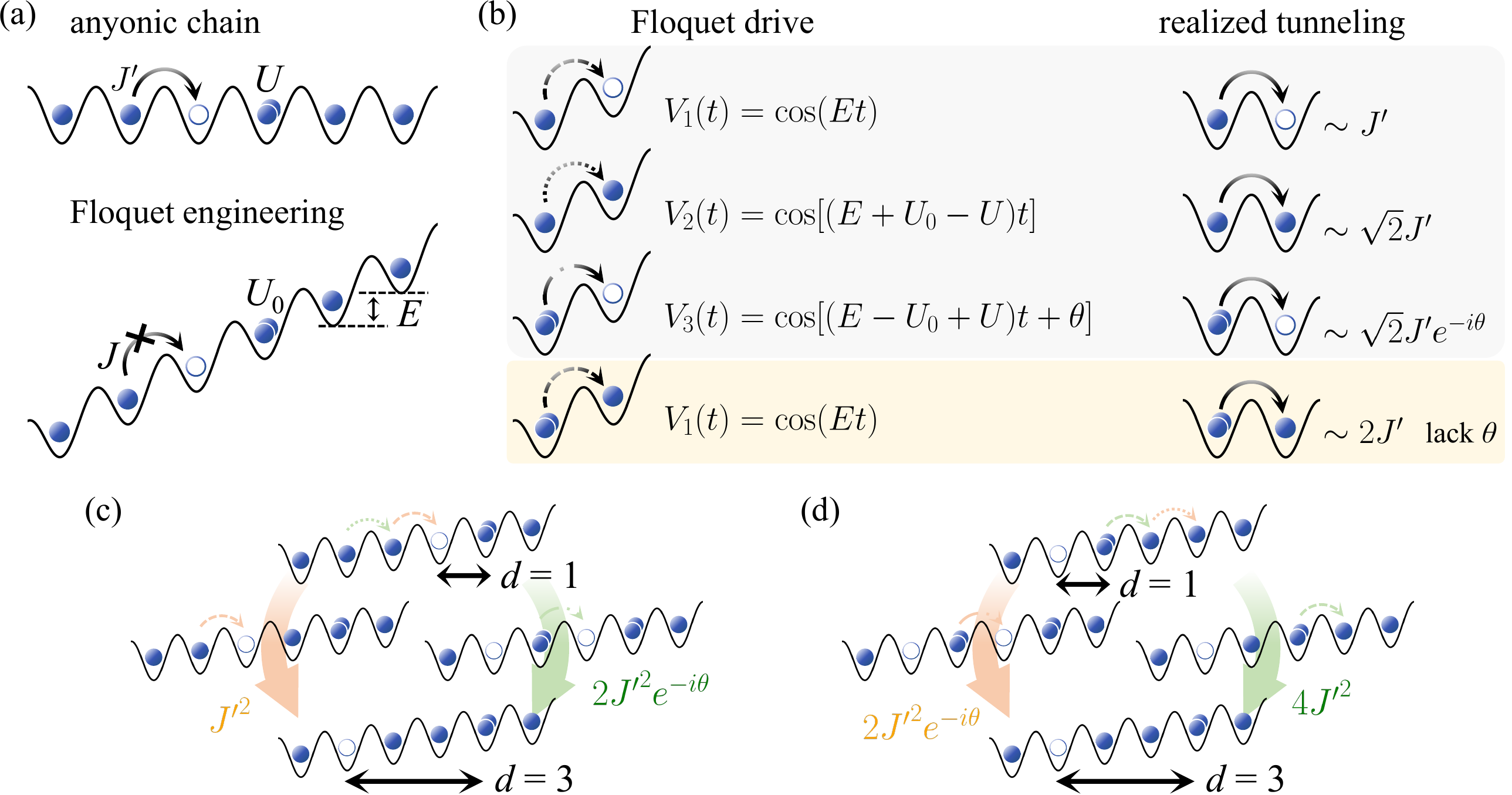}
    \caption{Floquet realization of the anyon Hubbard model. (a) Schematic of the anyonic chain and its implementation via Floquet engineering in a tilted Bose-Hubbard lattice. The tilt $E$ suppresses bare tunneling $J$, while periodic modulation restores resonant hopping. (b) Density-dependent tunneling processes addressed by the Floquet drive. The first three processes reproduce the desired tunneling amplitudes $J^\prime$, $\sqrt{2}J^\prime$, and $\sqrt{2}J^\prime e^{-i\theta}$ (gray block). At unit filling, the additional transition $|2,1\rangle\to|1,2\rangle$ is resonant with the same channel as $|1,0\rangle\to|0,1\rangle$ and therefore lacks the phase $\theta$ (orange block). (c) Interference between two processes for holon-doublon separation ($d=1\to3$): two particle sequential hopping (${J^\prime}^2$) and one particle hopping twice ($2{J^\prime}^2e^{-i\theta}$). (d) Similarly, two particle sequential hopping ($2{J^\prime}^2 e^{-i\theta}$) and one particle hopping twice ($4{J^\prime}^2$).}
    \label{fig:floquet}
\end{figure}

Following the experimental framework demonstrated in Ref.~\cite{kwan24a}, we discuss the realization of the anyon Hubbard model in an optical lattice. As shown in Fig.~\ref{fig:floquet}(a), the experimental implementation utilizes a tilted 1D Bose-Hubbard chain with interaction $U_0$, where a magnetic field gradient creates a static energy offset $E$ between adjacent sites. This potential tilt effectively suppresses the bare tunneling $J$. To restore particle tunneling with amplitude $J^\prime$, the lattice depth is modulated with resonant frequencies that compensate for the energy differences of specific hopping processes, a technique known as Floquet engineering.

This Floquet approach provides an effective route to realize density-dependent hopping phases, as illustrated in Fig.~\ref{fig:floquet}(b). The Hamiltonian is 
\begin{align}
\hat{H}^\prime(t) =& -J\sum_{j=1}^{L-1}\left(\hat{b}_j^\dagger\hat{b}_{j+1}+\text{h.c.}\right)+\frac{U_0}{2}\sum_{j=1}^L\hat{n}_j(\hat{n}_j-1)+E\sum_{j=1}^Lj\hat{n}_j+\sum_iV_i(t).
\end{align}
Defining the effective onsite interaction as $U$, the primary processes are: (i) $|1,0\rangle \to |0,1\rangle$, driven at frequency $\omega_1 = E$; (ii) $|1,1\rangle \to |0,2\rangle$, driven at $\omega_2 = E - (U_0 - U)$; (iii) $|2,0\rangle \to |1,1\rangle$, driven at $\omega_3 = E + (U_0 + U)$. To map these processes onto the anyon Hubbard model, an additional phase offset $\theta$ is imprinted on the drive for process (iii), while processes (i) and (ii) remain reference transitions [gray block in Fig.~\ref{fig:floquet}(b)]. This configuration is sufficient for low fillings; however, at unit filling, an additional process (iv) $|2,1\rangle \to |1,2\rangle$ emerges as an important additional contribution. Due to the energy offset $E$, this transition is unavoidably induced by the same Floquet channel as process (i) and lacks the required phase (orange block). 

We demonstrate that this imperfect higher-order term does not destroy the observation of holon-doublon propagation. As shown in Fig.~\ref{fig:floquet}(c), the expansion of a pair from distance $d=1$ to $d=3$ involves an interference between two major paths: (left) sequential particle hops that accumulate no phase, and (right) one particle hops twice, acquiring an anyonic phase $e^{-i\theta}$. The resulting $\theta$-dependent interference patterns remain robust, as process (iv) does not dominate the dynamics in this regime. Similarly, in Fig.~\ref{fig:floquet}(d), two processes contribute to pair expansion: (left) the sequential hopping process accumulates a phase $\theta$, whereas (right) twice hopping carries no phase in the present Floquet scheme. Although this differs from the ideal anyonic model, where the latter would acquire a phase $e^{2i\theta}$, the interference between the two paths remains sensitive to $\theta$. 

Our analysis confirms that the proposed Floquet scheme remains a powerful tool for observing the hallmarks of anyonic statistics, specifically the suppression of coherent transport and asymmetric expansion dynamics. The parameters used in the numerical simulation of Floquet engineering in the main text are $E/J = 30$, $U_0/J = 4$, $U/J = 0.2$, $K/J = 40$, with an effective tunneling rate $J^\prime/J \sim 0.5$.

\begin{thebibliography}{100}%
\makeatletter
\providecommand \@ifxundefined [1]{%
 \@ifx{#1\undefined}
}%
\providecommand \@ifnum [1]{%
 \ifnum #1\expandafter \@firstoftwo
 \else \expandafter \@secondoftwo
 \fi
}%
\providecommand \@ifx [1]{%
 \ifx #1\expandafter \@firstoftwo
 \else \expandafter \@secondoftwo
 \fi
}%
\providecommand \natexlab [1]{#1}%
\providecommand \enquote  [1]{``#1''}%
\providecommand \bibnamefont  [1]{#1}%
\providecommand \bibfnamefont [1]{#1}%
\providecommand \citenamefont [1]{#1}%
\providecommand \href@noop [0]{\@secondoftwo}%
\providecommand \href [0]{\begingroup \@sanitize@url \@href}%
\providecommand \@href[1]{\@@startlink{#1}\@@href}%
\providecommand \@@href[1]{\endgroup#1\@@endlink}%
\providecommand \@sanitize@url [0]{\catcode `\\12\catcode `\$12\catcode
  `\&12\catcode `\#12\catcode `\^12\catcode `\_12\catcode `\%12\relax}%
\providecommand \@@startlink[1]{}%
\providecommand \@@endlink[0]{}%
\providecommand \url  [0]{\begingroup\@sanitize@url \@url }%
\providecommand \@url [1]{\endgroup\@href {#1}{\urlprefix }}%
\providecommand \urlprefix  [0]{URL }%
\providecommand \Eprint [0]{\href }%
\providecommand \doibase [0]{https://doi.org/}%
\providecommand \selectlanguage [0]{\@gobble}%
\providecommand \bibinfo  [0]{\@secondoftwo}%
\providecommand \bibfield  [0]{\@secondoftwo}%
\providecommand \translation [1]{[#1]}%
\providecommand \BibitemOpen [0]{}%
\providecommand \bibitemStop [0]{}%
\providecommand \bibitemNoStop [0]{.\EOS\space}%
\providecommand \EOS [0]{\spacefactor3000\relax}%
\providecommand \BibitemShut  [1]{\csname bibitem#1\endcsname}%
\let\auto@bib@innerbib\@empty
\bibitem [{\citenamefont {Leinaas}\ and\ \citenamefont
  {Myrheim}(1977)}]{leinaas77}%
  \BibitemOpen
  \bibfield  {author} {\bibinfo {author} {\bibfnamefont {J.~M.}\ \bibnamefont
  {Leinaas}}\ and\ \bibinfo {author} {\bibfnamefont {J.}~\bibnamefont
  {Myrheim}},\ }\bibfield  {title} {\bibinfo {title} {On the theory of
  identical particles},\ }\href {https://doi.org/10.1007/BF02727953} {\bibfield
   {journal} {\bibinfo  {journal} {Il Nuovo Cimento B (1971-1996)}\ }\textbf
  {\bibinfo {volume} {37}},\ \bibinfo {pages} {1} (\bibinfo {year}
  {1977})}\BibitemShut {NoStop}%
\bibitem [{\citenamefont {Goldin}\ \emph {et~al.}(1981)\citenamefont {Goldin},
  \citenamefont {Menikoff},\ and\ \citenamefont {Sharp}}]{goldin81}%
  \BibitemOpen
  \bibfield  {author} {\bibinfo {author} {\bibfnamefont {G.~A.}\ \bibnamefont
  {Goldin}}, \bibinfo {author} {\bibfnamefont {R.}~\bibnamefont {Menikoff}},\
  and\ \bibinfo {author} {\bibfnamefont {D.~H.}\ \bibnamefont {Sharp}},\
  }\bibfield  {title} {\bibinfo {title} {Representations of a local current
  algebra in nonsimply connected space and the aharonov–bohm effect},\ }\href
  {https://doi.org/10.1063/1.525110} {\bibfield  {journal} {\bibinfo  {journal}
  {Journal of Mathematical Physics}\ }\textbf {\bibinfo {volume} {22}},\
  \bibinfo {pages} {1664} (\bibinfo {year} {1981})}\BibitemShut {NoStop}%
\bibitem [{\citenamefont {Wilczek}(1982)}]{wilczek82}%
  \BibitemOpen
  \bibfield  {author} {\bibinfo {author} {\bibfnamefont {F.}~\bibnamefont
  {Wilczek}},\ }\bibfield  {title} {\bibinfo {title} {Quantum mechanics of
  fractional-spin particles},\ }\href
  {https://doi.org/10.1103/PhysRevLett.49.957} {\bibfield  {journal} {\bibinfo
  {journal} {Phys. Rev. Lett.}\ }\textbf {\bibinfo {volume} {49}},\ \bibinfo
  {pages} {957} (\bibinfo {year} {1982})}\BibitemShut {NoStop}%
\bibitem [{\citenamefont {Khare}(1998)}]{khare98}%
  \BibitemOpen
  \bibfield  {author} {\bibinfo {author} {\bibfnamefont {A.}~\bibnamefont
  {Khare}},\ }\href {https://doi.org/10.1142/2988} {\emph {\bibinfo {title}
  {\textup{Fractional Statistics and Quantum Theory}}}}\ (\bibinfo  {publisher}
  {WORLD SCIENTIFIC},\ \bibinfo {year} {1998})\BibitemShut {NoStop}%
\bibitem [{\citenamefont {Greiter}\ and\ \citenamefont
  {Wilczek}(2024)}]{greiter24}%
  \BibitemOpen
  \bibfield  {author} {\bibinfo {author} {\bibfnamefont {M.}~\bibnamefont
  {Greiter}}\ and\ \bibinfo {author} {\bibfnamefont {F.}~\bibnamefont
  {Wilczek}},\ }\bibfield  {title} {\bibinfo {title} {Fractional statistics},\
  }\href {https://doi.org/10.1146/annurev-conmatphys-040423-014045} {\bibfield
  {journal} {\bibinfo  {journal} {Annual Review of Condensed Matter Physics}\
  }\textbf {\bibinfo {volume} {15}},\ \bibinfo {pages} {131} (\bibinfo {year}
  {2024})}\BibitemShut {NoStop}%
\bibitem [{\citenamefont {Halperin}(1984)}]{halperin84}%
  \BibitemOpen
  \bibfield  {author} {\bibinfo {author} {\bibfnamefont {B.~I.}\ \bibnamefont
  {Halperin}},\ }\bibfield  {title} {\bibinfo {title} {Statistics of
  quasiparticles and the hierarchy of fractional quantized hall states},\
  }\href {https://doi.org/10.1103/PhysRevLett.52.1583} {\bibfield  {journal}
  {\bibinfo  {journal} {Phys. Rev. Lett.}\ }\textbf {\bibinfo {volume} {52}},\
  \bibinfo {pages} {1583} (\bibinfo {year} {1984})}\BibitemShut {NoStop}%
\bibitem [{\citenamefont {Arovas}\ \emph {et~al.}(1984)\citenamefont {Arovas},
  \citenamefont {Schrieffer},\ and\ \citenamefont {Wilczek}}]{arovas84}%
  \BibitemOpen
  \bibfield  {author} {\bibinfo {author} {\bibfnamefont {D.}~\bibnamefont
  {Arovas}}, \bibinfo {author} {\bibfnamefont {J.~R.}\ \bibnamefont
  {Schrieffer}},\ and\ \bibinfo {author} {\bibfnamefont {F.}~\bibnamefont
  {Wilczek}},\ }\bibfield  {title} {\bibinfo {title} {Fractional statistics and
  the quantum hall effect},\ }\href
  {https://doi.org/10.1103/PhysRevLett.53.722} {\bibfield  {journal} {\bibinfo
  {journal} {Phys. Rev. Lett.}\ }\textbf {\bibinfo {volume} {53}},\ \bibinfo
  {pages} {722} (\bibinfo {year} {1984})}\BibitemShut {NoStop}%
\bibitem [{\citenamefont {Bartolomei}\ \emph {et~al.}(2020)\citenamefont
  {Bartolomei}, \citenamefont {Kumar}, \citenamefont {Bisognin}, \citenamefont
  {Marguerite}, \citenamefont {Berroir}, \citenamefont {Bocquillon},
  \citenamefont {Pla{\c{c}}ais}, \citenamefont {Cavanna}, \citenamefont {Dong},
  \citenamefont {Gennser}, \citenamefont {Jin},\ and\ \citenamefont
  {F{\`e}ve}}]{bartolomei20}%
  \BibitemOpen
  \bibfield  {author} {\bibinfo {author} {\bibfnamefont {H.}~\bibnamefont
  {Bartolomei}}, \bibinfo {author} {\bibfnamefont {M.}~\bibnamefont {Kumar}},
  \bibinfo {author} {\bibfnamefont {R.}~\bibnamefont {Bisognin}}, \bibinfo
  {author} {\bibfnamefont {A.}~\bibnamefont {Marguerite}}, \bibinfo {author}
  {\bibfnamefont {J.-M.}\ \bibnamefont {Berroir}}, \bibinfo {author}
  {\bibfnamefont {E.}~\bibnamefont {Bocquillon}}, \bibinfo {author}
  {\bibfnamefont {B.}~\bibnamefont {Pla{\c{c}}ais}}, \bibinfo {author}
  {\bibfnamefont {A.}~\bibnamefont {Cavanna}}, \bibinfo {author} {\bibfnamefont
  {Q.}~\bibnamefont {Dong}}, \bibinfo {author} {\bibfnamefont {U.}~\bibnamefont
  {Gennser}}, \bibinfo {author} {\bibfnamefont {Y.}~\bibnamefont {Jin}},\ and\
  \bibinfo {author} {\bibfnamefont {G.}~\bibnamefont {F{\`e}ve}},\ }\bibfield
  {title} {\bibinfo {title} {Fractional statistics in anyon collisions},\
  }\href {https://doi.org/10.1126/science.aaz5601} {\bibfield  {journal}
  {\bibinfo  {journal} {Science}\ }\textbf {\bibinfo {volume} {368}},\ \bibinfo
  {pages} {173} (\bibinfo {year} {2020})}\BibitemShut {NoStop}%
\bibitem [{\citenamefont {Nakamura}\ \emph {et~al.}(2020)\citenamefont
  {Nakamura}, \citenamefont {Liang}, \citenamefont {Gardner},\ and\
  \citenamefont {Manfra}}]{nakamura20}%
  \BibitemOpen
  \bibfield  {author} {\bibinfo {author} {\bibfnamefont {J.}~\bibnamefont
  {Nakamura}}, \bibinfo {author} {\bibfnamefont {S.}~\bibnamefont {Liang}},
  \bibinfo {author} {\bibfnamefont {G.~C.}\ \bibnamefont {Gardner}},\ and\
  \bibinfo {author} {\bibfnamefont {M.~J.}\ \bibnamefont {Manfra}},\ }\bibfield
   {title} {\bibinfo {title} {Direct observation of anyonic braiding
  statistics},\ }\href {https://doi.org/10.1038/s41567-020-1019-1} {\bibfield
  {journal} {\bibinfo  {journal} {Nature Physics}\ }\textbf {\bibinfo {volume}
  {16}},\ \bibinfo {pages} {931} (\bibinfo {year} {2020})}\BibitemShut
  {NoStop}%
\bibitem [{\citenamefont {Coldea}\ \emph {et~al.}(2001)\citenamefont {Coldea},
  \citenamefont {Tennant}, \citenamefont {Tsvelik},\ and\ \citenamefont
  {Tylczynski}}]{coldea01}%
  \BibitemOpen
  \bibfield  {author} {\bibinfo {author} {\bibfnamefont {R.}~\bibnamefont
  {Coldea}}, \bibinfo {author} {\bibfnamefont {D.~A.}\ \bibnamefont {Tennant}},
  \bibinfo {author} {\bibfnamefont {A.~M.}\ \bibnamefont {Tsvelik}},\ and\
  \bibinfo {author} {\bibfnamefont {Z.}~\bibnamefont {Tylczynski}},\ }\bibfield
   {title} {\bibinfo {title} {Experimental realization of a 2d fractional
  quantum spin liquid},\ }\href {https://doi.org/10.1103/PhysRevLett.86.1335}
  {\bibfield  {journal} {\bibinfo  {journal} {Phys. Rev. Lett.}\ }\textbf
  {\bibinfo {volume} {86}},\ \bibinfo {pages} {1335} (\bibinfo {year}
  {2001})}\BibitemShut {NoStop}%
\bibitem [{\citenamefont {Kitaev}(2006)}]{kitaev06}%
  \BibitemOpen
  \bibfield  {author} {\bibinfo {author} {\bibfnamefont {A.}~\bibnamefont
  {Kitaev}},\ }\bibfield  {title} {\bibinfo {title} {Anyons in an exactly
  solved model and beyond},\ }\href {https://doi.org/10.1016/j.aop.2005.10.005}
  {\bibfield  {journal} {\bibinfo  {journal} {Annals of Physics}\ }\textbf
  {\bibinfo {volume} {321}},\ \bibinfo {pages} {2} (\bibinfo {year}
  {2006})}\BibitemShut {NoStop}%
\bibitem [{\citenamefont {Semeghini}\ \emph {et~al.}(2021)\citenamefont
  {Semeghini}, \citenamefont {Levine}, \citenamefont {Keesling}, \citenamefont
  {Ebadi}, \citenamefont {Wang}, \citenamefont {Bluvstein}, \citenamefont
  {Verresen}, \citenamefont {Pichler}, \citenamefont {Kalinowski},
  \citenamefont {Samajdar}, \citenamefont {Omran}, \citenamefont {Sachdev},
  \citenamefont {Vishwanath}, \citenamefont {Greiner}, \citenamefont
  {Vuleti\'{c}},\ and\ \citenamefont {Lukin}}]{semeghini21}%
  \BibitemOpen
  \bibfield  {author} {\bibinfo {author} {\bibfnamefont {G.}~\bibnamefont
  {Semeghini}}, \bibinfo {author} {\bibfnamefont {H.}~\bibnamefont {Levine}},
  \bibinfo {author} {\bibfnamefont {A.}~\bibnamefont {Keesling}}, \bibinfo
  {author} {\bibfnamefont {S.}~\bibnamefont {Ebadi}}, \bibinfo {author}
  {\bibfnamefont {T.~T.}\ \bibnamefont {Wang}}, \bibinfo {author}
  {\bibfnamefont {D.}~\bibnamefont {Bluvstein}}, \bibinfo {author}
  {\bibfnamefont {R.}~\bibnamefont {Verresen}}, \bibinfo {author}
  {\bibfnamefont {H.}~\bibnamefont {Pichler}}, \bibinfo {author} {\bibfnamefont
  {M.}~\bibnamefont {Kalinowski}}, \bibinfo {author} {\bibfnamefont
  {R.}~\bibnamefont {Samajdar}}, \bibinfo {author} {\bibfnamefont
  {A.}~\bibnamefont {Omran}}, \bibinfo {author} {\bibfnamefont
  {S.}~\bibnamefont {Sachdev}}, \bibinfo {author} {\bibfnamefont
  {A.}~\bibnamefont {Vishwanath}}, \bibinfo {author} {\bibfnamefont
  {M.}~\bibnamefont {Greiner}}, \bibinfo {author} {\bibfnamefont
  {V.}~\bibnamefont {Vuleti\'{c}}},\ and\ \bibinfo {author} {\bibfnamefont
  {M.~D.}\ \bibnamefont {Lukin}},\ }\bibfield  {title} {\bibinfo {title}
  {Probing topological spin liquids on a programmable quantum simulator},\
  }\href {https://doi.org/10.1126/science.abi8794} {\bibfield  {journal}
  {\bibinfo  {journal} {Science}\ }\textbf {\bibinfo {volume} {374}},\ \bibinfo
  {pages} {1242} (\bibinfo {year} {2021})}\BibitemShut {NoStop}%
\bibitem [{\citenamefont {Kitaev}(2003)}]{kitaev03}%
  \BibitemOpen
  \bibfield  {author} {\bibinfo {author} {\bibfnamefont {A.~Y.}\ \bibnamefont
  {Kitaev}},\ }\bibfield  {title} {\bibinfo {title} {Fault-tolerant quantum
  computation by anyons},\ }\href
  {https://doi.org/10.1016/S0003-4916(02)00018-0} {\bibfield  {journal}
  {\bibinfo  {journal} {Annals of physics}\ }\textbf {\bibinfo {volume}
  {303}},\ \bibinfo {pages} {2} (\bibinfo {year} {2003})}\BibitemShut {NoStop}%
\bibitem [{\citenamefont {Bravyi}(2006)}]{bravyi06}%
  \BibitemOpen
  \bibfield  {author} {\bibinfo {author} {\bibfnamefont {S.}~\bibnamefont
  {Bravyi}},\ }\bibfield  {title} {\bibinfo {title} {Universal quantum
  computation with the $\nu$=5/2 fractional quantum hall state},\ }\href
  {https://doi.org/10.1103/PhysRevA.73.042313} {\bibfield  {journal} {\bibinfo
  {journal} {Phys. Rev. A}\ }\textbf {\bibinfo {volume} {73}},\ \bibinfo
  {pages} {042313} (\bibinfo {year} {2006})}\BibitemShut {NoStop}%
\bibitem [{\citenamefont {Nayak}\ \emph {et~al.}(2008)\citenamefont {Nayak},
  \citenamefont {Simon}, \citenamefont {Stern}, \citenamefont {Freedman},\ and\
  \citenamefont {Das~Sarma}}]{nayak08}%
  \BibitemOpen
  \bibfield  {author} {\bibinfo {author} {\bibfnamefont {C.}~\bibnamefont
  {Nayak}}, \bibinfo {author} {\bibfnamefont {S.~H.}\ \bibnamefont {Simon}},
  \bibinfo {author} {\bibfnamefont {A.}~\bibnamefont {Stern}}, \bibinfo
  {author} {\bibfnamefont {M.}~\bibnamefont {Freedman}},\ and\ \bibinfo
  {author} {\bibfnamefont {S.}~\bibnamefont {Das~Sarma}},\ }\bibfield  {title}
  {\bibinfo {title} {Non-abelian anyons and topological quantum computation},\
  }\href {https://doi.org/10.1103/RevModPhys.80.1083} {\bibfield  {journal}
  {\bibinfo  {journal} {Rev. Mod. Phys.}\ }\textbf {\bibinfo {volume} {80}},\
  \bibinfo {pages} {1083} (\bibinfo {year} {2008})}\BibitemShut {NoStop}%
\bibitem [{\citenamefont {AI}\ and\ \citenamefont
  {Collaborators}(2023)}]{google23}%
  \BibitemOpen
  \bibfield  {author} {\bibinfo {author} {\bibfnamefont {G.~Q.}\ \bibnamefont
  {AI}}\ and\ \bibinfo {author} {\bibnamefont {Collaborators}},\ }\bibfield
  {title} {\bibinfo {title} {Non-abelian braiding of graph vertices in a
  superconducting processor},\ }\href
  {https://doi.org/10.1038/s41586-023-05954-4} {\bibfield  {journal} {\bibinfo
  {journal} {Nature}\ }\textbf {\bibinfo {volume} {618}},\ \bibinfo {pages}
  {264} (\bibinfo {year} {2023})}\BibitemShut {NoStop}%
\bibitem [{\citenamefont {Haldane}(1988)}]{haldane88}%
  \BibitemOpen
  \bibfield  {author} {\bibinfo {author} {\bibfnamefont {F.~D.~M.}\
  \bibnamefont {Haldane}},\ }\bibfield  {title} {\bibinfo {title} {Exact
  jastrow-gutzwiller resonating-valence-bond ground state of the
  spin-$\frac{1}{2}$ antiferromagnetic heisenberg chain with
  1/${\mathrm{r}}^{2}$ exchange},\ }\href
  {https://doi.org/10.1103/PhysRevLett.60.635} {\bibfield  {journal} {\bibinfo
  {journal} {Phys. Rev. Lett.}\ }\textbf {\bibinfo {volume} {60}},\ \bibinfo
  {pages} {635} (\bibinfo {year} {1988})}\BibitemShut {NoStop}%
\bibitem [{\citenamefont {Shastry}(1988)}]{shastry88}%
  \BibitemOpen
  \bibfield  {author} {\bibinfo {author} {\bibfnamefont {B.~S.}\ \bibnamefont
  {Shastry}},\ }\bibfield  {title} {\bibinfo {title} {Exact solution of an
  s=1/2 heisenberg antiferromagnetic chain with long-ranged interactions},\
  }\href {https://doi.org/10.1103/PhysRevLett.60.639} {\bibfield  {journal}
  {\bibinfo  {journal} {Phys. Rev. Lett.}\ }\textbf {\bibinfo {volume} {60}},\
  \bibinfo {pages} {639} (\bibinfo {year} {1988})}\BibitemShut {NoStop}%
\bibitem [{\citenamefont {Haldane}(1991)}]{haldane91}%
  \BibitemOpen
  \bibfield  {author} {\bibinfo {author} {\bibfnamefont {F.~D.~M.}\
  \bibnamefont {Haldane}},\ }\bibfield  {title} {\bibinfo {title} {``fractional
  statistics'' in arbitrary dimensions: A generalization of the pauli
  principle},\ }\href {https://doi.org/10.1103/PhysRevLett.67.937} {\bibfield
  {journal} {\bibinfo  {journal} {Phys. Rev. Lett.}\ }\textbf {\bibinfo
  {volume} {67}},\ \bibinfo {pages} {937} (\bibinfo {year} {1991})}\BibitemShut
  {NoStop}%
\bibitem [{\citenamefont {Kundu}(1999)}]{kundu99}%
  \BibitemOpen
  \bibfield  {author} {\bibinfo {author} {\bibfnamefont {A.}~\bibnamefont
  {Kundu}},\ }\bibfield  {title} {\bibinfo {title} {Exact solution of double
  $\ensuremath{\delta}$ function bose gas through an interacting anyon gas},\
  }\href {https://doi.org/10.1103/PhysRevLett.83.1275} {\bibfield  {journal}
  {\bibinfo  {journal} {Phys. Rev. Lett.}\ }\textbf {\bibinfo {volume} {83}},\
  \bibinfo {pages} {1275} (\bibinfo {year} {1999})}\BibitemShut {NoStop}%
\bibitem [{\citenamefont {Harshman}\ and\ \citenamefont
  {Knapp}(2020)}]{harshman20}%
  \BibitemOpen
  \bibfield  {author} {\bibinfo {author} {\bibfnamefont {N.~L.}\ \bibnamefont
  {Harshman}}\ and\ \bibinfo {author} {\bibfnamefont {A.~C.}\ \bibnamefont
  {Knapp}},\ }\bibfield  {title} {\bibinfo {title} {Anyons from three-body
  hard-core interactions in one dimension},\ }\href
  {https://doi.org/10.1016/j.aop.2019.168003} {\bibfield  {journal} {\bibinfo
  {journal} {Annals of Physics}\ }\textbf {\bibinfo {volume} {412}},\ \bibinfo
  {pages} {168003} (\bibinfo {year} {2020})}\BibitemShut {NoStop}%
\bibitem [{\citenamefont {Bonkhoff}\ \emph {et~al.}(2021)\citenamefont
  {Bonkhoff}, \citenamefont {J\"agering}, \citenamefont {Eggert}, \citenamefont
  {Pelster}, \citenamefont {Thorwart},\ and\ \citenamefont
  {Posske}}]{bonkhoff21}%
  \BibitemOpen
  \bibfield  {author} {\bibinfo {author} {\bibfnamefont {M.}~\bibnamefont
  {Bonkhoff}}, \bibinfo {author} {\bibfnamefont {K.}~\bibnamefont
  {J\"agering}}, \bibinfo {author} {\bibfnamefont {S.}~\bibnamefont {Eggert}},
  \bibinfo {author} {\bibfnamefont {A.}~\bibnamefont {Pelster}}, \bibinfo
  {author} {\bibfnamefont {M.}~\bibnamefont {Thorwart}},\ and\ \bibinfo
  {author} {\bibfnamefont {T.}~\bibnamefont {Posske}},\ }\bibfield  {title}
  {\bibinfo {title} {Bosonic continuum theory of one-dimensional lattice
  anyons},\ }\href {https://doi.org/10.1103/PhysRevLett.126.163201} {\bibfield
  {journal} {\bibinfo  {journal} {Phys. Rev. Lett.}\ }\textbf {\bibinfo
  {volume} {126}},\ \bibinfo {pages} {163201} (\bibinfo {year}
  {2021})}\BibitemShut {NoStop}%
\bibitem [{\citenamefont {Keilmann}\ \emph {et~al.}(2011)\citenamefont
  {Keilmann}, \citenamefont {Lanzmich}, \citenamefont {McCulloch},\ and\
  \citenamefont {Roncaglia}}]{keilmann11}%
  \BibitemOpen
  \bibfield  {author} {\bibinfo {author} {\bibfnamefont {T.}~\bibnamefont
  {Keilmann}}, \bibinfo {author} {\bibfnamefont {S.}~\bibnamefont {Lanzmich}},
  \bibinfo {author} {\bibfnamefont {I.}~\bibnamefont {McCulloch}},\ and\
  \bibinfo {author} {\bibfnamefont {M.}~\bibnamefont {Roncaglia}},\ }\bibfield
  {title} {\bibinfo {title} {Statistically induced phase transitions and anyons
  in 1d optical lattices},\ }\href {https://doi.org/10.1038/ncomms1353}
  {\bibfield  {journal} {\bibinfo  {journal} {Nature communications}\ }\textbf
  {\bibinfo {volume} {2}},\ \bibinfo {pages} {361} (\bibinfo {year}
  {2011})}\BibitemShut {NoStop}%
\bibitem [{\citenamefont {Greschner}\ and\ \citenamefont
  {Santos}(2015)}]{greschner15}%
  \BibitemOpen
  \bibfield  {author} {\bibinfo {author} {\bibfnamefont {S.}~\bibnamefont
  {Greschner}}\ and\ \bibinfo {author} {\bibfnamefont {L.}~\bibnamefont
  {Santos}},\ }\bibfield  {title} {\bibinfo {title} {Anyon hubbard model in
  one-dimensional optical lattices},\ }\href
  {https://doi.org/10.1103/PhysRevLett.115.053002} {\bibfield  {journal}
  {\bibinfo  {journal} {Phys. Rev. Lett.}\ }\textbf {\bibinfo {volume} {115}},\
  \bibinfo {pages} {053002} (\bibinfo {year} {2015})}\BibitemShut {NoStop}%
\bibitem [{\citenamefont {Nagies}\ \emph {et~al.}(2024)\citenamefont {Nagies},
  \citenamefont {Wang}, \citenamefont {Knapp}, \citenamefont {Eckardt},\ and\
  \citenamefont {Harshman}}]{nagies24}%
  \BibitemOpen
  \bibfield  {author} {\bibinfo {author} {\bibfnamefont {S.}~\bibnamefont
  {Nagies}}, \bibinfo {author} {\bibfnamefont {B.}~\bibnamefont {Wang}},
  \bibinfo {author} {\bibfnamefont {A.~C.}\ \bibnamefont {Knapp}}, \bibinfo
  {author} {\bibfnamefont {A.}~\bibnamefont {Eckardt}},\ and\ \bibinfo {author}
  {\bibfnamefont {N.~L.}\ \bibnamefont {Harshman}},\ }\bibfield  {title}
  {\bibinfo {title} {Beyond braid statistics: Constructing a lattice model for
  anyons with exchange statistics intrinsic to one dimension},\ }\href
  {https://doi.org/10.21468/SciPostPhys.16.3.086} {\bibfield  {journal}
  {\bibinfo  {journal} {SciPost Physics}\ }\textbf {\bibinfo {volume} {16}},\
  \bibinfo {pages} {086} (\bibinfo {year} {2024})}\BibitemShut {NoStop}%
\bibitem [{\citenamefont {Wang}\ \emph {et~al.}(2025)\citenamefont {Wang},
  \citenamefont {Vashisht}, \citenamefont {Guo}, \citenamefont {Dhar},
  \citenamefont {Landini}, \citenamefont {N\"agerl},\ and\ \citenamefont
  {Goldman}}]{wang25}%
  \BibitemOpen
  \bibfield  {author} {\bibinfo {author} {\bibfnamefont {B.}~\bibnamefont
  {Wang}}, \bibinfo {author} {\bibfnamefont {A.}~\bibnamefont {Vashisht}},
  \bibinfo {author} {\bibfnamefont {Y.}~\bibnamefont {Guo}}, \bibinfo {author}
  {\bibfnamefont {S.}~\bibnamefont {Dhar}}, \bibinfo {author} {\bibfnamefont
  {M.}~\bibnamefont {Landini}}, \bibinfo {author} {\bibfnamefont {H.-C.}\
  \bibnamefont {N\"agerl}},\ and\ \bibinfo {author} {\bibfnamefont
  {N.}~\bibnamefont {Goldman}},\ }\bibfield  {title} {\bibinfo {title}
  {Anyonization of bosons in one dimension: An effective swap model},\ }\href
  {https://doi.org/10.1103/2np8-mp39} {\bibfield  {journal} {\bibinfo
  {journal} {Phys. Rev. Lett.}\ }\textbf {\bibinfo {volume} {135}},\ \bibinfo
  {pages} {253403} (\bibinfo {year} {2025})}\BibitemShut {NoStop}%
\bibitem [{\citenamefont {Hao}\ \emph {et~al.}(2008)\citenamefont {Hao},
  \citenamefont {Zhang},\ and\ \citenamefont {Chen}}]{hao08}%
  \BibitemOpen
  \bibfield  {author} {\bibinfo {author} {\bibfnamefont {Y.}~\bibnamefont
  {Hao}}, \bibinfo {author} {\bibfnamefont {Y.}~\bibnamefont {Zhang}},\ and\
  \bibinfo {author} {\bibfnamefont {S.}~\bibnamefont {Chen}},\ }\bibfield
  {title} {\bibinfo {title} {Ground-state properties of one-dimensional anyon
  gases},\ }\href {https://doi.org/10.1103/PhysRevA.78.023631} {\bibfield
  {journal} {\bibinfo  {journal} {Phys. Rev. A}\ }\textbf {\bibinfo {volume}
  {78}},\ \bibinfo {pages} {023631} (\bibinfo {year} {2008})}\BibitemShut
  {NoStop}%
\bibitem [{\citenamefont {Hao}\ \emph {et~al.}(2009)\citenamefont {Hao},
  \citenamefont {Zhang},\ and\ \citenamefont {Chen}}]{hao09}%
  \BibitemOpen
  \bibfield  {author} {\bibinfo {author} {\bibfnamefont {Y.}~\bibnamefont
  {Hao}}, \bibinfo {author} {\bibfnamefont {Y.}~\bibnamefont {Zhang}},\ and\
  \bibinfo {author} {\bibfnamefont {S.}~\bibnamefont {Chen}},\ }\bibfield
  {title} {\bibinfo {title} {Ground-state properties of hard-core anyons in
  one-dimensional optical lattices},\ }\href
  {https://doi.org/10.1103/PhysRevA.79.043633} {\bibfield  {journal} {\bibinfo
  {journal} {Phys. Rev. A}\ }\textbf {\bibinfo {volume} {79}},\ \bibinfo
  {pages} {043633} (\bibinfo {year} {2009})}\BibitemShut {NoStop}%
\bibitem [{\citenamefont {Tang}\ \emph {et~al.}(2015)\citenamefont {Tang},
  \citenamefont {Eggert},\ and\ \citenamefont {Pelster}}]{tang15}%
  \BibitemOpen
  \bibfield  {author} {\bibinfo {author} {\bibfnamefont {G.}~\bibnamefont
  {Tang}}, \bibinfo {author} {\bibfnamefont {S.}~\bibnamefont {Eggert}},\ and\
  \bibinfo {author} {\bibfnamefont {A.}~\bibnamefont {Pelster}},\ }\bibfield
  {title} {\bibinfo {title} {Ground-state properties of anyons in a
  one-dimensional lattice},\ }\href
  {https://doi.org/10.1088/1367-2630/17/12/123016} {\bibfield  {journal}
  {\bibinfo  {journal} {New Journal of Physics}\ }\textbf {\bibinfo {volume}
  {17}},\ \bibinfo {pages} {123016} (\bibinfo {year} {2015})}\BibitemShut
  {NoStop}%
\bibitem [{\citenamefont {Str\"ater}\ \emph {et~al.}(2016)\citenamefont
  {Str\"ater}, \citenamefont {Srivastava},\ and\ \citenamefont
  {Eckardt}}]{strater16}%
  \BibitemOpen
  \bibfield  {author} {\bibinfo {author} {\bibfnamefont {C.}~\bibnamefont
  {Str\"ater}}, \bibinfo {author} {\bibfnamefont {S.~C.~L.}\ \bibnamefont
  {Srivastava}},\ and\ \bibinfo {author} {\bibfnamefont {A.}~\bibnamefont
  {Eckardt}},\ }\bibfield  {title} {\bibinfo {title} {Floquet realization and
  signatures of one-dimensional anyons in an optical lattice},\ }\href
  {https://doi.org/10.1103/PhysRevLett.117.205303} {\bibfield  {journal}
  {\bibinfo  {journal} {Phys. Rev. Lett.}\ }\textbf {\bibinfo {volume} {117}},\
  \bibinfo {pages} {205303} (\bibinfo {year} {2016})}\BibitemShut {NoStop}%
\bibitem [{\citenamefont {Zhang}\ \emph {et~al.}(2017)\citenamefont {Zhang},
  \citenamefont {Greschner}, \citenamefont {Fan}, \citenamefont {Scott},\ and\
  \citenamefont {Zhang}}]{zhang17}%
  \BibitemOpen
  \bibfield  {author} {\bibinfo {author} {\bibfnamefont {W.}~\bibnamefont
  {Zhang}}, \bibinfo {author} {\bibfnamefont {S.}~\bibnamefont {Greschner}},
  \bibinfo {author} {\bibfnamefont {E.}~\bibnamefont {Fan}}, \bibinfo {author}
  {\bibfnamefont {T.~C.}\ \bibnamefont {Scott}},\ and\ \bibinfo {author}
  {\bibfnamefont {Y.}~\bibnamefont {Zhang}},\ }\bibfield  {title} {\bibinfo
  {title} {Ground-state properties of the one-dimensional unconstrained
  pseudo-anyon hubbard model},\ }\href
  {https://doi.org/10.1103/PhysRevA.95.053614} {\bibfield  {journal} {\bibinfo
  {journal} {Phys. Rev. A}\ }\textbf {\bibinfo {volume} {95}},\ \bibinfo
  {pages} {053614} (\bibinfo {year} {2017})}\BibitemShut {NoStop}%
\bibitem [{\citenamefont {Bonkhoff}\ \emph {et~al.}(2025)\citenamefont
  {Bonkhoff}, \citenamefont {J\"agering}, \citenamefont {Hu}, \citenamefont
  {Pelster}, \citenamefont {Eggert},\ and\ \citenamefont
  {Schneider}}]{bonkhoff25}%
  \BibitemOpen
  \bibfield  {author} {\bibinfo {author} {\bibfnamefont {M.}~\bibnamefont
  {Bonkhoff}}, \bibinfo {author} {\bibfnamefont {K.}~\bibnamefont
  {J\"agering}}, \bibinfo {author} {\bibfnamefont {S.}~\bibnamefont {Hu}},
  \bibinfo {author} {\bibfnamefont {A.}~\bibnamefont {Pelster}}, \bibinfo
  {author} {\bibfnamefont {S.}~\bibnamefont {Eggert}},\ and\ \bibinfo {author}
  {\bibfnamefont {I.}~\bibnamefont {Schneider}},\ }\bibfield  {title} {\bibinfo
  {title} {Anyonic phase transitions in the 1d extended hubbard model with
  fractional statistics},\ }\href {https://doi.org/10.1103/7n1c-vq2p}
  {\bibfield  {journal} {\bibinfo  {journal} {Phys. Rev. Lett.}\ }\textbf
  {\bibinfo {volume} {135}},\ \bibinfo {pages} {036601} (\bibinfo {year}
  {2025})}\BibitemShut {NoStop}%
\bibitem [{\citenamefont {Fr{\"o}lian}\ \emph {et~al.}(2022)\citenamefont
  {Fr{\"o}lian}, \citenamefont {Chisholm}, \citenamefont {Neri}, \citenamefont
  {Cabrera}, \citenamefont {Ramos}, \citenamefont {Celi},\ and\ \citenamefont
  {Tarruell}}]{frolian22}%
  \BibitemOpen
  \bibfield  {author} {\bibinfo {author} {\bibfnamefont {A.}~\bibnamefont
  {Fr{\"o}lian}}, \bibinfo {author} {\bibfnamefont {C.~S.}\ \bibnamefont
  {Chisholm}}, \bibinfo {author} {\bibfnamefont {E.}~\bibnamefont {Neri}},
  \bibinfo {author} {\bibfnamefont {C.~R.}\ \bibnamefont {Cabrera}}, \bibinfo
  {author} {\bibfnamefont {R.}~\bibnamefont {Ramos}}, \bibinfo {author}
  {\bibfnamefont {A.}~\bibnamefont {Celi}},\ and\ \bibinfo {author}
  {\bibfnamefont {L.}~\bibnamefont {Tarruell}},\ }\bibfield  {title} {\bibinfo
  {title} {Realizing a 1d topological gauge theory in an optically dressed
  bec},\ }\href {https://doi.org/10.1038/s41586-022-04943-3} {\bibfield
  {journal} {\bibinfo  {journal} {Nature}\ }\textbf {\bibinfo {volume} {608}},\
  \bibinfo {pages} {293} (\bibinfo {year} {2022})}\BibitemShut {NoStop}%
\bibitem [{\citenamefont {Chisholm}\ \emph {et~al.}(2022)\citenamefont
  {Chisholm}, \citenamefont {Fr\"olian}, \citenamefont {Neri}, \citenamefont
  {Ramos}, \citenamefont {Tarruell},\ and\ \citenamefont {Celi}}]{chisholm22}%
  \BibitemOpen
  \bibfield  {author} {\bibinfo {author} {\bibfnamefont {C.~S.}\ \bibnamefont
  {Chisholm}}, \bibinfo {author} {\bibfnamefont {A.}~\bibnamefont {Fr\"olian}},
  \bibinfo {author} {\bibfnamefont {E.}~\bibnamefont {Neri}}, \bibinfo {author}
  {\bibfnamefont {R.}~\bibnamefont {Ramos}}, \bibinfo {author} {\bibfnamefont
  {L.}~\bibnamefont {Tarruell}},\ and\ \bibinfo {author} {\bibfnamefont
  {A.}~\bibnamefont {Celi}},\ }\bibfield  {title} {\bibinfo {title} {Encoding a
  one-dimensional topological gauge theory in a raman-coupled bose-einstein
  condensate},\ }\href {https://doi.org/10.1103/PhysRevResearch.4.043088}
  {\bibfield  {journal} {\bibinfo  {journal} {Phys. Rev. Res.}\ }\textbf
  {\bibinfo {volume} {4}},\ \bibinfo {pages} {043088} (\bibinfo {year}
  {2022})}\BibitemShut {NoStop}%
\bibitem [{\citenamefont {Kwan}\ \emph {et~al.}(2024)\citenamefont {Kwan},
  \citenamefont {Segura}, \citenamefont {Li}, \citenamefont {Kim},
  \citenamefont {Gorshkov}, \citenamefont {Eckardt}, \citenamefont
  {Bakkali-Hassani},\ and\ \citenamefont {Greiner}}]{kwan24}%
  \BibitemOpen
  \bibfield  {author} {\bibinfo {author} {\bibfnamefont {J.}~\bibnamefont
  {Kwan}}, \bibinfo {author} {\bibfnamefont {P.}~\bibnamefont {Segura}},
  \bibinfo {author} {\bibfnamefont {Y.}~\bibnamefont {Li}}, \bibinfo {author}
  {\bibfnamefont {S.}~\bibnamefont {Kim}}, \bibinfo {author} {\bibfnamefont
  {A.~V.}\ \bibnamefont {Gorshkov}}, \bibinfo {author} {\bibfnamefont
  {A.}~\bibnamefont {Eckardt}}, \bibinfo {author} {\bibfnamefont
  {B.}~\bibnamefont {Bakkali-Hassani}},\ and\ \bibinfo {author} {\bibfnamefont
  {M.}~\bibnamefont {Greiner}},\ }\bibfield  {title} {\bibinfo {title}
  {Realization of one-dimensional anyons with arbitrary statistical phase},\
  }\href {https://doi.org/10.1126/science.adi3252} {\bibfield  {journal}
  {\bibinfo  {journal} {Science}\ }\textbf {\bibinfo {volume} {386}},\ \bibinfo
  {pages} {1055} (\bibinfo {year} {2024})}\BibitemShut {NoStop}%
\bibitem [{\citenamefont {Dhar}\ \emph {et~al.}(2025)\citenamefont {Dhar},
  \citenamefont {Wang}, \citenamefont {Horvath}, \citenamefont {Vashisht},
  \citenamefont {Zeng}, \citenamefont {Zvonarev}, \citenamefont {Goldman},
  \citenamefont {Guo}, \citenamefont {Landini},\ and\ \citenamefont
  {N{\"a}gerl}}]{dhar25}%
  \BibitemOpen
  \bibfield  {author} {\bibinfo {author} {\bibfnamefont {S.}~\bibnamefont
  {Dhar}}, \bibinfo {author} {\bibfnamefont {B.}~\bibnamefont {Wang}}, \bibinfo
  {author} {\bibfnamefont {M.}~\bibnamefont {Horvath}}, \bibinfo {author}
  {\bibfnamefont {A.}~\bibnamefont {Vashisht}}, \bibinfo {author}
  {\bibfnamefont {Y.}~\bibnamefont {Zeng}}, \bibinfo {author} {\bibfnamefont
  {M.~B.}\ \bibnamefont {Zvonarev}}, \bibinfo {author} {\bibfnamefont
  {N.}~\bibnamefont {Goldman}}, \bibinfo {author} {\bibfnamefont
  {Y.}~\bibnamefont {Guo}}, \bibinfo {author} {\bibfnamefont {M.}~\bibnamefont
  {Landini}},\ and\ \bibinfo {author} {\bibfnamefont {H.-C.}\ \bibnamefont
  {N{\"a}gerl}},\ }\bibfield  {title} {\bibinfo {title} {Observing anyonization
  of bosons in a quantum gas},\ }\href
  {https://doi.org/10.1038/s41586-025-09016-9} {\bibfield  {journal} {\bibinfo
  {journal} {Nature}\ }\textbf {\bibinfo {volume} {642}},\ \bibinfo {pages}
  {53} (\bibinfo {year} {2025})}\BibitemShut {NoStop}%
\bibitem [{\citenamefont {Bakkali-Hassani}\ \emph {et~al.}()\citenamefont
  {Bakkali-Hassani}, \citenamefont {Kwan}, \citenamefont {Segura},
  \citenamefont {Li}, \citenamefont {Tesfaye}, \citenamefont
  {Valent{\'\i}-Rojas}, \citenamefont {Eckardt},\ and\ \citenamefont
  {Greiner}}]{bakkali26}%
  \BibitemOpen
  \bibfield  {author} {\bibinfo {author} {\bibfnamefont {B.}~\bibnamefont
  {Bakkali-Hassani}}, \bibinfo {author} {\bibfnamefont {J.}~\bibnamefont
  {Kwan}}, \bibinfo {author} {\bibfnamefont {P.}~\bibnamefont {Segura}},
  \bibinfo {author} {\bibfnamefont {Y.}~\bibnamefont {Li}}, \bibinfo {author}
  {\bibfnamefont {I.}~\bibnamefont {Tesfaye}}, \bibinfo {author} {\bibfnamefont
  {G.}~\bibnamefont {Valent{\'\i}-Rojas}}, \bibinfo {author} {\bibfnamefont
  {A.}~\bibnamefont {Eckardt}},\ and\ \bibinfo {author} {\bibfnamefont
  {M.}~\bibnamefont {Greiner}},\ }\bibfield  {title} {\bibinfo {title}
  {{Revealing pseudo-fermionization and chiral binding of one-dimensional
  anyons using adiabatic state preparation}},\ }\href
  {https://arxiv.org/abs/2602.20421} {\bibinfo  {journal} {arXiv:2602.20421}\
  }\BibitemShut {NoStop}%
\bibitem [{\citenamefont {del Campo}(2008)}]{delcampo08}%
  \BibitemOpen
\bibfield  {journal} {  }\bibfield  {author} {\bibinfo {author} {\bibfnamefont
  {A.}~\bibnamefont {del Campo}},\ }\bibfield  {title} {\bibinfo {title}
  {Fermionization and bosonization of expanding one-dimensional anyonic
  fluids},\ }\href {https://doi.org/10.1103/PhysRevA.78.045602} {\bibfield
  {journal} {\bibinfo  {journal} {Phys. Rev. A}\ }\textbf {\bibinfo {volume}
  {78}},\ \bibinfo {pages} {045602} (\bibinfo {year} {2008})}\BibitemShut
  {NoStop}%
\bibitem [{\citenamefont {Wang}\ \emph {et~al.}(2014)\citenamefont {Wang},
  \citenamefont {Wang},\ and\ \citenamefont {Zhang}}]{wang14}%
  \BibitemOpen
  \bibfield  {author} {\bibinfo {author} {\bibfnamefont {L.}~\bibnamefont
  {Wang}}, \bibinfo {author} {\bibfnamefont {L.}~\bibnamefont {Wang}},\ and\
  \bibinfo {author} {\bibfnamefont {Y.}~\bibnamefont {Zhang}},\ }\bibfield
  {title} {\bibinfo {title} {Quantum walks of two interacting anyons in
  one-dimensional optical lattices},\ }\href
  {https://doi.org/10.1103/PhysRevA.90.063618} {\bibfield  {journal} {\bibinfo
  {journal} {Phys. Rev. A}\ }\textbf {\bibinfo {volume} {90}},\ \bibinfo
  {pages} {063618} (\bibinfo {year} {2014})}\BibitemShut {NoStop}%
\bibitem [{\citenamefont {Wright}\ \emph {et~al.}(2014)\citenamefont {Wright},
  \citenamefont {Rigol}, \citenamefont {Davis},\ and\ \citenamefont
  {Kheruntsyan}}]{wright14}%
  \BibitemOpen
  \bibfield  {author} {\bibinfo {author} {\bibfnamefont {T.~M.}\ \bibnamefont
  {Wright}}, \bibinfo {author} {\bibfnamefont {M.}~\bibnamefont {Rigol}},
  \bibinfo {author} {\bibfnamefont {M.~J.}\ \bibnamefont {Davis}},\ and\
  \bibinfo {author} {\bibfnamefont {K.~V.}\ \bibnamefont {Kheruntsyan}},\
  }\bibfield  {title} {\bibinfo {title} {Nonequilibrium dynamics of
  one-dimensional hard-core anyons following a quench: Complete relaxation of
  one-body observables},\ }\href
  {https://doi.org/10.1103/PhysRevLett.113.050601} {\bibfield  {journal}
  {\bibinfo  {journal} {Phys. Rev. Lett.}\ }\textbf {\bibinfo {volume} {113}},\
  \bibinfo {pages} {050601} (\bibinfo {year} {2014})}\BibitemShut {NoStop}%
\bibitem [{\citenamefont {Piroli}\ and\ \citenamefont
  {Calabrese}(2017)}]{piroli17}%
  \BibitemOpen
  \bibfield  {author} {\bibinfo {author} {\bibfnamefont {L.}~\bibnamefont
  {Piroli}}\ and\ \bibinfo {author} {\bibfnamefont {P.}~\bibnamefont
  {Calabrese}},\ }\bibfield  {title} {\bibinfo {title} {Exact dynamics
  following an interaction quench in a one-dimensional anyonic gas},\ }\href
  {https://doi.org/10.1103/PhysRevA.96.023611} {\bibfield  {journal} {\bibinfo
  {journal} {Phys. Rev. A}\ }\textbf {\bibinfo {volume} {96}},\ \bibinfo
  {pages} {023611} (\bibinfo {year} {2017})}\BibitemShut {NoStop}%
\bibitem [{\citenamefont {Liu}\ \emph {et~al.}(2018)\citenamefont {Liu},
  \citenamefont {Garrison}, \citenamefont {Deng}, \citenamefont {Gong},\ and\
  \citenamefont {Gorshkov}}]{liu18}%
  \BibitemOpen
  \bibfield  {author} {\bibinfo {author} {\bibfnamefont {F.}~\bibnamefont
  {Liu}}, \bibinfo {author} {\bibfnamefont {J.~R.}\ \bibnamefont {Garrison}},
  \bibinfo {author} {\bibfnamefont {D.-L.}\ \bibnamefont {Deng}}, \bibinfo
  {author} {\bibfnamefont {Z.-X.}\ \bibnamefont {Gong}},\ and\ \bibinfo
  {author} {\bibfnamefont {A.~V.}\ \bibnamefont {Gorshkov}},\ }\bibfield
  {title} {\bibinfo {title} {Asymmetric particle transport and light-cone
  dynamics induced by anyonic statistics},\ }\href
  {https://doi.org/10.1103/PhysRevLett.121.250404} {\bibfield  {journal}
  {\bibinfo  {journal} {Phys. Rev. Lett.}\ }\textbf {\bibinfo {volume} {121}},\
  \bibinfo {pages} {250404} (\bibinfo {year} {2018})}\BibitemShut {NoStop}%
\bibitem [{\citenamefont {Chen}\ \emph {et~al.}()\citenamefont {Chen},
  \citenamefont {Huang}, \citenamefont {Zhang},\ and\ \citenamefont
  {Zhang}}]{chen25}%
  \BibitemOpen
  \bibfield  {author} {\bibinfo {author} {\bibfnamefont {R.-J.}\ \bibnamefont
  {Chen}}, \bibinfo {author} {\bibfnamefont {Y.-X.}\ \bibnamefont {Huang}},
  \bibinfo {author} {\bibfnamefont {G.-Q.}\ \bibnamefont {Zhang}},\ and\
  \bibinfo {author} {\bibfnamefont {D.-W.}\ \bibnamefont {Zhang}},\ }\bibfield
  {title} {\bibinfo {title} {{Asymmetric and chiral dynamics of two-component
  anyons with synthetic gauge flux}},\ }\href
  {https://arxiv.org/abs/2512.19139} {\bibinfo  {journal} {arXiv:2512.19139}\
  }\BibitemShut {NoStop}%
\bibitem [{\citenamefont {Kinoshita}\ \emph {et~al.}(2006)\citenamefont
  {Kinoshita}, \citenamefont {Wenger},\ and\ \citenamefont
  {Weiss}}]{kinoshita06}%
  \BibitemOpen
\bibfield  {journal} {  }\bibfield  {author} {\bibinfo {author} {\bibfnamefont
  {T.}~\bibnamefont {Kinoshita}}, \bibinfo {author} {\bibfnamefont
  {T.}~\bibnamefont {Wenger}},\ and\ \bibinfo {author} {\bibfnamefont {D.~S.}\
  \bibnamefont {Weiss}},\ }\bibfield  {title} {\bibinfo {title} {A quantum
  newton's cradle},\ }\href {https://doi.org/10.1038/nature04693} {\bibfield
  {journal} {\bibinfo  {journal} {Nature}\ }\textbf {\bibinfo {volume} {440}},\
  \bibinfo {pages} {900} (\bibinfo {year} {2006})}\BibitemShut {NoStop}%
\bibitem [{\citenamefont {Trotzky}\ \emph {et~al.}(2012)\citenamefont
  {Trotzky}, \citenamefont {Chen}, \citenamefont {Flesch}, \citenamefont
  {McCulloch}, \citenamefont {Schollw{\"o}ck}, \citenamefont {Eisert},\ and\
  \citenamefont {Bloch}}]{trotzky12}%
  \BibitemOpen
  \bibfield  {author} {\bibinfo {author} {\bibfnamefont {S.}~\bibnamefont
  {Trotzky}}, \bibinfo {author} {\bibfnamefont {Y.-A.}\ \bibnamefont {Chen}},
  \bibinfo {author} {\bibfnamefont {A.}~\bibnamefont {Flesch}}, \bibinfo
  {author} {\bibfnamefont {I.~P.}\ \bibnamefont {McCulloch}}, \bibinfo {author}
  {\bibfnamefont {U.}~\bibnamefont {Schollw{\"o}ck}}, \bibinfo {author}
  {\bibfnamefont {J.}~\bibnamefont {Eisert}},\ and\ \bibinfo {author}
  {\bibfnamefont {I.}~\bibnamefont {Bloch}},\ }\bibfield  {title} {\bibinfo
  {title} {Probing the relaxation towards equilibrium in an isolated strongly
  correlated one-dimensional bose gas},\ }\href
  {https://doi.org/10.1038/nphys2232} {\bibfield  {journal} {\bibinfo
  {journal} {Nature physics}\ }\textbf {\bibinfo {volume} {8}},\ \bibinfo
  {pages} {325} (\bibinfo {year} {2012})}\BibitemShut {NoStop}%
\bibitem [{\citenamefont {Gring}\ \emph {et~al.}(2012)\citenamefont {Gring},
  \citenamefont {Kuhnert}, \citenamefont {Langen}, \citenamefont {Kitagawa},
  \citenamefont {Rauer}, \citenamefont {Schreitl}, \citenamefont {Mazets},
  \citenamefont {Smith}, \citenamefont {Demler},\ and\ \citenamefont
  {Schmiedmayer}}]{gring12}%
  \BibitemOpen
  \bibfield  {author} {\bibinfo {author} {\bibfnamefont {M.}~\bibnamefont
  {Gring}}, \bibinfo {author} {\bibfnamefont {M.}~\bibnamefont {Kuhnert}},
  \bibinfo {author} {\bibfnamefont {T.}~\bibnamefont {Langen}}, \bibinfo
  {author} {\bibfnamefont {T.}~\bibnamefont {Kitagawa}}, \bibinfo {author}
  {\bibfnamefont {B.}~\bibnamefont {Rauer}}, \bibinfo {author} {\bibfnamefont
  {M.}~\bibnamefont {Schreitl}}, \bibinfo {author} {\bibfnamefont
  {I.}~\bibnamefont {Mazets}}, \bibinfo {author} {\bibfnamefont {D.~A.}\
  \bibnamefont {Smith}}, \bibinfo {author} {\bibfnamefont {E.}~\bibnamefont
  {Demler}},\ and\ \bibinfo {author} {\bibfnamefont {J.}~\bibnamefont
  {Schmiedmayer}},\ }\bibfield  {title} {\bibinfo {title} {Relaxation and
  prethermalization in an isolated quantum system},\ }\href
  {https://doi.org/10.1126/science.1224953} {\bibfield  {journal} {\bibinfo
  {journal} {Science}\ }\textbf {\bibinfo {volume} {337}},\ \bibinfo {pages}
  {1318} (\bibinfo {year} {2012})}\BibitemShut {NoStop}%
\bibitem [{\citenamefont {Choi}\ \emph {et~al.}(2016)\citenamefont {Choi},
  \citenamefont {Hild}, \citenamefont {Zeiher}, \citenamefont {Schau{\ss}},
  \citenamefont {Rubio-Abadal}, \citenamefont {Yefsah}, \citenamefont
  {Khemani}, \citenamefont {Huse}, \citenamefont {Bloch},\ and\ \citenamefont
  {Gross}}]{choi16}%
  \BibitemOpen
  \bibfield  {author} {\bibinfo {author} {\bibfnamefont {J.-y.}\ \bibnamefont
  {Choi}}, \bibinfo {author} {\bibfnamefont {S.}~\bibnamefont {Hild}}, \bibinfo
  {author} {\bibfnamefont {J.}~\bibnamefont {Zeiher}}, \bibinfo {author}
  {\bibfnamefont {P.}~\bibnamefont {Schau{\ss}}}, \bibinfo {author}
  {\bibfnamefont {A.}~\bibnamefont {Rubio-Abadal}}, \bibinfo {author}
  {\bibfnamefont {T.}~\bibnamefont {Yefsah}}, \bibinfo {author} {\bibfnamefont
  {V.}~\bibnamefont {Khemani}}, \bibinfo {author} {\bibfnamefont {D.~A.}\
  \bibnamefont {Huse}}, \bibinfo {author} {\bibfnamefont {I.}~\bibnamefont
  {Bloch}},\ and\ \bibinfo {author} {\bibfnamefont {C.}~\bibnamefont {Gross}},\
  }\bibfield  {title} {\bibinfo {title} {Exploring the many-body localization
  transition in two dimensions},\ }\href
  {https://doi.org/10.1126/science.aaf8834} {\bibfield  {journal} {\bibinfo
  {journal} {Science}\ }\textbf {\bibinfo {volume} {352}},\ \bibinfo {pages}
  {1547} (\bibinfo {year} {2016})}\BibitemShut {NoStop}%
\bibitem [{\citenamefont {Pr{\"u}fer}\ \emph {et~al.}(2018)\citenamefont
  {Pr{\"u}fer}, \citenamefont {Kunkel}, \citenamefont {Strobel}, \citenamefont
  {Lannig}, \citenamefont {Linnemann}, \citenamefont {Schmied}, \citenamefont
  {Berges}, \citenamefont {Gasenzer},\ and\ \citenamefont
  {Oberthaler}}]{prufer18}%
  \BibitemOpen
  \bibfield  {author} {\bibinfo {author} {\bibfnamefont {M.}~\bibnamefont
  {Pr{\"u}fer}}, \bibinfo {author} {\bibfnamefont {P.}~\bibnamefont {Kunkel}},
  \bibinfo {author} {\bibfnamefont {H.}~\bibnamefont {Strobel}}, \bibinfo
  {author} {\bibfnamefont {S.}~\bibnamefont {Lannig}}, \bibinfo {author}
  {\bibfnamefont {D.}~\bibnamefont {Linnemann}}, \bibinfo {author}
  {\bibfnamefont {C.-M.}\ \bibnamefont {Schmied}}, \bibinfo {author}
  {\bibfnamefont {J.}~\bibnamefont {Berges}}, \bibinfo {author} {\bibfnamefont
  {T.}~\bibnamefont {Gasenzer}},\ and\ \bibinfo {author} {\bibfnamefont
  {M.~K.}\ \bibnamefont {Oberthaler}},\ }\bibfield  {title} {\bibinfo {title}
  {Observation of universal dynamics in a spinor bose gas far from
  equilibrium},\ }\href {https://doi.org/10.1038/s41586-018-0659-0} {\bibfield
  {journal} {\bibinfo  {journal} {Nature}\ }\textbf {\bibinfo {volume} {563}},\
  \bibinfo {pages} {217} (\bibinfo {year} {2018})}\BibitemShut {NoStop}%
\bibitem [{\citenamefont {Erne}\ \emph {et~al.}(2018)\citenamefont {Erne},
  \citenamefont {B{\"u}cker}, \citenamefont {Gasenzer}, \citenamefont
  {Berges},\ and\ \citenamefont {Schmiedmayer}}]{erne18}%
  \BibitemOpen
  \bibfield  {author} {\bibinfo {author} {\bibfnamefont {S.}~\bibnamefont
  {Erne}}, \bibinfo {author} {\bibfnamefont {R.}~\bibnamefont {B{\"u}cker}},
  \bibinfo {author} {\bibfnamefont {T.}~\bibnamefont {Gasenzer}}, \bibinfo
  {author} {\bibfnamefont {J.}~\bibnamefont {Berges}},\ and\ \bibinfo {author}
  {\bibfnamefont {J.}~\bibnamefont {Schmiedmayer}},\ }\bibfield  {title}
  {\bibinfo {title} {Universal dynamics in an isolated one-dimensional bose gas
  far from equilibrium},\ }\href {https://doi.org/10.1038/s41586-018-0667-0}
  {\bibfield  {journal} {\bibinfo  {journal} {Nature}\ }\textbf {\bibinfo
  {volume} {563}},\ \bibinfo {pages} {225} (\bibinfo {year}
  {2018})}\BibitemShut {NoStop}%
\bibitem [{\citenamefont {Polkovnikov}\ \emph {et~al.}(2011)\citenamefont
  {Polkovnikov}, \citenamefont {Sengupta}, \citenamefont {Silva},\ and\
  \citenamefont {Vengalattore}}]{polkovnikov11}%
  \BibitemOpen
  \bibfield  {author} {\bibinfo {author} {\bibfnamefont {A.}~\bibnamefont
  {Polkovnikov}}, \bibinfo {author} {\bibfnamefont {K.}~\bibnamefont
  {Sengupta}}, \bibinfo {author} {\bibfnamefont {A.}~\bibnamefont {Silva}},\
  and\ \bibinfo {author} {\bibfnamefont {M.}~\bibnamefont {Vengalattore}},\
  }\bibfield  {title} {\bibinfo {title} {Colloquium: Nonequilibrium dynamics of
  closed interacting quantum systems},\ }\href
  {https://doi.org/10.1103/RevModPhys.83.863} {\bibfield  {journal} {\bibinfo
  {journal} {Rev. Mod. Phys.}\ }\textbf {\bibinfo {volume} {83}},\ \bibinfo
  {pages} {863} (\bibinfo {year} {2011})}\BibitemShut {NoStop}%
\bibitem [{\citenamefont {Vasseur}\ and\ \citenamefont
  {Moore}(2016)}]{vasseur16}%
  \BibitemOpen
  \bibfield  {author} {\bibinfo {author} {\bibfnamefont {R.}~\bibnamefont
  {Vasseur}}\ and\ \bibinfo {author} {\bibfnamefont {J.~E.}\ \bibnamefont
  {Moore}},\ }\bibfield  {title} {\bibinfo {title} {Nonequilibrium quantum
  dynamics and transport: from integrability to many-body localization},\
  }\href {https://doi.org/10.1088/1742-5468/2016/06/064010} {\bibfield
  {journal} {\bibinfo  {journal} {Journal of Statistical Mechanics: Theory and
  Experiment}\ }\textbf {\bibinfo {volume} {2016}},\ \bibinfo {pages} {064010}
  (\bibinfo {year} {2016})}\BibitemShut {NoStop}%
\bibitem [{\citenamefont {D'Alessio}\ \emph {et~al.}(2016)\citenamefont
  {D'Alessio}, \citenamefont {Kafri}, \citenamefont {Polkovnikov},\ and\
  \citenamefont {Rigol}}]{alessio16}%
  \BibitemOpen
  \bibfield  {author} {\bibinfo {author} {\bibfnamefont {L.}~\bibnamefont
  {D'Alessio}}, \bibinfo {author} {\bibfnamefont {Y.}~\bibnamefont {Kafri}},
  \bibinfo {author} {\bibfnamefont {A.}~\bibnamefont {Polkovnikov}},\ and\
  \bibinfo {author} {\bibfnamefont {M.}~\bibnamefont {Rigol}},\ }\bibfield
  {title} {\bibinfo {title} {From quantum chaos and eigenstate thermalization
  to statistical mechanics and thermodynamics},\ }\href
  {https://doi.org/10.1080/00018732.2016.1198134} {\bibfield  {journal}
  {\bibinfo  {journal} {Advances in Physics}\ }\textbf {\bibinfo {volume}
  {65}},\ \bibinfo {pages} {239} (\bibinfo {year} {2016})}\BibitemShut
  {NoStop}%
\bibitem [{\citenamefont {Essler}\ and\ \citenamefont
  {Fagotti}(2016)}]{essler16}%
  \BibitemOpen
  \bibfield  {author} {\bibinfo {author} {\bibfnamefont {F.~H.}\ \bibnamefont
  {Essler}}\ and\ \bibinfo {author} {\bibfnamefont {M.}~\bibnamefont
  {Fagotti}},\ }\bibfield  {title} {\bibinfo {title} {Quench dynamics and
  relaxation in isolated integrable quantum spin chains},\ }\href
  {https://doi.org/10.1088/1742-5468/2016/06/064002} {\bibfield  {journal}
  {\bibinfo  {journal} {Journal of Statistical Mechanics: Theory and
  Experiment}\ }\textbf {\bibinfo {volume} {2016}},\ \bibinfo {pages} {064002}
  (\bibinfo {year} {2016})}\BibitemShut {NoStop}%
\bibitem [{\citenamefont {Ueda}(2020)}]{ueda20}%
  \BibitemOpen
  \bibfield  {author} {\bibinfo {author} {\bibfnamefont {M.}~\bibnamefont
  {Ueda}},\ }\bibfield  {title} {\bibinfo {title} {Quantum equilibration,
  thermalization and prethermalization in ultracold atoms},\ }\href
  {https://doi.org/10.1038/s42254-020-0237-x} {\bibfield  {journal} {\bibinfo
  {journal} {Nature Reviews Physics}\ }\textbf {\bibinfo {volume} {2}},\
  \bibinfo {pages} {669} (\bibinfo {year} {2020})}\BibitemShut {NoStop}%
\bibitem [{\citenamefont {Bertini}\ \emph {et~al.}(2021)\citenamefont
  {Bertini}, \citenamefont {Heidrich-Meisner}, \citenamefont {Karrasch},
  \citenamefont {Prosen}, \citenamefont {Steinigeweg},\ and\ \citenamefont
  {\ifmmode \check{Z}\else \v{Z}\fi{}nidari\ifmmode~\check{c}\else
  \v{c}\fi{}}}]{bertini21}%
  \BibitemOpen
  \bibfield  {author} {\bibinfo {author} {\bibfnamefont {B.}~\bibnamefont
  {Bertini}}, \bibinfo {author} {\bibfnamefont {F.}~\bibnamefont
  {Heidrich-Meisner}}, \bibinfo {author} {\bibfnamefont {C.}~\bibnamefont
  {Karrasch}}, \bibinfo {author} {\bibfnamefont {T.}~\bibnamefont {Prosen}},
  \bibinfo {author} {\bibfnamefont {R.}~\bibnamefont {Steinigeweg}},\ and\
  \bibinfo {author} {\bibfnamefont {M.}~\bibnamefont {\ifmmode \check{Z}\else
  \v{Z}\fi{}nidari\ifmmode~\check{c}\else \v{c}\fi{}}},\ }\bibfield  {title}
  {\bibinfo {title} {Finite-temperature transport in one-dimensional quantum
  lattice models},\ }\href {https://doi.org/10.1103/RevModPhys.93.025003}
  {\bibfield  {journal} {\bibinfo  {journal} {Rev. Mod. Phys.}\ }\textbf
  {\bibinfo {volume} {93}},\ \bibinfo {pages} {025003} (\bibinfo {year}
  {2021})}\BibitemShut {NoStop}%
\bibitem [{\citenamefont {Nichols}\ \emph {et~al.}(2019)\citenamefont
  {Nichols}, \citenamefont {Cheuk}, \citenamefont {Okan}, \citenamefont
  {Hartke}, \citenamefont {Mendez}, \citenamefont {Senthil}, \citenamefont
  {Khatami}, \citenamefont {Zhang},\ and\ \citenamefont
  {Zwierlein}}]{nichols19}%
  \BibitemOpen
  \bibfield  {author} {\bibinfo {author} {\bibfnamefont {M.~A.}\ \bibnamefont
  {Nichols}}, \bibinfo {author} {\bibfnamefont {L.~W.}\ \bibnamefont {Cheuk}},
  \bibinfo {author} {\bibfnamefont {M.}~\bibnamefont {Okan}}, \bibinfo {author}
  {\bibfnamefont {T.~R.}\ \bibnamefont {Hartke}}, \bibinfo {author}
  {\bibfnamefont {E.}~\bibnamefont {Mendez}}, \bibinfo {author} {\bibfnamefont
  {T.}~\bibnamefont {Senthil}}, \bibinfo {author} {\bibfnamefont
  {E.}~\bibnamefont {Khatami}}, \bibinfo {author} {\bibfnamefont
  {H.}~\bibnamefont {Zhang}},\ and\ \bibinfo {author} {\bibfnamefont {M.~W.}\
  \bibnamefont {Zwierlein}},\ }\bibfield  {title} {\bibinfo {title} {Spin
  transport in a mott insulator of ultracold fermions},\ }\href
  {https://doi.org/10.1126/science.aat4387} {\bibfield  {journal} {\bibinfo
  {journal} {Science}\ }\textbf {\bibinfo {volume} {363}},\ \bibinfo {pages}
  {383} (\bibinfo {year} {2019})}\BibitemShut {NoStop}%
\bibitem [{\citenamefont {Jepsen}\ \emph {et~al.}(2020)\citenamefont {Jepsen},
  \citenamefont {Amato-Grill}, \citenamefont {Dimitrova}, \citenamefont {Ho},
  \citenamefont {Demler},\ and\ \citenamefont {Ketterle}}]{jepsen20}%
  \BibitemOpen
  \bibfield  {author} {\bibinfo {author} {\bibfnamefont {P.~N.}\ \bibnamefont
  {Jepsen}}, \bibinfo {author} {\bibfnamefont {J.}~\bibnamefont {Amato-Grill}},
  \bibinfo {author} {\bibfnamefont {I.}~\bibnamefont {Dimitrova}}, \bibinfo
  {author} {\bibfnamefont {W.~W.}\ \bibnamefont {Ho}}, \bibinfo {author}
  {\bibfnamefont {E.}~\bibnamefont {Demler}},\ and\ \bibinfo {author}
  {\bibfnamefont {W.}~\bibnamefont {Ketterle}},\ }\bibfield  {title} {\bibinfo
  {title} {Spin transport in a tunable heisenberg model realized with ultracold
  atoms},\ }\href {https://doi.org/10.1038/s41586-020-3033-y} {\bibfield
  {journal} {\bibinfo  {journal} {Nature}\ }\textbf {\bibinfo {volume} {588}},\
  \bibinfo {pages} {403} (\bibinfo {year} {2020})}\BibitemShut {NoStop}%
\bibitem [{\citenamefont {Joshi}\ \emph {et~al.}(2022)\citenamefont {Joshi},
  \citenamefont {Kranzl}, \citenamefont {Schuckert}, \citenamefont {Lovas},
  \citenamefont {Maier}, \citenamefont {Blatt}, \citenamefont {Knap},\ and\
  \citenamefont {Roos}}]{joshi22}%
  \BibitemOpen
  \bibfield  {author} {\bibinfo {author} {\bibfnamefont {M.~K.}\ \bibnamefont
  {Joshi}}, \bibinfo {author} {\bibfnamefont {F.}~\bibnamefont {Kranzl}},
  \bibinfo {author} {\bibfnamefont {A.}~\bibnamefont {Schuckert}}, \bibinfo
  {author} {\bibfnamefont {I.}~\bibnamefont {Lovas}}, \bibinfo {author}
  {\bibfnamefont {C.}~\bibnamefont {Maier}}, \bibinfo {author} {\bibfnamefont
  {R.}~\bibnamefont {Blatt}}, \bibinfo {author} {\bibfnamefont
  {M.}~\bibnamefont {Knap}},\ and\ \bibinfo {author} {\bibfnamefont {C.~F.}\
  \bibnamefont {Roos}},\ }\bibfield  {title} {\bibinfo {title} {Observing
  emergent hydrodynamics in a long-range quantum magnet},\ }\href
  {https://doi.org/10.1126/science.abk2400} {\bibfield  {journal} {\bibinfo
  {journal} {Science}\ }\textbf {\bibinfo {volume} {376}},\ \bibinfo {pages}
  {720} (\bibinfo {year} {2022})}\BibitemShut {NoStop}%
\bibitem [{\citenamefont {Wienand}\ \emph {et~al.}(2024)\citenamefont
  {Wienand}, \citenamefont {Karch}, \citenamefont {Impertro}, \citenamefont
  {Schweizer}, \citenamefont {McCulloch}, \citenamefont {Vasseur},
  \citenamefont {Gopalakrishnan}, \citenamefont {Aidelsburger},\ and\
  \citenamefont {Bloch}}]{wienand24}%
  \BibitemOpen
  \bibfield  {author} {\bibinfo {author} {\bibfnamefont {J.~F.}\ \bibnamefont
  {Wienand}}, \bibinfo {author} {\bibfnamefont {S.}~\bibnamefont {Karch}},
  \bibinfo {author} {\bibfnamefont {A.}~\bibnamefont {Impertro}}, \bibinfo
  {author} {\bibfnamefont {C.}~\bibnamefont {Schweizer}}, \bibinfo {author}
  {\bibfnamefont {E.}~\bibnamefont {McCulloch}}, \bibinfo {author}
  {\bibfnamefont {R.}~\bibnamefont {Vasseur}}, \bibinfo {author} {\bibfnamefont
  {S.}~\bibnamefont {Gopalakrishnan}}, \bibinfo {author} {\bibfnamefont
  {M.}~\bibnamefont {Aidelsburger}},\ and\ \bibinfo {author} {\bibfnamefont
  {I.}~\bibnamefont {Bloch}},\ }\bibfield  {title} {\bibinfo {title} {Emergence
  of fluctuating hydrodynamics in chaotic quantum systems},\ }\href
  {https://doi.org/10.1038/s41567-024-02611-z} {\bibfield  {journal} {\bibinfo
  {journal} {Nature Physics}\ }\textbf {\bibinfo {volume} {20}},\ \bibinfo
  {pages} {1732} (\bibinfo {year} {2024})}\BibitemShut {NoStop}%
\bibitem [{\citenamefont {Ljubotina}\ \emph {et~al.}(2019)\citenamefont
  {Ljubotina}, \citenamefont {\ifmmode \check{Z}\else
  \v{Z}\fi{}nidari\ifmmode~\check{c}\else \v{c}\fi{}},\ and\ \citenamefont
  {Prosen}}]{ljubotina19}%
  \BibitemOpen
  \bibfield  {author} {\bibinfo {author} {\bibfnamefont {M.}~\bibnamefont
  {Ljubotina}}, \bibinfo {author} {\bibfnamefont {M.}~\bibnamefont {\ifmmode
  \check{Z}\else \v{Z}\fi{}nidari\ifmmode~\check{c}\else \v{c}\fi{}}},\ and\
  \bibinfo {author} {\bibfnamefont {T.~c.~v.}\ \bibnamefont {Prosen}},\
  }\bibfield  {title} {\bibinfo {title} {Kardar-parisi-zhang physics in the
  quantum heisenberg magnet},\ }\href
  {https://doi.org/10.1103/PhysRevLett.122.210602} {\bibfield  {journal}
  {\bibinfo  {journal} {Phys. Rev. Lett.}\ }\textbf {\bibinfo {volume} {122}},\
  \bibinfo {pages} {210602} (\bibinfo {year} {2019})}\BibitemShut {NoStop}%
\bibitem [{\citenamefont {Das}\ \emph {et~al.}(2019)\citenamefont {Das},
  \citenamefont {Kulkarni}, \citenamefont {Spohn},\ and\ \citenamefont
  {Dhar}}]{das19}%
  \BibitemOpen
  \bibfield  {author} {\bibinfo {author} {\bibfnamefont {A.}~\bibnamefont
  {Das}}, \bibinfo {author} {\bibfnamefont {M.}~\bibnamefont {Kulkarni}},
  \bibinfo {author} {\bibfnamefont {H.}~\bibnamefont {Spohn}},\ and\ \bibinfo
  {author} {\bibfnamefont {A.}~\bibnamefont {Dhar}},\ }\bibfield  {title}
  {\bibinfo {title} {Kardar-parisi-zhang scaling for an integrable lattice
  landau-lifshitz spin chain},\ }\href
  {https://doi.org/10.1103/PhysRevE.100.042116} {\bibfield  {journal} {\bibinfo
   {journal} {Phys. Rev. E}\ }\textbf {\bibinfo {volume} {100}},\ \bibinfo
  {pages} {042116} (\bibinfo {year} {2019})}\BibitemShut {NoStop}%
\bibitem [{\citenamefont {Gopalakrishnan}\ and\ \citenamefont
  {Vasseur}(2019)}]{gopalakrishnan19}%
  \BibitemOpen
  \bibfield  {author} {\bibinfo {author} {\bibfnamefont {S.}~\bibnamefont
  {Gopalakrishnan}}\ and\ \bibinfo {author} {\bibfnamefont {R.}~\bibnamefont
  {Vasseur}},\ }\bibfield  {title} {\bibinfo {title} {Kinetic theory of spin
  diffusion and superdiffusion in $xxz$ spin chains},\ }\href
  {https://doi.org/10.1103/PhysRevLett.122.127202} {\bibfield  {journal}
  {\bibinfo  {journal} {Phys. Rev. Lett.}\ }\textbf {\bibinfo {volume} {122}},\
  \bibinfo {pages} {127202} (\bibinfo {year} {2019})}\BibitemShut {NoStop}%
\bibitem [{\citenamefont {De~Nardis}\ \emph {et~al.}(2019)\citenamefont
  {De~Nardis}, \citenamefont {Medenjak}, \citenamefont {Karrasch},\ and\
  \citenamefont {Ilievski}}]{nardis19}%
  \BibitemOpen
  \bibfield  {author} {\bibinfo {author} {\bibfnamefont {J.}~\bibnamefont
  {De~Nardis}}, \bibinfo {author} {\bibfnamefont {M.}~\bibnamefont {Medenjak}},
  \bibinfo {author} {\bibfnamefont {C.}~\bibnamefont {Karrasch}},\ and\
  \bibinfo {author} {\bibfnamefont {E.}~\bibnamefont {Ilievski}},\ }\bibfield
  {title} {\bibinfo {title} {Anomalous spin diffusion in one-dimensional
  antiferromagnets},\ }\href {https://doi.org/10.1103/PhysRevLett.123.186601}
  {\bibfield  {journal} {\bibinfo  {journal} {Phys. Rev. Lett.}\ }\textbf
  {\bibinfo {volume} {123}},\ \bibinfo {pages} {186601} (\bibinfo {year}
  {2019})}\BibitemShut {NoStop}%
\bibitem [{\citenamefont {Ye}\ \emph {et~al.}(2022)\citenamefont {Ye},
  \citenamefont {Machado}, \citenamefont {Kemp}, \citenamefont {Hutson},\ and\
  \citenamefont {Yao}}]{ye22}%
  \BibitemOpen
  \bibfield  {author} {\bibinfo {author} {\bibfnamefont {B.}~\bibnamefont
  {Ye}}, \bibinfo {author} {\bibfnamefont {F.}~\bibnamefont {Machado}},
  \bibinfo {author} {\bibfnamefont {J.}~\bibnamefont {Kemp}}, \bibinfo {author}
  {\bibfnamefont {R.~B.}\ \bibnamefont {Hutson}},\ and\ \bibinfo {author}
  {\bibfnamefont {N.~Y.}\ \bibnamefont {Yao}},\ }\bibfield  {title} {\bibinfo
  {title} {Universal kardar-parisi-zhang dynamics in integrable quantum
  systems},\ }\href {https://doi.org/10.1103/PhysRevLett.129.230602} {\bibfield
   {journal} {\bibinfo  {journal} {Phys. Rev. Lett.}\ }\textbf {\bibinfo
  {volume} {129}},\ \bibinfo {pages} {230602} (\bibinfo {year}
  {2022})}\BibitemShut {NoStop}%
\bibitem [{\citenamefont {Wei}\ \emph {et~al.}(2022)\citenamefont {Wei},
  \citenamefont {Rubio-Abadal}, \citenamefont {Ye}, \citenamefont {Machado},
  \citenamefont {Kemp}, \citenamefont {Srakaew}, \citenamefont {Hollerith},
  \citenamefont {Rui}, \citenamefont {Gopalakrishnan}, \citenamefont {Yao},
  \citenamefont {Bloch},\ and\ \citenamefont {Zeiher}}]{wei22}%
  \BibitemOpen
  \bibfield  {author} {\bibinfo {author} {\bibfnamefont {D.}~\bibnamefont
  {Wei}}, \bibinfo {author} {\bibfnamefont {A.}~\bibnamefont {Rubio-Abadal}},
  \bibinfo {author} {\bibfnamefont {B.}~\bibnamefont {Ye}}, \bibinfo {author}
  {\bibfnamefont {F.}~\bibnamefont {Machado}}, \bibinfo {author} {\bibfnamefont
  {J.}~\bibnamefont {Kemp}}, \bibinfo {author} {\bibfnamefont {K.}~\bibnamefont
  {Srakaew}}, \bibinfo {author} {\bibfnamefont {S.}~\bibnamefont {Hollerith}},
  \bibinfo {author} {\bibfnamefont {J.}~\bibnamefont {Rui}}, \bibinfo {author}
  {\bibfnamefont {S.}~\bibnamefont {Gopalakrishnan}}, \bibinfo {author}
  {\bibfnamefont {N.~Y.}\ \bibnamefont {Yao}}, \bibinfo {author} {\bibfnamefont
  {I.}~\bibnamefont {Bloch}},\ and\ \bibinfo {author} {\bibfnamefont
  {J.}~\bibnamefont {Zeiher}},\ }\bibfield  {title} {\bibinfo {title} {Quantum
  gas microscopy of kardar-parisi-zhang superdiffusion},\ }\href
  {https://doi.org/10.1126/science.abk2397} {\bibfield  {journal} {\bibinfo
  {journal} {Science}\ }\textbf {\bibinfo {volume} {376}},\ \bibinfo {pages}
  {716} (\bibinfo {year} {2022})}\BibitemShut {NoStop}%
\bibitem [{\citenamefont {Rosenberg}\ \emph {et~al.}(2024)\citenamefont
  {Rosenberg}, \citenamefont {Andersen}, \citenamefont {Samajdar},
  \citenamefont {Petukhov}, \citenamefont {Hoke}, \citenamefont {Abanin},
  \citenamefont {Bengtsson}, \citenamefont {Drozdov}, \citenamefont {Erickson},
  \citenamefont {Klimov},\ and\ \citenamefont {et~al}}]{rosenberg24}%
  \BibitemOpen
  \bibfield  {author} {\bibinfo {author} {\bibfnamefont {E.}~\bibnamefont
  {Rosenberg}}, \bibinfo {author} {\bibfnamefont {T.}~\bibnamefont {Andersen}},
  \bibinfo {author} {\bibfnamefont {R.}~\bibnamefont {Samajdar}}, \bibinfo
  {author} {\bibfnamefont {A.}~\bibnamefont {Petukhov}}, \bibinfo {author}
  {\bibfnamefont {J.}~\bibnamefont {Hoke}}, \bibinfo {author} {\bibfnamefont
  {D.}~\bibnamefont {Abanin}}, \bibinfo {author} {\bibfnamefont
  {A.}~\bibnamefont {Bengtsson}}, \bibinfo {author} {\bibfnamefont
  {I.}~\bibnamefont {Drozdov}}, \bibinfo {author} {\bibfnamefont
  {C.}~\bibnamefont {Erickson}}, \bibinfo {author} {\bibfnamefont
  {P.}~\bibnamefont {Klimov}},\ and\ \bibinfo {author} {\bibnamefont {et~al}},\
  }\bibfield  {title} {\bibinfo {title} {Dynamics of magnetization at infinite
  temperature in a heisenberg spin chain},\ }\href
  {https://doi.org/10.1126/science.adi7877} {\bibfield  {journal} {\bibinfo
  {journal} {Science}\ }\textbf {\bibinfo {volume} {384}},\ \bibinfo {pages}
  {48} (\bibinfo {year} {2024})}\BibitemShut {NoStop}%
\bibitem [{\citenamefont {Batchelor}\ \emph {et~al.}(2006)\citenamefont
  {Batchelor}, \citenamefont {Guan},\ and\ \citenamefont
  {Oelkers}}]{batchelor06}%
  \BibitemOpen
  \bibfield  {author} {\bibinfo {author} {\bibfnamefont {M.~T.}\ \bibnamefont
  {Batchelor}}, \bibinfo {author} {\bibfnamefont {X.-W.}\ \bibnamefont
  {Guan}},\ and\ \bibinfo {author} {\bibfnamefont {N.}~\bibnamefont
  {Oelkers}},\ }\bibfield  {title} {\bibinfo {title} {One-dimensional
  interacting anyon gas: Low-energy properties and haldane exclusion
  statistics},\ }\href {https://doi.org/10.1103/PhysRevLett.96.210402}
  {\bibfield  {journal} {\bibinfo  {journal} {Phys. Rev. Lett.}\ }\textbf
  {\bibinfo {volume} {96}},\ \bibinfo {pages} {210402} (\bibinfo {year}
  {2006})}\BibitemShut {NoStop}%
\bibitem [{\citenamefont {Wilczek}(1990)}]{wilczek90}%
  \BibitemOpen
  \bibfield  {author} {\bibinfo {author} {\bibfnamefont {F.}~\bibnamefont
  {Wilczek}},\ }\href {https://doi.org/10.1142/0961} {\emph {\bibinfo {title}
  {\textup{Fractional statistics and anyon superconductivity}}}},\
  Vol.~\bibinfo {volume} {5}\ (\bibinfo  {publisher} {World scientific},\
  \bibinfo {year} {1990})\BibitemShut {NoStop}%
\bibitem [{\citenamefont {Clark}\ \emph {et~al.}(2018)\citenamefont {Clark},
  \citenamefont {Anderson}, \citenamefont {Feng}, \citenamefont {Gaj},
  \citenamefont {Levin},\ and\ \citenamefont {Chin}}]{clark18}%
  \BibitemOpen
  \bibfield  {author} {\bibinfo {author} {\bibfnamefont {L.~W.}\ \bibnamefont
  {Clark}}, \bibinfo {author} {\bibfnamefont {B.~M.}\ \bibnamefont {Anderson}},
  \bibinfo {author} {\bibfnamefont {L.}~\bibnamefont {Feng}}, \bibinfo {author}
  {\bibfnamefont {A.}~\bibnamefont {Gaj}}, \bibinfo {author} {\bibfnamefont
  {K.}~\bibnamefont {Levin}},\ and\ \bibinfo {author} {\bibfnamefont
  {C.}~\bibnamefont {Chin}},\ }\bibfield  {title} {\bibinfo {title}
  {Observation of density-dependent gauge fields in a bose-einstein condensate
  based on micromotion control in a shaken two-dimensional lattice},\ }\href
  {https://doi.org/10.1103/PhysRevLett.121.030402} {\bibfield  {journal}
  {\bibinfo  {journal} {Phys. Rev. Lett.}\ }\textbf {\bibinfo {volume} {121}},\
  \bibinfo {pages} {030402} (\bibinfo {year} {2018})}\BibitemShut {NoStop}%
\bibitem [{\citenamefont {G{\"o}rg}\ \emph {et~al.}(2019)\citenamefont
  {G{\"o}rg}, \citenamefont {Sandholzer}, \citenamefont {Minguzzi},
  \citenamefont {Desbuquois}, \citenamefont {Messer},\ and\ \citenamefont
  {Esslinger}}]{gorg19}%
  \BibitemOpen
  \bibfield  {author} {\bibinfo {author} {\bibfnamefont {F.}~\bibnamefont
  {G{\"o}rg}}, \bibinfo {author} {\bibfnamefont {K.}~\bibnamefont
  {Sandholzer}}, \bibinfo {author} {\bibfnamefont {J.}~\bibnamefont
  {Minguzzi}}, \bibinfo {author} {\bibfnamefont {R.}~\bibnamefont
  {Desbuquois}}, \bibinfo {author} {\bibfnamefont {M.}~\bibnamefont {Messer}},\
  and\ \bibinfo {author} {\bibfnamefont {T.}~\bibnamefont {Esslinger}},\
  }\bibfield  {title} {\bibinfo {title} {Realization of density-dependent
  peierls phases to engineer quantized gauge fields coupled to ultracold
  matter},\ }\href {https://doi.org/10.1038/s41567-019-0615-4} {\bibfield
  {journal} {\bibinfo  {journal} {Nature Physics}\ }\textbf {\bibinfo {volume}
  {15}},\ \bibinfo {pages} {1161} (\bibinfo {year} {2019})}\BibitemShut
  {NoStop}%
\bibitem [{\citenamefont {Schweizer}\ \emph {et~al.}(2019)\citenamefont
  {Schweizer}, \citenamefont {Grusdt}, \citenamefont {Berngruber},
  \citenamefont {Barbiero}, \citenamefont {Demler}, \citenamefont {Goldman},
  \citenamefont {Bloch},\ and\ \citenamefont {Aidelsburger}}]{schweizer19}%
  \BibitemOpen
  \bibfield  {author} {\bibinfo {author} {\bibfnamefont {C.}~\bibnamefont
  {Schweizer}}, \bibinfo {author} {\bibfnamefont {F.}~\bibnamefont {Grusdt}},
  \bibinfo {author} {\bibfnamefont {M.}~\bibnamefont {Berngruber}}, \bibinfo
  {author} {\bibfnamefont {L.}~\bibnamefont {Barbiero}}, \bibinfo {author}
  {\bibfnamefont {E.}~\bibnamefont {Demler}}, \bibinfo {author} {\bibfnamefont
  {N.}~\bibnamefont {Goldman}}, \bibinfo {author} {\bibfnamefont
  {I.}~\bibnamefont {Bloch}},\ and\ \bibinfo {author} {\bibfnamefont
  {M.}~\bibnamefont {Aidelsburger}},\ }\bibfield  {title} {\bibinfo {title}
  {Floquet approach to $\mathbb{Z}_2$ lattice gauge theories with ultracold
  atoms in optical lattices},\ }\href
  {https://doi.org/10.1038/s41567-019-0649-7} {\bibfield  {journal} {\bibinfo
  {journal} {Nature Physics}\ }\textbf {\bibinfo {volume} {15}},\ \bibinfo
  {pages} {1168} (\bibinfo {year} {2019})}\BibitemShut {NoStop}%
\bibitem [{\citenamefont {Haegeman}\ \emph {et~al.}(2011)\citenamefont
  {Haegeman}, \citenamefont {Cirac}, \citenamefont {Osborne}, \citenamefont
  {Pi\ifmmode~\check{z}\else \v{z}\fi{}orn}, \citenamefont {Verschelde},\ and\
  \citenamefont {Verstraete}}]{haegeman11}%
  \BibitemOpen
  \bibfield  {author} {\bibinfo {author} {\bibfnamefont {J.}~\bibnamefont
  {Haegeman}}, \bibinfo {author} {\bibfnamefont {J.~I.}\ \bibnamefont {Cirac}},
  \bibinfo {author} {\bibfnamefont {T.~J.}\ \bibnamefont {Osborne}}, \bibinfo
  {author} {\bibfnamefont {I.}~\bibnamefont {Pi\ifmmode~\check{z}\else
  \v{z}\fi{}orn}}, \bibinfo {author} {\bibfnamefont {H.}~\bibnamefont
  {Verschelde}},\ and\ \bibinfo {author} {\bibfnamefont {F.}~\bibnamefont
  {Verstraete}},\ }\bibfield  {title} {\bibinfo {title} {{Time-dependent
  variational principle for quantum lattices}},\ }\href
  {https://doi.org/10.1103/PhysRevLett.107.070601} {\bibfield  {journal}
  {\bibinfo  {journal} {Phys. Rev. Lett.}\ }\textbf {\bibinfo {volume} {107}},\
  \bibinfo {pages} {070601} (\bibinfo {year} {2011})}\BibitemShut {NoStop}%
\bibitem [{\citenamefont {Hauschild}\ \emph {et~al.}(2024)\citenamefont
  {Hauschild}, \citenamefont {Unfried}, \citenamefont {Anand}, \citenamefont
  {Andrews}, \citenamefont {Bintz}, \citenamefont {Borla}, \citenamefont
  {Divic}, \citenamefont {Drescher}, \citenamefont {Geiger}, \citenamefont
  {Hefel}, \citenamefont {Hémery}, \citenamefont {Kadow}, \citenamefont
  {Kemp}, \citenamefont {Kirchner}, \citenamefont {Liu}, \citenamefont
  {Möller}, \citenamefont {Parker}, \citenamefont {Rader}, \citenamefont
  {Romen}, \citenamefont {Scalet}, \citenamefont {Schoonderwoerd},
  \citenamefont {Schulz}, \citenamefont {Soejima}, \citenamefont {Thoma},
  \citenamefont {Wu}, \citenamefont {Zechmann}, \citenamefont {Zweng},
  \citenamefont {Mong}, \citenamefont {Zaletel},\ and\ \citenamefont
  {Pollmann}}]{tenpy24}%
  \BibitemOpen
  \bibfield  {author} {\bibinfo {author} {\bibfnamefont {J.}~\bibnamefont
  {Hauschild}}, \bibinfo {author} {\bibfnamefont {J.}~\bibnamefont {Unfried}},
  \bibinfo {author} {\bibfnamefont {S.}~\bibnamefont {Anand}}, \bibinfo
  {author} {\bibfnamefont {B.}~\bibnamefont {Andrews}}, \bibinfo {author}
  {\bibfnamefont {M.}~\bibnamefont {Bintz}}, \bibinfo {author} {\bibfnamefont
  {U.}~\bibnamefont {Borla}}, \bibinfo {author} {\bibfnamefont
  {S.}~\bibnamefont {Divic}}, \bibinfo {author} {\bibfnamefont
  {M.}~\bibnamefont {Drescher}}, \bibinfo {author} {\bibfnamefont
  {J.}~\bibnamefont {Geiger}}, \bibinfo {author} {\bibfnamefont
  {M.}~\bibnamefont {Hefel}}, \bibinfo {author} {\bibfnamefont
  {K.}~\bibnamefont {Hémery}}, \bibinfo {author} {\bibfnamefont
  {W.}~\bibnamefont {Kadow}}, \bibinfo {author} {\bibfnamefont
  {J.}~\bibnamefont {Kemp}}, \bibinfo {author} {\bibfnamefont {N.}~\bibnamefont
  {Kirchner}}, \bibinfo {author} {\bibfnamefont {V.~S.}\ \bibnamefont {Liu}},
  \bibinfo {author} {\bibfnamefont {G.}~\bibnamefont {Möller}}, \bibinfo
  {author} {\bibfnamefont {D.}~\bibnamefont {Parker}}, \bibinfo {author}
  {\bibfnamefont {M.}~\bibnamefont {Rader}}, \bibinfo {author} {\bibfnamefont
  {A.}~\bibnamefont {Romen}}, \bibinfo {author} {\bibfnamefont
  {S.}~\bibnamefont {Scalet}}, \bibinfo {author} {\bibfnamefont
  {L.}~\bibnamefont {Schoonderwoerd}}, \bibinfo {author} {\bibfnamefont
  {M.}~\bibnamefont {Schulz}}, \bibinfo {author} {\bibfnamefont
  {T.}~\bibnamefont {Soejima}}, \bibinfo {author} {\bibfnamefont
  {P.}~\bibnamefont {Thoma}}, \bibinfo {author} {\bibfnamefont
  {Y.}~\bibnamefont {Wu}}, \bibinfo {author} {\bibfnamefont {P.}~\bibnamefont
  {Zechmann}}, \bibinfo {author} {\bibfnamefont {L.}~\bibnamefont {Zweng}},
  \bibinfo {author} {\bibfnamefont {R.~S.~K.}\ \bibnamefont {Mong}}, \bibinfo
  {author} {\bibfnamefont {M.~P.}\ \bibnamefont {Zaletel}},\ and\ \bibinfo
  {author} {\bibfnamefont {F.}~\bibnamefont {Pollmann}},\ }\bibfield  {title}
  {\bibinfo {title} {{Tensor network Python (TeNPy) version 1}},\ }\href
  {https://doi.org/10.21468/SciPostPhysCodeb.41} {\bibfield  {journal}
  {\bibinfo  {journal} {SciPost Phys. Codebases}\ ,\ \bibinfo {pages} {41}}
  (\bibinfo {year} {2024})}\BibitemShut {NoStop}%
\bibitem [{\citenamefont {Rispoli}\ \emph {et~al.}(2019)\citenamefont
  {Rispoli}, \citenamefont {Lukin}, \citenamefont {Schittko}, \citenamefont
  {Kim}, \citenamefont {Tai}, \citenamefont {L{\'e}onard},\ and\ \citenamefont
  {Greiner}}]{rispoli19}%
  \BibitemOpen
  \bibfield  {author} {\bibinfo {author} {\bibfnamefont {M.}~\bibnamefont
  {Rispoli}}, \bibinfo {author} {\bibfnamefont {A.}~\bibnamefont {Lukin}},
  \bibinfo {author} {\bibfnamefont {R.}~\bibnamefont {Schittko}}, \bibinfo
  {author} {\bibfnamefont {S.}~\bibnamefont {Kim}}, \bibinfo {author}
  {\bibfnamefont {M.~E.}\ \bibnamefont {Tai}}, \bibinfo {author} {\bibfnamefont
  {J.}~\bibnamefont {L{\'e}onard}},\ and\ \bibinfo {author} {\bibfnamefont
  {M.}~\bibnamefont {Greiner}},\ }\bibfield  {title} {\bibinfo {title} {Quantum
  critical behaviour at the many-body localization transition},\ }\href
  {https://doi.org/10.1038/s41586-019-1527-2} {\bibfield  {journal} {\bibinfo
  {journal} {Nature}\ }\textbf {\bibinfo {volume} {573}},\ \bibinfo {pages}
  {385} (\bibinfo {year} {2019})}\BibitemShut {NoStop}%
\bibitem [{\citenamefont {Due\~nas}\ \emph {et~al.}(2025)\citenamefont
  {Due\~nas}, \citenamefont {Pe\~na},\ and\ \citenamefont
  {Rodr\'{\i}guez}}]{duenas25}%
  \BibitemOpen
  \bibfield  {author} {\bibinfo {author} {\bibfnamefont {O.}~\bibnamefont
  {Due\~nas}}, \bibinfo {author} {\bibfnamefont {D.}~\bibnamefont {Pe\~na}},\
  and\ \bibinfo {author} {\bibfnamefont {A.}~\bibnamefont {Rodr\'{\i}guez}},\
  }\bibfield  {title} {\bibinfo {title} {Propagation of two-particle
  correlations across the chaotic phase for interacting bosons},\ }\href
  {https://doi.org/10.1103/PhysRevResearch.7.L012031} {\bibfield  {journal}
  {\bibinfo  {journal} {Phys. Rev. Res.}\ }\textbf {\bibinfo {volume} {7}},\
  \bibinfo {pages} {L012031} (\bibinfo {year} {2025})}\BibitemShut {NoStop}%
\bibitem [{\citenamefont {Bhakuni}\ and\ \citenamefont
  {Lev}(2024)}]{bhakuni24}%
  \BibitemOpen
  \bibfield  {author} {\bibinfo {author} {\bibfnamefont {D.~S.}\ \bibnamefont
  {Bhakuni}}\ and\ \bibinfo {author} {\bibfnamefont {Y.~B.}\ \bibnamefont
  {Lev}},\ }\bibfield  {title} {\bibinfo {title} {Dynamic scaling relation in
  quantum many-body systems},\ }\href
  {https://doi.org/10.1103/PhysRevB.110.014203} {\bibfield  {journal} {\bibinfo
   {journal} {Phys. Rev. B}\ }\textbf {\bibinfo {volume} {110}},\ \bibinfo
  {pages} {014203} (\bibinfo {year} {2024})}\BibitemShut {NoStop}%
\bibitem [{\citenamefont {Kwon}\ \emph {et~al.}()\citenamefont {Kwon},
  \citenamefont {Fujimoto}, \citenamefont {Hur}, \citenamefont {Lee},
  \citenamefont {Hwang}, \citenamefont {Kim}, \citenamefont {Hamazaki},
  \citenamefont {Kawaguchi},\ and\ \citenamefont {yoon Choi}}]{kwon26}%
  \BibitemOpen
  \bibfield  {author} {\bibinfo {author} {\bibfnamefont {K.}~\bibnamefont
  {Kwon}}, \bibinfo {author} {\bibfnamefont {K.}~\bibnamefont {Fujimoto}},
  \bibinfo {author} {\bibfnamefont {J.}~\bibnamefont {Hur}}, \bibinfo {author}
  {\bibfnamefont {B.}~\bibnamefont {Lee}}, \bibinfo {author} {\bibfnamefont
  {S.}~\bibnamefont {Hwang}}, \bibinfo {author} {\bibfnamefont
  {S.}~\bibnamefont {Kim}}, \bibinfo {author} {\bibfnamefont {R.}~\bibnamefont
  {Hamazaki}}, \bibinfo {author} {\bibfnamefont {Y.}~\bibnamefont
  {Kawaguchi}},\ and\ \bibinfo {author} {\bibfnamefont {J.}~\bibnamefont {yoon
  Choi}},\ }\bibfield  {title} {\bibinfo {title} {{Universal Family-Vicsek
  scaling in quantum gases far from equilibrium}},\ }\href
  {https://arxiv.org/abs/2603.09060} {\bibinfo  {journal} {arXiv:2603.09060}\
  }\BibitemShut {NoStop}%
\bibitem [{\citenamefont {Edwards}\ and\ \citenamefont
  {Wilkinson}(1982)}]{edwards82}%
  \BibitemOpen
\bibfield  {journal} {  }\bibfield  {author} {\bibinfo {author} {\bibfnamefont
  {S.~F.}\ \bibnamefont {Edwards}}\ and\ \bibinfo {author} {\bibfnamefont
  {D.~R.}\ \bibnamefont {Wilkinson}},\ }\bibfield  {title} {\bibinfo {title}
  {The surface statistics of a granular aggregate},\ }\href
  {https://doi.org/10.1098/rspa.1982.0056} {\bibfield  {journal} {\bibinfo
  {journal} {Proceedings of the Royal Society A: Mathematical, Physical and
  Engineering Sciences}\ ,\ \bibinfo {pages} {17}} (\bibinfo {year}
  {1982})}\BibitemShut {NoStop}%
\bibitem [{\citenamefont {Vicsek}\ and\ \citenamefont
  {Family}(1984)}]{vicsek84}%
  \BibitemOpen
  \bibfield  {author} {\bibinfo {author} {\bibfnamefont {T.}~\bibnamefont
  {Vicsek}}\ and\ \bibinfo {author} {\bibfnamefont {F.}~\bibnamefont
  {Family}},\ }\bibfield  {title} {\bibinfo {title} {Dynamic scaling for
  aggregation of clusters},\ }\href
  {https://doi.org/10.1103/PhysRevLett.52.1669} {\bibfield  {journal} {\bibinfo
   {journal} {Phys. Rev. Lett.}\ }\textbf {\bibinfo {volume} {52}},\ \bibinfo
  {pages} {1669} (\bibinfo {year} {1984})}\BibitemShut {NoStop}%
\bibitem [{\citenamefont {Family}\ and\ \citenamefont
  {Vicsek}(1985)}]{family85}%
  \BibitemOpen
  \bibfield  {author} {\bibinfo {author} {\bibfnamefont {F.}~\bibnamefont
  {Family}}\ and\ \bibinfo {author} {\bibfnamefont {T.}~\bibnamefont
  {Vicsek}},\ }\bibfield  {title} {\bibinfo {title} {Scaling of the active zone
  in the eden process on percolation networks and the ballistic deposition
  model},\ }\href {https://doi.org/10.1088/0305-4470/18/2/005} {\bibfield
  {journal} {\bibinfo  {journal} {Journal of Physics A: Mathematical and
  General}\ }\textbf {\bibinfo {volume} {18}},\ \bibinfo {pages} {L75}
  (\bibinfo {year} {1985})}\BibitemShut {NoStop}%
\bibitem [{\citenamefont {Kardar}\ \emph {et~al.}(1986)\citenamefont {Kardar},
  \citenamefont {Parisi},\ and\ \citenamefont {Zhang}}]{kardar86}%
  \BibitemOpen
  \bibfield  {author} {\bibinfo {author} {\bibfnamefont {M.}~\bibnamefont
  {Kardar}}, \bibinfo {author} {\bibfnamefont {G.}~\bibnamefont {Parisi}},\
  and\ \bibinfo {author} {\bibfnamefont {Y.-C.}\ \bibnamefont {Zhang}},\
  }\bibfield  {title} {\bibinfo {title} {Dynamic scaling of growing
  interfaces},\ }\href {https://doi.org/10.1103/PhysRevLett.56.889} {\bibfield
  {journal} {\bibinfo  {journal} {Phys. Rev. Lett.}\ }\textbf {\bibinfo
  {volume} {56}},\ \bibinfo {pages} {889} (\bibinfo {year} {1986})}\BibitemShut
  {NoStop}%
\bibitem [{\citenamefont {Fujimoto}\ \emph {et~al.}(2020)\citenamefont
  {Fujimoto}, \citenamefont {Hamazaki},\ and\ \citenamefont
  {Kawaguchi}}]{fujimoto20}%
  \BibitemOpen
  \bibfield  {author} {\bibinfo {author} {\bibfnamefont {K.}~\bibnamefont
  {Fujimoto}}, \bibinfo {author} {\bibfnamefont {R.}~\bibnamefont {Hamazaki}},\
  and\ \bibinfo {author} {\bibfnamefont {Y.}~\bibnamefont {Kawaguchi}},\
  }\bibfield  {title} {\bibinfo {title} {Family-vicsek scaling of roughness
  growth in a strongly interacting bose gas},\ }\href
  {https://doi.org/10.1103/PhysRevLett.124.210604} {\bibfield  {journal}
  {\bibinfo  {journal} {Phys. Rev. Lett.}\ }\textbf {\bibinfo {volume} {124}},\
  \bibinfo {pages} {210604} (\bibinfo {year} {2020})}\BibitemShut {NoStop}%
\bibitem [{SM()}]{SM}%
  \BibitemOpen
  \href@noop {} {}\bibinfo {note} {See Supplemental Material for CTD analyses,
  numerical convergence, extraction of growth exponent, robustness to initial
  states and disorder, entanglement-entropy scaling analyses, additional
  holon–doublon propagation mechanisms, and experimental
  realization.}\BibitemShut {Stop}%
\bibitem [{\citenamefont {Fujimoto}\ \emph {et~al.}(2022)\citenamefont
  {Fujimoto}, \citenamefont {Hamazaki},\ and\ \citenamefont
  {Kawaguchi}}]{fujimoto22}%
  \BibitemOpen
  \bibfield  {author} {\bibinfo {author} {\bibfnamefont {K.}~\bibnamefont
  {Fujimoto}}, \bibinfo {author} {\bibfnamefont {R.}~\bibnamefont {Hamazaki}},\
  and\ \bibinfo {author} {\bibfnamefont {Y.}~\bibnamefont {Kawaguchi}},\
  }\bibfield  {title} {\bibinfo {title} {Impact of dissipation on universal
  fluctuation dynamics in open quantum systems},\ }\href
  {https://doi.org/10.1103/PhysRevLett.129.110403} {\bibfield  {journal}
  {\bibinfo  {journal} {Phys. Rev. Lett.}\ }\textbf {\bibinfo {volume} {129}},\
  \bibinfo {pages} {110403} (\bibinfo {year} {2022})}\BibitemShut {NoStop}%
\bibitem [{\citenamefont {Fujimoto}\ \emph {et~al.}(2021)\citenamefont
  {Fujimoto}, \citenamefont {Hamazaki},\ and\ \citenamefont
  {Kawaguchi}}]{PhysRevLett.127.090601}%
  \BibitemOpen
  \bibfield  {author} {\bibinfo {author} {\bibfnamefont {K.}~\bibnamefont
  {Fujimoto}}, \bibinfo {author} {\bibfnamefont {R.}~\bibnamefont {Hamazaki}},\
  and\ \bibinfo {author} {\bibfnamefont {Y.}~\bibnamefont {Kawaguchi}},\
  }\bibfield  {title} {\bibinfo {title} {Dynamical scaling of surface roughness
  and entanglement entropy in disordered fermion models},\ }\href
  {https://doi.org/10.1103/PhysRevLett.127.090601} {\bibfield  {journal}
  {\bibinfo  {journal} {Phys. Rev. Lett.}\ }\textbf {\bibinfo {volume} {127}},\
  \bibinfo {pages} {090601} (\bibinfo {year} {2021})}\BibitemShut {NoStop}%
\bibitem [{\citenamefont {\ifmmode \check{Z}\else
  \v{Z}\fi{}nidari\ifmmode~\check{c}\else \v{c}\fi{}}\ \emph
  {et~al.}(2008)\citenamefont {\ifmmode \check{Z}\else
  \v{Z}\fi{}nidari\ifmmode~\check{c}\else \v{c}\fi{}}, \citenamefont {Prosen},\
  and\ \citenamefont {Prelov\ifmmode~\check{s}\else
  \v{s}\fi{}ek}}]{PhysRevB.77.064426}%
  \BibitemOpen
  \bibfield  {author} {\bibinfo {author} {\bibfnamefont {M.}~\bibnamefont
  {\ifmmode \check{Z}\else \v{Z}\fi{}nidari\ifmmode~\check{c}\else
  \v{c}\fi{}}}, \bibinfo {author} {\bibfnamefont {T.~c.~v.}\ \bibnamefont
  {Prosen}},\ and\ \bibinfo {author} {\bibfnamefont {P.}~\bibnamefont
  {Prelov\ifmmode~\check{s}\else \v{s}\fi{}ek}},\ }\bibfield  {title} {\bibinfo
  {title} {Many-body localization in the heisenberg $xxz$ magnet in a random
  field},\ }\href {https://doi.org/10.1103/PhysRevB.77.064426} {\bibfield
  {journal} {\bibinfo  {journal} {Phys. Rev. B}\ }\textbf {\bibinfo {volume}
  {77}},\ \bibinfo {pages} {064426} (\bibinfo {year} {2008})}\BibitemShut
  {NoStop}%
\bibitem [{\citenamefont {Bardarson}\ \emph {et~al.}(2012)\citenamefont
  {Bardarson}, \citenamefont {Pollmann},\ and\ \citenamefont
  {Moore}}]{bardarson12}%
  \BibitemOpen
  \bibfield  {author} {\bibinfo {author} {\bibfnamefont {J.~H.}\ \bibnamefont
  {Bardarson}}, \bibinfo {author} {\bibfnamefont {F.}~\bibnamefont
  {Pollmann}},\ and\ \bibinfo {author} {\bibfnamefont {J.~E.}\ \bibnamefont
  {Moore}},\ }\bibfield  {title} {\bibinfo {title} {Unbounded growth of
  entanglement in models of many-body localization},\ }\href
  {https://doi.org/10.1103/PhysRevLett.109.017202} {\bibfield  {journal}
  {\bibinfo  {journal} {Phys. Rev. Lett.}\ }\textbf {\bibinfo {volume} {109}},\
  \bibinfo {pages} {017202} (\bibinfo {year} {2012})}\BibitemShut {NoStop}%
\bibitem [{\citenamefont {Huse}\ \emph {et~al.}(2014)\citenamefont {Huse},
  \citenamefont {Nandkishore},\ and\ \citenamefont {Oganesyan}}]{huse14}%
  \BibitemOpen
  \bibfield  {author} {\bibinfo {author} {\bibfnamefont {D.~A.}\ \bibnamefont
  {Huse}}, \bibinfo {author} {\bibfnamefont {R.}~\bibnamefont {Nandkishore}},\
  and\ \bibinfo {author} {\bibfnamefont {V.}~\bibnamefont {Oganesyan}},\
  }\bibfield  {title} {\bibinfo {title} {Phenomenology of fully
  many-body-localized systems},\ }\href
  {https://doi.org/10.1103/PhysRevB.90.174202} {\bibfield  {journal} {\bibinfo
  {journal} {Phys. Rev. B}\ }\textbf {\bibinfo {volume} {90}},\ \bibinfo
  {pages} {174202} (\bibinfo {year} {2014})}\BibitemShut {NoStop}%
\bibitem [{\citenamefont {Lukin}\ \emph {et~al.}(2019)\citenamefont {Lukin},
  \citenamefont {Rispoli}, \citenamefont {Schittko}, \citenamefont {Tai},
  \citenamefont {Kaufman}, \citenamefont {Choi}, \citenamefont {Khemani},
  \citenamefont {L{\'e}onard},\ and\ \citenamefont {Greiner}}]{lukin19}%
  \BibitemOpen
  \bibfield  {author} {\bibinfo {author} {\bibfnamefont {A.}~\bibnamefont
  {Lukin}}, \bibinfo {author} {\bibfnamefont {M.}~\bibnamefont {Rispoli}},
  \bibinfo {author} {\bibfnamefont {R.}~\bibnamefont {Schittko}}, \bibinfo
  {author} {\bibfnamefont {M.~E.}\ \bibnamefont {Tai}}, \bibinfo {author}
  {\bibfnamefont {A.~M.}\ \bibnamefont {Kaufman}}, \bibinfo {author}
  {\bibfnamefont {S.}~\bibnamefont {Choi}}, \bibinfo {author} {\bibfnamefont
  {V.}~\bibnamefont {Khemani}}, \bibinfo {author} {\bibfnamefont
  {J.}~\bibnamefont {L{\'e}onard}},\ and\ \bibinfo {author} {\bibfnamefont
  {M.}~\bibnamefont {Greiner}},\ }\bibfield  {title} {\bibinfo {title} {Probing
  entanglement in a many-body–localized system},\ }\href
  {https://doi.org/10.1126/science.aau0818} {\bibfield  {journal} {\bibinfo
  {journal} {Science}\ }\textbf {\bibinfo {volume} {364}},\ \bibinfo {pages}
  {256} (\bibinfo {year} {2019})}\BibitemShut {NoStop}%
\bibitem [{\citenamefont {Kiefer-Emmanouilidis}\ \emph
  {et~al.}(2020)\citenamefont {Kiefer-Emmanouilidis}, \citenamefont {Unanyan},
  \citenamefont {Fleischhauer},\ and\ \citenamefont
  {Sirker}}]{emmanouilidis20}%
  \BibitemOpen
  \bibfield  {author} {\bibinfo {author} {\bibfnamefont {M.}~\bibnamefont
  {Kiefer-Emmanouilidis}}, \bibinfo {author} {\bibfnamefont {R.}~\bibnamefont
  {Unanyan}}, \bibinfo {author} {\bibfnamefont {M.}~\bibnamefont
  {Fleischhauer}},\ and\ \bibinfo {author} {\bibfnamefont {J.}~\bibnamefont
  {Sirker}},\ }\bibfield  {title} {\bibinfo {title} {Evidence for unbounded
  growth of the number entropy in many-body localized phases},\ }\href
  {https://doi.org/10.1103/PhysRevLett.124.243601} {\bibfield  {journal}
  {\bibinfo  {journal} {Phys. Rev. Lett.}\ }\textbf {\bibinfo {volume} {124}},\
  \bibinfo {pages} {243601} (\bibinfo {year} {2020})}\BibitemShut {NoStop}%
\bibitem [{\citenamefont {Luitz}\ and\ \citenamefont {Lev}(2020)}]{luitz20}%
  \BibitemOpen
  \bibfield  {author} {\bibinfo {author} {\bibfnamefont {D.~J.}\ \bibnamefont
  {Luitz}}\ and\ \bibinfo {author} {\bibfnamefont {Y.~B.}\ \bibnamefont
  {Lev}},\ }\bibfield  {title} {\bibinfo {title} {Absence of slow particle
  transport in the many-body localized phase},\ }\href
  {https://doi.org/10.1103/PhysRevB.102.100202} {\bibfield  {journal} {\bibinfo
   {journal} {Phys. Rev. B}\ }\textbf {\bibinfo {volume} {102}},\ \bibinfo
  {pages} {100202} (\bibinfo {year} {2020})}\BibitemShut {NoStop}%
\bibitem [{\citenamefont {Chen}\ \emph {et~al.}(2025)\citenamefont {Chen},
  \citenamefont {Chen},\ and\ \citenamefont {Wang}}]{chenj25}%
  \BibitemOpen
  \bibfield  {author} {\bibinfo {author} {\bibfnamefont {J.}~\bibnamefont
  {Chen}}, \bibinfo {author} {\bibfnamefont {C.}~\bibnamefont {Chen}},\ and\
  \bibinfo {author} {\bibfnamefont {X.}~\bibnamefont {Wang}},\ }\bibfield
  {title} {\bibinfo {title} {Symmetry- and energy-resolved entanglement
  dynamics in a disordered bose-hubbard model},\ }\href
  {https://doi.org/10.1103/88sq-cm27} {\bibfield  {journal} {\bibinfo
  {journal} {Phys. Rev. B}\ }\textbf {\bibinfo {volume} {112}},\ \bibinfo
  {pages} {094201} (\bibinfo {year} {2025})}\BibitemShut {NoStop}%
\bibitem [{\citenamefont {Cheneau}\ \emph {et~al.}(2012)\citenamefont
  {Cheneau}, \citenamefont {Barmettler}, \citenamefont {Poletti}, \citenamefont
  {Endres}, \citenamefont {Schau{\ss}}, \citenamefont {Fukuhara}, \citenamefont
  {Gross}, \citenamefont {Bloch}, \citenamefont {Kollath},\ and\ \citenamefont
  {Kuhr}}]{cheneau12}%
  \BibitemOpen
  \bibfield  {author} {\bibinfo {author} {\bibfnamefont {M.}~\bibnamefont
  {Cheneau}}, \bibinfo {author} {\bibfnamefont {P.}~\bibnamefont {Barmettler}},
  \bibinfo {author} {\bibfnamefont {D.}~\bibnamefont {Poletti}}, \bibinfo
  {author} {\bibfnamefont {M.}~\bibnamefont {Endres}}, \bibinfo {author}
  {\bibfnamefont {P.}~\bibnamefont {Schau{\ss}}}, \bibinfo {author}
  {\bibfnamefont {T.}~\bibnamefont {Fukuhara}}, \bibinfo {author}
  {\bibfnamefont {C.}~\bibnamefont {Gross}}, \bibinfo {author} {\bibfnamefont
  {I.}~\bibnamefont {Bloch}}, \bibinfo {author} {\bibfnamefont
  {C.}~\bibnamefont {Kollath}},\ and\ \bibinfo {author} {\bibfnamefont
  {S.}~\bibnamefont {Kuhr}},\ }\bibfield  {title} {\bibinfo {title}
  {Light-cone-like spreading of correlations in a quantum many-body system},\
  }\href {https://doi.org/10.1038/nature10748} {\bibfield  {journal} {\bibinfo
  {journal} {Nature}\ }\textbf {\bibinfo {volume} {481}},\ \bibinfo {pages}
  {484} (\bibinfo {year} {2012})}\BibitemShut {NoStop}%
\bibitem [{\citenamefont {Endres}\ \emph {et~al.}(2011)\citenamefont {Endres},
  \citenamefont {Cheneau}, \citenamefont {Fukuhara}, \citenamefont
  {Weitenberg}, \citenamefont {Schauss}, \citenamefont {Gross}, \citenamefont
  {Mazza}, \citenamefont {Banuls}, \citenamefont {Pollet}, \citenamefont
  {Bloch}, ,\ and\ \citenamefont {Kuhr}}]{endres11}%
  \BibitemOpen
  \bibfield  {author} {\bibinfo {author} {\bibfnamefont {M.}~\bibnamefont
  {Endres}}, \bibinfo {author} {\bibfnamefont {M.}~\bibnamefont {Cheneau}},
  \bibinfo {author} {\bibfnamefont {T.}~\bibnamefont {Fukuhara}}, \bibinfo
  {author} {\bibfnamefont {C.}~\bibnamefont {Weitenberg}}, \bibinfo {author}
  {\bibfnamefont {P.}~\bibnamefont {Schauss}}, \bibinfo {author} {\bibfnamefont
  {C.}~\bibnamefont {Gross}}, \bibinfo {author} {\bibfnamefont
  {L.}~\bibnamefont {Mazza}}, \bibinfo {author} {\bibfnamefont {M.~C.}\
  \bibnamefont {Banuls}}, \bibinfo {author} {\bibfnamefont {L.}~\bibnamefont
  {Pollet}}, \bibinfo {author} {\bibfnamefont {I.}~\bibnamefont {Bloch}}, ,\
  and\ \bibinfo {author} {\bibfnamefont {S.}~\bibnamefont {Kuhr}},\ }\bibfield
  {title} {\bibinfo {title} {Observation of correlated particle-hole pairs and
  string order in low-dimensional mott insulators},\ }\href
  {https://doi.org/10.1126/science.1209284} {\bibfield  {journal} {\bibinfo
  {journal} {Science}\ }\textbf {\bibinfo {volume} {334}},\ \bibinfo {pages}
  {200} (\bibinfo {year} {2011})}\BibitemShut {NoStop}%
\bibitem [{\citenamefont {Schweigler}\ \emph {et~al.}(2017)\citenamefont
  {Schweigler}, \citenamefont {Kasper}, \citenamefont {Erne}, \citenamefont
  {Mazets}, \citenamefont {Rauer}, \citenamefont {Cataldini}, \citenamefont
  {Langen}, \citenamefont {Gasenzer}, \citenamefont {Berges},\ and\
  \citenamefont {Schmiedmayer}}]{schweigler17}%
  \BibitemOpen
  \bibfield  {author} {\bibinfo {author} {\bibfnamefont {T.}~\bibnamefont
  {Schweigler}}, \bibinfo {author} {\bibfnamefont {V.}~\bibnamefont {Kasper}},
  \bibinfo {author} {\bibfnamefont {S.}~\bibnamefont {Erne}}, \bibinfo {author}
  {\bibfnamefont {I.}~\bibnamefont {Mazets}}, \bibinfo {author} {\bibfnamefont
  {B.}~\bibnamefont {Rauer}}, \bibinfo {author} {\bibfnamefont
  {F.}~\bibnamefont {Cataldini}}, \bibinfo {author} {\bibfnamefont
  {T.}~\bibnamefont {Langen}}, \bibinfo {author} {\bibfnamefont
  {T.}~\bibnamefont {Gasenzer}}, \bibinfo {author} {\bibfnamefont
  {J.}~\bibnamefont {Berges}},\ and\ \bibinfo {author} {\bibfnamefont
  {J.}~\bibnamefont {Schmiedmayer}},\ }\bibfield  {title} {\bibinfo {title}
  {Experimental characterization of a quantum many-body system via higher-order
  correlations},\ }\href {https://doi.org/10.1038/nature22310} {\bibfield
  {journal} {\bibinfo  {journal} {Nature}\ }\textbf {\bibinfo {volume} {545}},\
  \bibinfo {pages} {323} (\bibinfo {year} {2017})}\BibitemShut {NoStop}%
\bibitem [{\citenamefont {Islam}\ \emph {et~al.}(2015)\citenamefont {Islam},
  \citenamefont {Ma}, \citenamefont {Preiss}, \citenamefont {Eric~Tai},
  \citenamefont {Lukin}, \citenamefont {Rispoli},\ and\ \citenamefont
  {Greiner}}]{islam15}%
  \BibitemOpen
  \bibfield  {author} {\bibinfo {author} {\bibfnamefont {R.}~\bibnamefont
  {Islam}}, \bibinfo {author} {\bibfnamefont {R.}~\bibnamefont {Ma}}, \bibinfo
  {author} {\bibfnamefont {P.~M.}\ \bibnamefont {Preiss}}, \bibinfo {author}
  {\bibfnamefont {M.}~\bibnamefont {Eric~Tai}}, \bibinfo {author}
  {\bibfnamefont {A.}~\bibnamefont {Lukin}}, \bibinfo {author} {\bibfnamefont
  {M.}~\bibnamefont {Rispoli}},\ and\ \bibinfo {author} {\bibfnamefont
  {M.}~\bibnamefont {Greiner}},\ }\bibfield  {title} {\bibinfo {title}
  {Measuring entanglement entropy in a quantum many-body system},\ }\href
  {https://doi.org/10.1038/nature15750} {\bibfield  {journal} {\bibinfo
  {journal} {Nature}\ }\textbf {\bibinfo {volume} {528}},\ \bibinfo {pages}
  {77} (\bibinfo {year} {2015})}\BibitemShut {NoStop}%
\bibitem [{\citenamefont {Kaufman}\ \emph {et~al.}(2016)\citenamefont
  {Kaufman}, \citenamefont {Tai}, \citenamefont {Lukin}, \citenamefont
  {Rispoli}, \citenamefont {Schittko}, \citenamefont {Preiss},\ and\
  \citenamefont {Greiner}}]{kaufman16}%
  \BibitemOpen
  \bibfield  {author} {\bibinfo {author} {\bibfnamefont {A.~M.}\ \bibnamefont
  {Kaufman}}, \bibinfo {author} {\bibfnamefont {M.~E.}\ \bibnamefont {Tai}},
  \bibinfo {author} {\bibfnamefont {A.}~\bibnamefont {Lukin}}, \bibinfo
  {author} {\bibfnamefont {M.}~\bibnamefont {Rispoli}}, \bibinfo {author}
  {\bibfnamefont {R.}~\bibnamefont {Schittko}}, \bibinfo {author}
  {\bibfnamefont {P.~M.}\ \bibnamefont {Preiss}},\ and\ \bibinfo {author}
  {\bibfnamefont {M.}~\bibnamefont {Greiner}},\ }\bibfield  {title} {\bibinfo
  {title} {Quantum thermalization through entanglement in an isolated many-body
  system},\ }\href {https://doi.org/10.1126/science.aaf6725} {\bibfield
  {journal} {\bibinfo  {journal} {Science}\ }\textbf {\bibinfo {volume}
  {353}},\ \bibinfo {pages} {794} (\bibinfo {year} {2016})}\BibitemShut
  {NoStop}%
\bibitem [{\citenamefont {Mazza}\ \emph {et~al.}(2018)\citenamefont {Mazza},
  \citenamefont {Viti}, \citenamefont {Carrega}, \citenamefont {Rossini},\ and\
  \citenamefont {De~Luca}}]{mazza18}%
  \BibitemOpen
  \bibfield  {author} {\bibinfo {author} {\bibfnamefont {L.}~\bibnamefont
  {Mazza}}, \bibinfo {author} {\bibfnamefont {J.}~\bibnamefont {Viti}},
  \bibinfo {author} {\bibfnamefont {M.}~\bibnamefont {Carrega}}, \bibinfo
  {author} {\bibfnamefont {D.}~\bibnamefont {Rossini}},\ and\ \bibinfo {author}
  {\bibfnamefont {A.}~\bibnamefont {De~Luca}},\ }\bibfield  {title} {\bibinfo
  {title} {Energy transport in an integrable parafermionic chain via
  generalized hydrodynamics},\ }\href
  {https://doi.org/10.1103/PhysRevB.98.075421} {\bibfield  {journal} {\bibinfo
  {journal} {Phys. Rev. B}\ }\textbf {\bibinfo {volume} {98}},\ \bibinfo
  {pages} {075421} (\bibinfo {year} {2018})}\BibitemShut {NoStop}%
\bibitem [{\citenamefont {Rossini}\ \emph {et~al.}(2019)\citenamefont
  {Rossini}, \citenamefont {Carrega}, \citenamefont {Calvanese~Strinati},\ and\
  \citenamefont {Mazza}}]{rossini19}%
  \BibitemOpen
  \bibfield  {author} {\bibinfo {author} {\bibfnamefont {D.}~\bibnamefont
  {Rossini}}, \bibinfo {author} {\bibfnamefont {M.}~\bibnamefont {Carrega}},
  \bibinfo {author} {\bibfnamefont {M.}~\bibnamefont {Calvanese~Strinati}},\
  and\ \bibinfo {author} {\bibfnamefont {L.}~\bibnamefont {Mazza}},\ }\bibfield
   {title} {\bibinfo {title} {Anyonic tight-binding models of parafermions and
  of fractionalized fermions},\ }\href
  {https://doi.org/10.1103/PhysRevB.99.085113} {\bibfield  {journal} {\bibinfo
  {journal} {Phys. Rev. B}\ }\textbf {\bibinfo {volume} {99}},\ \bibinfo
  {pages} {085113} (\bibinfo {year} {2019})}\BibitemShut {NoStop}%
\bibitem [{\citenamefont {Helluin}\ \emph {et~al.}(2025)\citenamefont
  {Helluin}, \citenamefont {Pinto-Dias}, \citenamefont {Fontaine},
  \citenamefont {Ravets}, \citenamefont {Bloch}, \citenamefont {Minguzzi},\
  and\ \citenamefont {Canet}}]{helluin25}%
  \BibitemOpen
  \bibfield  {author} {\bibinfo {author} {\bibfnamefont {F.}~\bibnamefont
  {Helluin}}, \bibinfo {author} {\bibfnamefont {D.}~\bibnamefont {Pinto-Dias}},
  \bibinfo {author} {\bibfnamefont {Q.}~\bibnamefont {Fontaine}}, \bibinfo
  {author} {\bibfnamefont {S.}~\bibnamefont {Ravets}}, \bibinfo {author}
  {\bibfnamefont {J.}~\bibnamefont {Bloch}}, \bibinfo {author} {\bibfnamefont
  {A.}~\bibnamefont {Minguzzi}},\ and\ \bibinfo {author} {\bibfnamefont
  {L.}~\bibnamefont {Canet}},\ }\bibfield  {title} {\bibinfo {title} {Phase
  diagram and universal scaling regimes of two-dimensional exciton-polariton
  bose-einstein condensates},\ }\href {https://doi.org/10.1103/3gmk-xccn}
  {\bibfield  {journal} {\bibinfo  {journal} {Phys. Rev. Res.}\ }\textbf
  {\bibinfo {volume} {7}},\ \bibinfo {pages} {033103} (\bibinfo {year}
  {2025})}\BibitemShut {NoStop}%
\end{thebibliography}

\begin{thebibliography}{2}%
\makeatletter
\providecommand \@ifxundefined [1]{%
 \@ifx{#1\undefined}
}%
\providecommand \@ifnum [1]{%
 \ifnum #1\expandafter \@firstoftwo
 \else \expandafter \@secondoftwo
 \fi
}%
\providecommand \@ifx [1]{%
 \ifx #1\expandafter \@firstoftwo
 \else \expandafter \@secondoftwo
 \fi
}%
\providecommand \natexlab [1]{#1}%
\providecommand \enquote  [1]{``#1''}%
\providecommand \bibnamefont  [1]{#1}%
\providecommand \bibfnamefont [1]{#1}%
\providecommand \citenamefont [1]{#1}%
\providecommand \href@noop [0]{\@secondoftwo}%
\providecommand \href [0]{\begingroup \@sanitize@url \@href}%
\providecommand \@href[1]{\@@startlink{#1}\@@href}%
\providecommand \@@href[1]{\endgroup#1\@@endlink}%
\providecommand \@sanitize@url [0]{\catcode `\\12\catcode `\$12\catcode
  `\&12\catcode `\#12\catcode `\^12\catcode `\_12\catcode `\%12\relax}%
\providecommand \@@startlink[1]{}%
\providecommand \@@endlink[0]{}%
\providecommand \url  [0]{\begingroup\@sanitize@url \@url }%
\providecommand \@url [1]{\endgroup\@href {#1}{\urlprefix }}%
\providecommand \urlprefix  [0]{URL }%
\providecommand \Eprint [0]{\href }%
\providecommand \doibase [0]{https://doi.org/}%
\providecommand \selectlanguage [0]{\@gobble}%
\providecommand \bibinfo  [0]{\@secondoftwo}%
\providecommand \bibfield  [0]{\@secondoftwo}%
\providecommand \translation [1]{[#1]}%
\providecommand \BibitemOpen [0]{}%
\providecommand \bibitemStop [0]{}%
\providecommand \bibitemNoStop [0]{.\EOS\space}%
\providecommand \EOS [0]{\spacefactor3000\relax}%
\providecommand \BibitemShut  [1]{\csname bibitem#1\endcsname}%
\let\auto@bib@innerbib\@empty
\bibitem [{\citenamefont {Barmettler}\ \emph {et~al.}(2012)\citenamefont
  {Barmettler}, \citenamefont {Poletti}, \citenamefont {Cheneau},\ and\
  \citenamefont {Kollath}}]{barmettler12}%
  \BibitemOpen
  \bibfield  {author} {\bibinfo {author} {\bibfnamefont {P.}~\bibnamefont
  {Barmettler}}, \bibinfo {author} {\bibfnamefont {D.}~\bibnamefont {Poletti}},
  \bibinfo {author} {\bibfnamefont {M.}~\bibnamefont {Cheneau}},\ and\ \bibinfo
  {author} {\bibfnamefont {C.}~\bibnamefont {Kollath}},\ }\bibfield  {title}
  {\bibinfo {title} {Propagation front of correlations in an interacting bose
  gas},\ }\href {https://doi.org/10.1103/PhysRevA.85.053625} {\bibfield
  {journal} {\bibinfo  {journal} {Phys. Rev. A}\ }\textbf {\bibinfo {volume}
  {85}},\ \bibinfo {pages} {053625} (\bibinfo {year} {2012})}\BibitemShut
  {NoStop}%
\bibitem [{\citenamefont {Kwan}\ \emph {et~al.}(2024)\citenamefont {Kwan},
  \citenamefont {Segura}, \citenamefont {Li}, \citenamefont {Kim},
  \citenamefont {Gorshkov}, \citenamefont {Eckardt}, \citenamefont
  {Bakkali-Hassani},\ and\ \citenamefont {Greiner}}]{kwan24a}%
  \BibitemOpen
  \bibfield  {author} {\bibinfo {author} {\bibfnamefont {J.}~\bibnamefont
  {Kwan}}, \bibinfo {author} {\bibfnamefont {P.}~\bibnamefont {Segura}},
  \bibinfo {author} {\bibfnamefont {Y.}~\bibnamefont {Li}}, \bibinfo {author}
  {\bibfnamefont {S.}~\bibnamefont {Kim}}, \bibinfo {author} {\bibfnamefont
  {A.~V.}\ \bibnamefont {Gorshkov}}, \bibinfo {author} {\bibfnamefont
  {A.}~\bibnamefont {Eckardt}}, \bibinfo {author} {\bibfnamefont
  {B.}~\bibnamefont {Bakkali-Hassani}},\ and\ \bibinfo {author} {\bibfnamefont
  {M.}~\bibnamefont {Greiner}},\ }\bibfield  {title} {\bibinfo {title}
  {Realization of one-dimensional anyons with arbitrary statistical phase},\
  }\href {https://doi.org/10.1126/science.adi3252} {\bibfield  {journal}
  {\bibinfo  {journal} {Science}\ }\textbf {\bibinfo {volume} {386}},\ \bibinfo
  {pages} {1055} (\bibinfo {year} {2024})}\BibitemShut {NoStop}%
\end{thebibliography}
\end{document}